\documentclass[%
 aip,
 jmp,%
 amsmath,amssymb,
 reprint,%
]{revtex4-1}

\usepackage{graphicx}
\graphicspath{{Figures/}}
\usepackage{dcolumn}
\usepackage{bm}
\usepackage{subfigure}

\usepackage{algorithm}
\usepackage{algpseudocode}
\usepackage{amsmath,amssymb,amsfonts}
\usepackage{graphics}
\usepackage{color}

\begin{document}


\title{A dynamic closure modeling framework for model order reduction of geophysical flows}

\author{Sk. M. Rahman}
\author{S. E. Ahmed}
\author{O. San}%
 \email{osan@okstate.edu}
\affiliation{ 
School of Mechanical \& Aerospace Engineering, Oklahoma State University, Stillwater, Oklahoma - 74078, USA.
}%

\date{\today}

\begin{abstract}
In this paper, a dynamic closure modeling approach has been derived to stabilize the projection-based reduced order models in the long-term evolution of forced-dissipative dynamical systems. To simplify our derivation without losing generalizability, the proposed reduced order modeling (ROM) framework is first constructed by Galerkin projection of the single-layer quasi-geostrophic equation, a standard prototype of large-scale general circulation models, onto a set of dominant proper orthogonal decomposition (POD) modes. We then propose an eddy viscosity closure approach to stabilize the resulting surrogate model considering the analogy between large eddy simulation (LES) and truncated modal projection. Our efforts, in particular, include the translation of the dynamic subgrid-scale model into our ROM setting by defining a test truncation similar to the test filtering in LES.  The a posteriori analysis shows that our approach is remarkably accurate, allowing us to integrate simulations over long time intervals at a nominally small computational overhead.    
\end{abstract}

\maketitle


\section{Introduction} 
\label{sec:intro}

High-fidelity numerical simulations are crucial for reliable predictions, control and diagnostics. Thanks to the huge advancement in computational power, in terms of speed (e.g., number of arithmetic operations per second) and memory, computational fluid dynamics has witnessed substantial development during the last century. This includes using much finer numerical resolution and fewer approximations. However, there are still some situations in which computational resources cannot meet the requirements for feasible simulations. This is evident, especially in numerical weather predictions, where accurate simulations are solely applicable to regional weather models, while solution of global models is still restricted to relatively coarse grids \cite{kalnay2003atmospheric,powers2017weather}.  

Although the progress in computer power and performance has been adequately following Moore's law \cite{moore1965,moore1975} during the past decades, it is becoming increasingly obvious in worldwide semiconductor industry that it is nearing its end \cite{powell2008quantum,kumar2015fundamental,waldrop2016chips,kumar2018end}. Therefore, the development of efficient algorithms that elevate the maximum attainable quality of numerical simulations with the available resources, or at least reduce the computational cost of traditional simulations, has become a must. The latter is particularly important when multiple forward simulations are required, like those encountered in inverse problems \cite{daescu2007efficiency,navon2009data,cao2007reduced,he2011use,houtekamer1998data,houtekamer2001sequential,bennett2005inverse,evensen2009data,law2012evaluating,buljak2011inverse}. Reduced order modeling (ROM), also known as model order reduction, is such a way of representing high-dimensional systems with much lower-dimensional (but dense) systems, resulting in substantial reduction in computational cost while keeping output quality within acceptable range \cite{quarteroni2015reduced, taira2017modal}. This is feasible due to the fact that most high-dimensional complex systems, basically follow low-dimensional characteristic dynamics. For example, complex fluid flows often consist of superposition of spatially or temporally developing coherent structures, either growing/decaying with a specific rate, oscillating with constant frequency or containing the largest possible kinetic energy. The evolution of such structures is responsible for the bulk mass, momentum, and energy transfer. Therefore, the development of reduced order surrogate models through extracting these underlying characteristics, would be an effective way to reduce the computational cost of numerical simulations and address more complex problems.

Among different ROM techniques, snapshot-based projection methods are particularly important where the time response of a system, either recorded from experiments or high-fidelity numerical simulations and given a certain input, is assumed to contain the essential behavior of that system. Proper orthogonal decomposition (POD), also known as principal component analysis, is a widely popular technique for snapshot-based ROMs \cite{antoulas2000survey,chinesta2011short,benner2015survey,rowley2017model}, and it has been introduced to the fluid dynamics community as a mathematical technique to extract coherent structures from turbulent flow fields \cite{lumley1967structure}.  

In general, POD computes a set of orthonormal basis vectors which describe the main directions (modes), by which the given dataset is characterized, in the $L_2$ sense \cite{berkooz1993proper}. Based on the energy cascade between different modes, the most energetic POD modes are selected to generate a reduced order system. POD coupled with the Galerkin projection has been considered an efficient approach to generate ROMs for linear and nonlinear systems \cite{ito1998reduced,iollo2000stability,rowley2004model,pinnau2008model,luchtenburg2009introduction,sachs2010pod,puzyrev2019pyrom}. It has been applied in a large number of problems involving fluid flows during the past few decades where the governing equations are projected onto these selected modes. While the high-dimensional approximation using standard discretization techniques often generates spaces with millions of degrees of freedoms, the reduced spaces spanned by ROMs are typically of order 100 or smaller \cite{milk2016pymor}. In practice, ROM approximation can lead to speedups of several orders of magnitude as well as great reduction in memory requirements. In fluid dynamics applications, the resulting dense system consists of triadic interactions due to the quadratic nonlinearity with an order of $O(R^3)$ computational load, where $R$ refers to the retained number of modes.  

However, it has been noted that truncated modes often contribute to the evolving dynamics of complex multidimensional turbulent flows, especially encountered in geophysical systems \cite{lassila2014model}, resulting in diverging POD-based solution. This leads to either increasing the number of selected modes to better embed the underlying system, or sacrificing results quality. The latter is, of course, unacceptable and would ruin the reliability and applicability of such models. On the other hand, increasing number of modes beyond some threshold, would increase the computational cost of solving ROMs in a way that approaches, or even exceeds, the cost of the original full-order model. Moreover, Rempfer \cite{rempfer2000low} has shown that a complete set of POD modes is not sufficient for a POD-Galerkin model to reproduce the full order dynamics accurately, even for many non-turbulent flows. He found that adding just small perturbations to the flow field (like those created by a small numerical error in the integration of the ODEs) would result in a ROM that, in general, may not faithfully represent the full order model anymore, and the dynamics of the POD-Galerkin model could show instabilities while the true dynamics of the system is stable. Noack et al. \cite{noack2003hierarchy} also reported the same observations. They proposed including an extra 'shift-mode' that represents the shift of short-term averaged flow away from the POD space such that the Galerkin approximation also includes an accurate representation of the unstable steady solution.

Generally, POD modes resolve the production much better than the dissipation, leading to an excess production of turbulent kinetic energy in the POD subspace \cite{cordier2013identification}. Several studies have approached this weak dissipation through the introduction of eddy-viscosity terms. Sirisup and Karniadakis \cite{sirisup2004spectral} proposed a dissipative model based on a spectral viscosity diffusion convolution operator to improve the long-term predictions of Galerkin-based ROMs. In their approach viscosity amplitude decreases with the mode number, and it was shown to guarantee a non-oscillatory behavior except for some negligible bounded oscillations. Cordier et al. \cite{cordier2013identification} reported that constant eddy viscosity, even carefully calibrated, is nonphysical and would lead to incorrect scaling characteristics. They proposed an improved nonlinear eddy-viscosity model using data assimilation techniques (4D-Var in particular).

In this paper, we aim at proposing an automated framework to produce stable Galerkin projection based reduced order models, using a sufficiently small number of modes, without sacrificing much accuracy through the introduction of closure ideas \cite{couplet2003intermodal,kalb2007intrinsic,kalashnikova2010stability,wang2011two,wang2012proper,amsallem2012stabilization,balajewicz2012stabilization,wells2017evolve,borggaard2011artificial,akhtar2012new,san2013proper,san2015stabilized}. Dynamic eddy viscosity based closure models have been applied in large eddy simulation (LES) area to provide numerical stabilization as well as statistical fidelity preservation using an explicit test filtering procedure \cite{germano1991dynamic, vreman2004eddy}. Using dynamic LES and ROM analogy, the chief novelty of this paper is to introduce a ``test truncation" mechanism to generate stable, self-adapted, dynamic ROMs for estimating long term dynamics of forced-dissipative turbulent flows.

To the authors' knowledge, the application of such dynamic closure models in ROMs is very limited, and the current work is an effort for such incorporation. Following LES ideology, the eddy viscosity concept is thought to provide an efficient framework to account for the unrepresented scales due to intense mode truncation. Eddy viscosity is computed on the fly using test truncation idea, similar to test filtering in LES. To assess our idea, the barotropic vorticity equation (BVE) is selected as our test bed. It is a widely used mathematical model to study the forced-dissipative large scale ocean circulation problems, also known as the single-layer quasi-geostrophic (QG) model, first introduced by Jule Charney \cite{charney1948scale, charney1949physical}. We found that the proposed dynamic closure approach gives stabilized results over longer time intervals, compared with regular ROMs, with a negligible computational overhead. 

The rest of the paper is organized as follows: Sec.~\ref{sec:bve} describes the governing equations briefly for the adopted test bed to generate snapshots; Sec.~\ref{sec:DROM} is devoted to the description and derivation of the dynamic eddy viscosity closure in ROM; in Sec.~\ref{sec:results}, we present and discuss our results of the proposed framework; and Sec.~\ref{sec:con} provides a summary of this study and the conclusions drawn from it. The numerical discretization schemes used for spatial and temporal derivative approximations as well as the generation and selection criteria of POD modes are specified in the Appendix at the end of this manuscript.

\section{Barotropic Vorticity Equation Model}
\label{sec:bve}

Atmospheric and oceanographic flows often take place over horizontal length scales, much larger than their vertical length scale. Therefore, they can be adequately described by using the shallow water equations (SWE). The single layer two-dimensional QG equation is an approximation of the SWE, filtering the inertia-gravity waves, under the following assumptions \cite{vallis2006atmospheric}:
\begin{itemize}
	\item Rossby number, Ro is small, such that inertial forces are an order of magnitude smaller than the Coriolis and pressure forces,
	\item Horizontal scale of motion is the same order of magnitude as the deformation scale, implying that the variations in fluid depth are small compared to its total depth,
	\item Variations in the Coriolis parameter are small,
	\item The timescale is the advective timescale, hence the gravity waves, which evolve on a short timescale, are filtered out. 
\end{itemize}

Much of the world’s ocean circulation is wind-driven in large-scale. Therefore, wind-driven flows of mid-latitude ocean basins have been studied by modelers using idealized single- and double-gyre wind forcing which helps in understanding various aspects of ocean dynamics, including the role of mesoscale eddies and their effect on mean circulation. The BVE describing the single-layer QG equation with dissipative and forcing terms is one of the most commonly used models for the double-gyre wind-driven geophysical flows \cite{majda2006nonlinear}. 

The BVE model is a simplified version of the more general primitive equations used in operational weather forecast centers \cite{kalnay2003atmospheric}, making it a suitable model for testing new ideas. Detailed discussions on different underlying mechanisms and formulations have been presented in literature \cite{holland1980example, munk1982observing, griffa1989wind, cummins1992inertial,greatbatch2000four,nadiga2001dispersive}. In dimensionless vorticity-streamfunction formulation, using $\beta$-plane assumption reasonable for most oceanic flows, the forced-dissipative BVE can be written as follows:
\begin{equation}\label{eq:nbve}
\frac{\partial \omega}{\partial t} + J(\omega,\psi) -\frac{1}{\mbox{Ro}}\frac{\partial \psi}{\partial x} = \frac{1}{\mbox{Re}}\nabla^2 \omega + \frac{1}{\mbox{Ro}}\sin(\pi y),
\end{equation}
where $\nabla^2$ refers to the Laplacian in two-dimensions, $\omega$ and $\psi$ are the kinematic vorticity and streamfunction, respectively, defined as:
\begin{equation} \label{eq:omegadfn}
\omega = \nabla \times \mathbf{u}, 
\end{equation}
\begin{equation} \label{eq:psidfn}
\mathbf{u} = \nabla \times \psi \hat{k},
\end{equation}
where $\mathbf{u}$ is the two-dimensional velocity field and $\hat{k}$ refers to the unit vector perpendicular to the horizontal plane.
The nonlinear advection term in Eq.~(\ref{eq:nbve}) is given by the Jacobian
\begin{equation}\label{eq:jac}
J(\omega,\psi) =  \frac{\partial \psi}{\partial y}\frac{\partial \omega}{\partial x} - \frac{\partial \psi}{\partial x}\frac{\partial \omega}{\partial y}.
\end{equation}

Eq.~(\ref{eq:nbve}) has two dimensionless parameters, Reynolds number, Re and Rossby number, Ro, which are related to the characteristic length and velocity scales in the following way:
\begin{equation}\label{eq:ReRo1}
\mbox{Re} = \frac{V L}{\nu}, \quad \mbox{Ro} = \frac{V}{\beta L^2},
\end{equation}
where $\nu$ is the horizontal eddy viscosity of the BVE model and $\beta$ is the gradient of the Coriolis parameter at the basin center ($y = 0$). $L$ is the basin length scale and $V$ is the velocity scale, also known as the Sverdrup velocity\cite{sverdrup1947wind}, and is given by
\begin{equation}\label{eq:sverdrup}
V = \frac{\tau_0}{\rho H}\frac{\pi}{\beta L},
\end{equation}
where $\tau_0$ is the maximum amplitude of the double-gyre wind stress, $\rho$ is the mean fluid density, and $H$ is the mean depth of the ocean basin.

Despite not being explicitly represented in Eq.~(\ref{eq:nbve}), there are two important relevant physical parameters, the Rhines scale, $\delta_I$ and the Munk scale, $\delta_M$ which are the nondimensional boundary layer
thicknesses for the inertial and viscous (Munk) layer of the basin geometry, respectively. As a physical interpretation of these parameters in BVE model,  $\delta_I$ accounts for the strength of nonlinearity and $\delta_M$ is a measure of dissipation strength. $\delta_I$ and $\delta_M$ can be defined as 
\begin{equation}
\dfrac{\delta_I}{L}=\left(\dfrac{V}{\beta L^2}\right)^{\frac{1}{2}} , \quad
\dfrac{\delta_M}{L}=\left(\dfrac{\nu}{\beta L^3}\right)^{\frac{1}{3}}
\end{equation}
and are related to Ro and Re by the following relations
\begin{equation}
\dfrac{\delta_I}{L}=\left(\mbox{Ro}\right)^{\frac{1}{2}} , \quad \dfrac{\delta_M}{L} = \left(\dfrac{\mbox{Ro}}{\mbox{Re}}\right)^{\frac{1}{3}}.
\end{equation}
Finally, in order to satisfy the incompressibility constraint, the  vorticity and streamfunction are related through the following Poisson equation:
\begin{equation}\label{eq:pois}
\nabla^2 \psi = -\omega.
\end{equation}

Following \cite{holm2003modeling,san2011approximate}, a four-gyre circulation problem is considered as a benchmark for oceanic flow to generate numerical data. Since ocean circulation models where the Munk and Rhines scales are close to each other, like the QG model, remain time dependent rather being converged to a steady state as time approaches to infinity \cite{medjo2000numerical}, numerical computations of these models are conducted in a statistically steady state, also known as the quasi-stationary state. Hence, in our study, we utilize numerical schemes suited for simulation of such type of ocean models and for long-time integration. In our full order model (FOM) simulations, we use a second-order accurate kinetic energy and enstrophy conserving Arakawa finite difference scheme \cite{arakawa1966computational}. The derivatives in the linear terms are also approximated using the standard second-order finite differences. Our time advancement scheme is given by the classical total variation diminishing third-order accurate Runge-Kutta scheme \cite{gottlieb1998total}. Details of the numerical schemes are given in Appendix. 

To close the problem, boundary and initial conditions need to be specified. Following previous studies, we use slip boundary condition for the velocity, which implies homogeneous Dirichlet boundary condition for the vorticity. Also, the impermeability boundary condition forces homogeneous Dirichlet boundary condition for the streamfunction:
\begin{equation}\label{eq:boundary}
\omega|_\Gamma = \psi|_\Gamma = 0,
\end{equation}
where $\Gamma$ refers to all boundary coordinates. As an initial state, we start our computations from a quiescent state (i.e., $\omega_{t=0} = \psi|_{t=0} = 0$) and integrate the model until a statistically steady state is obtained, i.e., the wind forcing, dissipation, and Jacobian (eddy flux of potential vorticity) balance each other.

\section{Dynamic Closure Modeling for Reduced Order Models}
\label{sec:DROM}

The main idea of this paper is that the effect of truncated modes can be approximated dynamically. To illustrate this online ROM closure idea, we first rewrite the governing equation as 
\begin{equation}\label{eq:nbve1}
\frac{\partial \omega}{\partial t} =- J(\omega,\psi) +\frac{1}{\mbox{Ro}}\frac{\partial \psi}{\partial x} + \frac{1}{\mbox{Re}}\nabla^2 \omega + \frac{1}{\mbox{Ro}}\sin(\pi y),
\end{equation}
where we approximate the prognostic variable (i.e., kinematic vorticity in this case) using the most energetic $R$ POD modes 
\begin{equation}\label{eq:ew}
    \omega(x,y,t) = \bar{\omega}(x,y) + \sum_{k=1}^{R} \alpha_{k}(t)\phi_{k}(x,y),
\end{equation}
where $\bar{\omega}(x,y)$ refers to the mean vorticity field of the training data set of snapshots, $\alpha_{k}(t)$ is the $k$th time-dependent coefficient and $\phi_{k}(x,y)$ is the $k$th spatial POD mode for the fluctuating vorticity field. The derivation of such POD modes is detailed in Appendix. We note that the POD modes are orthonormal (both orthogonal and normalized), i.e.,
\begin{equation}\label{eq:inner1}
    \int_{\Omega} \phi_{i}(x,y) \phi_{j}(x,y) dx dy = \delta_{ij},
\end{equation}
where $\Omega$ is the entire spatial domain and $\delta_{ij}$ is the Kronecker delta defined by
\begin{equation}\label{eq:inner2}
 \delta_{ij} = \begin{cases} 1, & \text{if } i=j,\\ 0, & \text{if } i\neq j. \end{cases}
\end{equation}
To simplify our notation we use the following angle-parenthesis definition for the inner product
\begin{equation}\label{eq:inner}
    \int_{\Omega} f(x,y) g(x,y) dx dy = \langle f; g\rangle,
\end{equation}
and hence $ \langle \phi_{i}; \phi_{j}\rangle = \delta_{ij}$. 
Since the vorticity and stream function are related through the kinematic relationship given by Eq.~(\ref{eq:pois}), we can expand the stream function using the same time-dependent coefficient,
\begin{equation}\label{eq:epsi}
    \psi(x,y,t) = \bar{\psi}(x,y) + \sum_{k=1}^{R} \alpha_{k}(t)\theta_{k}(x,y),
\end{equation}
where $\bar{\psi}(x,y)$ refers to the mean streamfunction field of the training data set of snapshots, and $\theta_{k}(x,y)$ is the $k$th spatial POD mode for the streamfunction, which can be obtained through solving the following Poisson equations (offline computing):
\begin{align}\label{eq:psi}
    \nabla^2 \bar{\psi}(x,y) &= -\bar{\omega}(x,y), \\
    \nabla^2 \theta_{k}(x,y) &= -\phi_{k}(x,y), \quad k=1,2, ..., R.
\end{align}
We note that the POD modes for streamfunction doesn't need to be orthonormal since the streamfunction is not a prognostic variable. Substituting Eq.~(\ref{eq:ew}) and Eq.~(\ref{eq:epsi}) into Eq.~(\ref{eq:nbve1}), an orthogonal Galerkin projection is then performed by multiplying Eq.~(\ref{eq:nbve1}) with the spatial POD modes $\phi_k(x,y)$, and integrating over the entire domain $\Omega$. The resulting dense dynamical system for $\alpha_k$ can be written as
\begin{equation} \label{eq:grom1}
  \frac{d \alpha_k}{d t} = \mathfrak{B}_{k} + \sum_{i=1}^{R} \mathfrak{L}^{i}_{k}\alpha_{i} + \sum_{i=1}^{R}\sum_{j=1}^{R} \mathfrak{N}^{ij}_{k}\alpha_{i}\alpha_{j}, \quad k=1,2, ..., R,
\end{equation}
where the predetermined model coefficients can be computed by the following numerical integration (offline computing) 
\begin{eqnarray}
  & & \mathfrak{B}_{k} = \big\langle  - J(\bar{\omega},\bar{\psi}) +\frac{1}{\mbox{Ro}}\frac{\partial \bar{\psi}}{\partial x} + \frac{1}{\mbox{Re}}\nabla^2 \bar{\omega} + \frac{1}{\mbox{Ro}}\sin(\pi y) ; \phi_{k} \big\rangle , \nonumber \\
  & & \mathfrak{L}^{i}_{k} = \big\langle - J(\bar{\omega},\theta_{i}) -  J(\phi_{i},\bar{\psi}) +  \frac{1}{\mbox{Ro}}\frac{\partial \theta_{i}}{\partial x} + \frac{1}{\mbox{Re}}\nabla^2 \phi_{i}; \phi_{k} \big\rangle  , \nonumber\\
  & &  \mathfrak{N}^{ij}_{k} =  \big\langle -J(\phi_{i},\theta_{j}); \phi_{k} \big\rangle. \label{eq:roma7}
\end{eqnarray}
To complete the dynamical system given by Eq.~(\ref{eq:grom1}), the initial conditions for $\alpha_k(t)$ may be obtained by the following projection
\begin{equation}\label{eq:ic1}
\alpha_{k}(t_{0}) = \big\langle \omega(x,y,t_{0}) - \bar{\omega}(x,y); \phi_{k} \big\rangle,
\end{equation}
where $\omega(x,y,t_{0})$ is the vorticity field specified at initial time $t_{0}$. The standard Galerkin projection ROM given by Eq.~(\ref{eq:grom1}), denoted as ROM-G in this study, often yields unstable solutions when the largest $R$ modes might not adequately capture the system's dynamics \cite{couplet2003intermodal,kalb2007intrinsic,bergmann2009enablers,kalashnikova2010stability,wang2011two,carlberg2011efficient,wang2012proper,amsallem2012stabilization,balajewicz2012stabilization,balajewicz2013low,lassila2013model,san2014proper,baiges2015reduced,wells2017evolve,xie2017approximate}. As we will illustrate in our numerical examples, this is particularly true for turbulent flows. Using an eddy viscosity approach, the stabilization of the ROM, using the analogy between ROM and LES, can be achieved by adding a regularization term to the governing equation \cite{san2015stabilized}
\begin{equation}\label{eq:nbve2}
\frac{\partial \omega}{\partial t} =- J(\omega,\psi) +\frac{1}{\mbox{Ro}}\frac{\partial \psi}{\partial x} + \frac{1}{\mbox{Re}}\nabla^2 \omega + \frac{1}{\mbox{Ro}}\sin(\pi y) + \nu_e \nabla^2 \omega,
\end{equation}
where $\nu_e$ refers to the eddy viscosity. To account for the effects of the truncated modes, following the similar Galerkin projection approach we may obtain a regularized ROM model
\begin{equation} \label{eq:srom2}
  \frac{d \alpha_k}{d t} = \mathfrak{B}_{k} +\tilde{\mathfrak{B}}_{k}+ \sum_{i=1}^{R} (\mathfrak{L}^{i}_{k}+\tilde{\mathfrak{L}}^{i}_{k})\alpha_{i} + \sum_{i=1}^{R}\sum_{j=1}^{R} \mathfrak{N}^{ij}_{k}\alpha_{i}\alpha_{j},
\end{equation}
where the additional two terms can be written as
\begin{align}\label{EVstab4}
  \tilde{\mathfrak{B}}_{k} &= \langle \nu_{e} \nabla^2 \bar{\omega};\phi_k\rangle , \nonumber\\
  \tilde{\mathfrak{L}}^{i}_{k} &= \langle\nu_{e}  \nabla^2 \phi_i; \phi_k\rangle .
\end{align}
The free stabilization parameter $\nu_e$ may be simply considered as a given empirical constant \cite{aubry1988dynamics,wang2012proper}. This empirical eddy viscosity idea may improve by supposing that the amount of dissipation is not identical for all the POD modes \cite{rempfer1991,cazemier1997,san2014proper}. It has been, however, shown that finding an optimal value for this parameter significantly improves the predictive performance of ROMs \cite{san2014proper,san2015stabilized}. Therefore, the main novelty of the present study is
the derivation of an automated approach to estimate this $\nu_e$ parameter dynamically at each time step (online computing). An alternative dynamic determination of $\nu_e$ has been presented by San and Maulik using a supervised neural network approach \cite{san2018extreme}. However, our effort in this paper aims at developing a mathematical model based on the idea of the ``test truncation", translating the idea of the test filtering approach \cite{germano1991dynamic} in dynamic LES subgrid-scale models into the ROM setting. 

To demonstrate our approach, we first utilize the test truncation to Eq.~(\ref{eq:srom2}) considering less number of modes $\tilde{R}$ (i.e., $\tilde{R} < R $)  
\begin{equation} \label{eq:rom3}
  \frac{d \tilde{\alpha}_k}{d t} = \mathfrak{B}_{k} +\tilde{\mathfrak{B}}_{k}+ \sum_{i=1}^{\tilde{R} } (\mathfrak{L}^{i}_{k}+\tilde{\mathfrak{L}}^{i}_{k})\tilde{\alpha}_{i} + \sum_{i=1}^{\tilde{R}}\sum_{j=1}^{\tilde{R} } \mathfrak{N}^{ij}_{k}\tilde{\alpha}_{i}\tilde{\alpha}_{j},
\end{equation}
where the $\tilde{\alpha}_k$ refers to our approximation for $k^{th}$ time-dependent coefficient on a test truncated space and we subtract the resulting model given by Eq.~(\ref{eq:rom3}) from the original model given by Eq.~(\ref{eq:srom2}) to yield the difference equation at each $k^{th}$ mode
\begin{align} \label{eq:dif}
 & \sum_{i=1}^{R} (\mathfrak{L}^{i}_{k}+\tilde{\mathfrak{L}}^{i}_{k})\alpha_{i} + \sum_{i=1}^{R}\sum_{j=1}^{R} \mathfrak{N}^{ij}_{k}\alpha_{i}\alpha_{j}  \nonumber \\ - & \sum_{i=1}^{\tilde{R} } (\mathfrak{L}^{i}_{k}+\tilde{\mathfrak{L}}^{i}_{k})\tilde{\alpha}_{i} -  \sum_{i=1}^{\tilde{R}}\sum_{j=1}^{\tilde{R} } \mathfrak{N}^{ij}_{k}\tilde{\alpha}_{i}\tilde{\alpha}_{j} =   \frac{d \alpha_k}{d t} - \frac{d \tilde{\alpha}_k}{d t}  ,
\end{align}
and we can approximate Eq.~(\ref{eq:dif}) by the modal scale similarity hypothesis
\begin{align} \label{eq:dif2}
 & \sum_{i=1}^{R} \mathfrak{L}^{i}_{k} + \sum_{i=1}^{R}\sum_{j=1}^{R} \mathfrak{N}^{ij}_{k}\alpha_{i}\alpha_{j}  -  \sum_{i=1}^{\tilde{R} } \mathfrak{L}^{i}_{k} -  \sum_{i=1}^{\tilde{R}}\sum_{j=1}^{\tilde{R} } \mathfrak{N}^{ij}_{k}\alpha_{i}\alpha_{j} \nonumber \\ & =-\sum_{i=1}^{R}  \tilde{\mathfrak{L}}^{i}_{k}\alpha_{i} + \sum_{i=1}^{\tilde{R}}\tilde{\mathfrak{L}}^{i}_{k}\alpha_{i} ,
\end{align}
where we assume that $\tilde{\alpha}_k \approx \alpha_k$ when we use the eddy viscosity closure. Therefore, using the definitions given by Eq.~(\ref{EVstab4}), we can rewrite Eq.~(\ref{eq:dif2}) as
\begin{align} \label{eq:dif3}
H_k = \nu_e M_k ,
\end{align}
where
\begin{align} \label{eq:dif4}
H_k & = \sum_{i=\tilde{R}+1}^{R} \mathfrak{L}^{i}_{k} \alpha_{i} + \sum_{i=1}^{R}\sum_{j=1}^{R} \mathfrak{N}^{ij}_{k}\alpha_{i}\alpha_{j}   -  \sum_{i=1}^{\tilde{R}}\sum_{j=1}^{\tilde{R} } \mathfrak{N}^{ij}_{k}\alpha_{i}\alpha_{j} \nonumber \\
M_k & =  - \sum_{i=\tilde{R}+1}^{R} \tilde{L}^{i}_{k}\alpha_{i}  ,
\end{align}
in which the predetermined coefficients of $\tilde{L}^{i}_{k}$ can be given (offline computing) 
\begin{align}\label{eq:c5}
  \tilde{L}^{i}_{k} = \langle \nabla^2 \phi_i; \phi_k\rangle ,
\end{align}
where we assume that $\nu_e$ is treated as constant locally in Eq.~(\ref{EVstab4}), i.e., frozen eddy viscosity field hypothesis expressed by $\tilde{\mathfrak{L}}^{i}_{k} = \nu_e \tilde{L}^{i}_{k}$. Similar to the approach driven by Lilly \cite{lilly1992proposed} for LES, we propose a least-squares based estimation for $\nu_e$ in Eq.~(\ref{eq:dif3}) where we define the error at each mode, $E_k=H_k- \nu_e M_k$. Once we square this error term
\begin{align}\label{eq:dyn0}
  E_{k}^{2} = H_{k}^{2} - 2 \nu_e H_k M_k + \nu_{e}^{2}M_{k}^{2},
\end{align}
then the eddy viscosity coefficient in our ROM model can be computed by minimizing the sum of square errors with respect to the free model parameter $\nu_e$ to obtain
\begin{align}\label{eq:dyn}
  \frac{\partial (\sum_{k=1}^{\tilde{R}}E_{k}^{2})}{\partial (\nu_e)} = -2 \sum_{k=1}^{\tilde{R}} H_k M_k + 2\nu_e \sum_{k=1}^{\tilde{R}} M_k M_k.
\end{align}
The right-hand side of the above equation becomes zero when the error is minimized to give us finally the following expression for the eddy viscosity coefficient
\begin{align}\label{eq:dyn1}
  \nu_e= \frac{\sum_{k=1}^{\tilde{R}} H_k M_k}{\sum_{k=1}^{\tilde{R}} M_k M_k},
\end{align}
where $H_k$ and $M_k$ are computed by Eq.~(\ref{eq:dif4}) at each time step (online computing). To provide always a positive eddy viscosity in our ROM simulations, we have also applied the following clipping rule \cite{sagaut2006large} 
\begin{align}\label{eq:dyn2}
  \nu_e= \max\Big(0,\frac{\sum_{k=1}^{\tilde{R}} H_k M_k}{\sum_{k=1}^{\tilde{R}} M_k M_k}\Big).
\end{align}
This completes the derivation of our dynamic closure model for ROM settings. We denote our proposed model as ROM-D (i.e., Eq.~(\ref{eq:srom2}) equipped with  Eq.~(\ref{eq:dyn2})).

\section{Numerical experiments}
\label{sec:results}

In this section, we primarily focus on the comparative performance of the standard ROM-G model and our proposed ROM-D model in estimating the flow behavior at different flow conditions. In addition, we also assess the robustness and prediction capability of the ROM-D model through the extrapolatory prediction and sensitivity tests. To produce the ideal data set and simulation results for our evaluations, we select the single-layer QG model as our benchmark test case which has been appeared in numerous studies as test problem \cite{cushman2011introduction,holm2003modeling,san2011approximate,cummins1992inertial,greatbatch2000four}. Indeed, the QG test problem comes with a great challenge of capturing wide range of scales and complex flow behavior on coarse spatial resolutions, for example, resolving the four-gyre circulation (in the time mean) \cite{greatbatch2000four}, which makes this problem a suitable test bed to evaluate the capability of ROMs. 

In the current study, we present our performance evaluations of the aforementioned reduced order methodologies based upon four distinct numerical experiments. In the first experiment, we investigate the predictive performance of both modeling frameworks at lower (Re, Ro) combination using the data snapshots extracted from a $256 \times 512$ resolution FOM simulation at the same flow condition. Next, we perform the similar analysis for comparatively higher (Re, Ro) combination in the second experiment. Finally, in the last two experiments, we test the extrapolatory predictive performance of the models using the data snapshots collected at higher/lower (Re, Ro) combination to predict a flow field at lower/higher (Re, Ro) combination, respectively. Throughout the analyses, we utilize the true projection of the FOM simulations as the baseline or reference for all the relative comparisons between ROMs. 

\subsection{FOM simulation and data snapshots collection}

\begin{figure*}[htbp]
\centering
\mbox{
\subfigure{\includegraphics[width=0.7\textwidth]{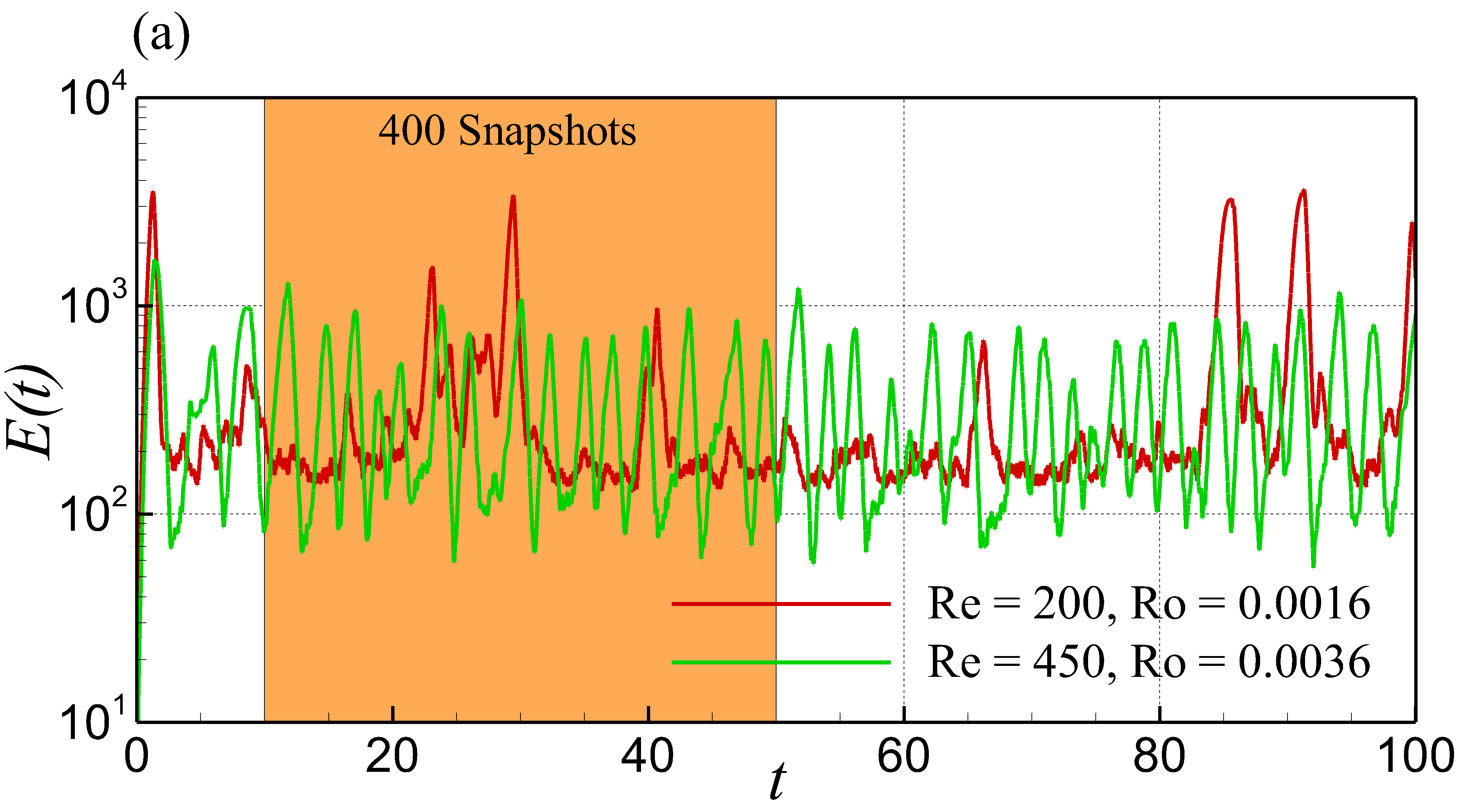}}
}\\
\mbox{
\subfigure{\includegraphics[width=0.7\textwidth]{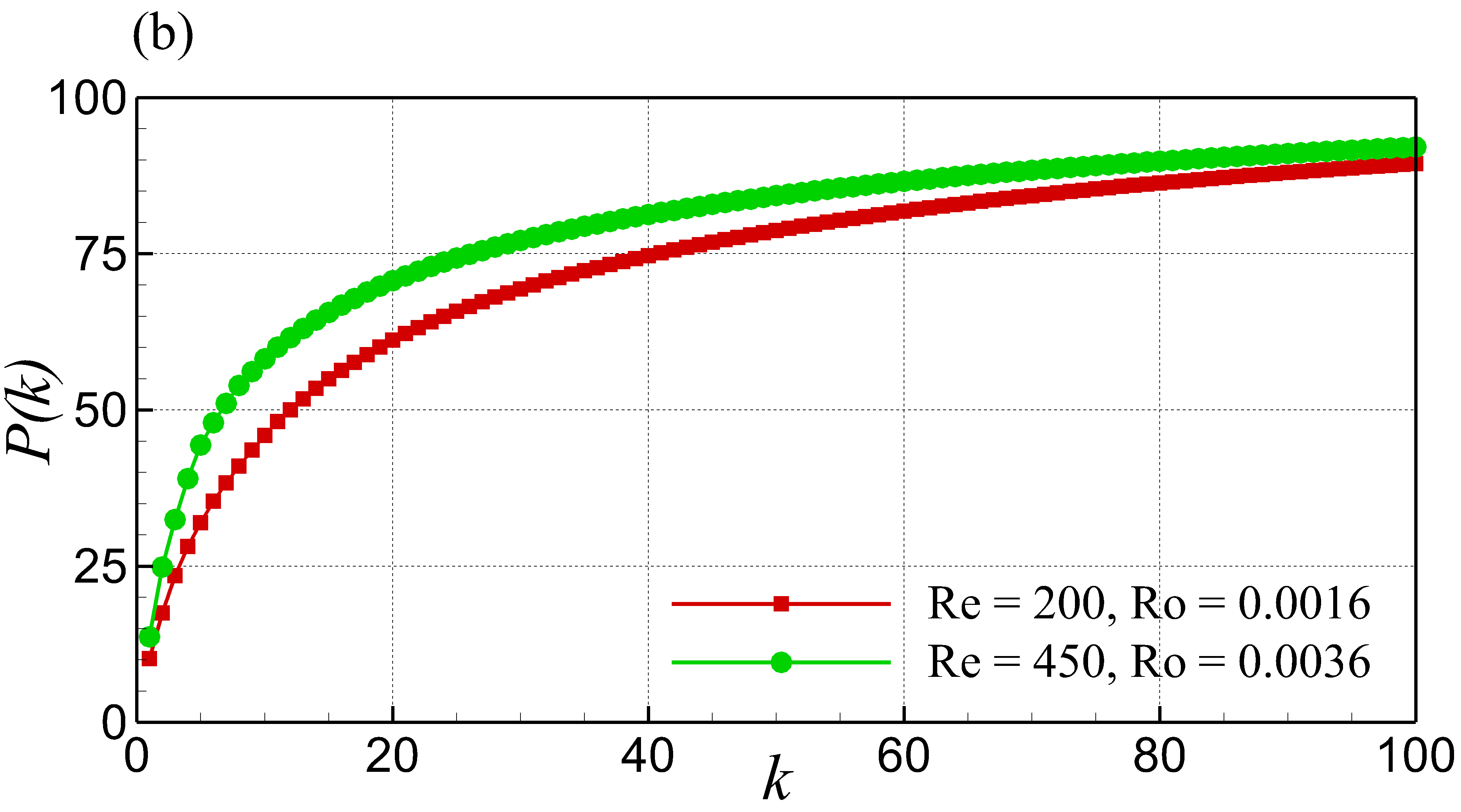}}
}
\caption{Graphical representation of (a) the snapshots selection (shaded in orange) from the time histories of total kinetic energy for various Re and Ro combinations and (b) the POD percentage energy accumulation with respect to modal index.}
\label{fig:1}
\end{figure*}

The computational domain of our test problem is $(x,y) \in [0, 1]\times[-1, 1]$. The FOM simulation is conducted starting from $t = 0$ to $t = 100$ using a fixed time step of $\Delta t = 2.5\times10^{-5}$ on a Munk layer resolving $256\times512$ grid resolution (i.e., consisting of about four grid points in the Munk scale, i.e., $\delta_M/L = 0.02$). In FIG.~\ref{fig:1}, we present the plots of time histories of the basin integrated total kinetic energy for both flow configurations, (i) Re $=200$, Ro $=0.0016$ and (ii) Re $=450$, Ro $=0.0036$, which can be calculated by:
\begin{align}
  E(t) = \frac{1}{2} \int_{\Omega} \left(\Big(\frac{\partial \psi}{\partial x}\Big)^2 + \Big(\frac{\partial \psi}{\partial y}\Big)^2 \right) dx dy.
\end{align}

\begin{figure*}[htbp]
\centering
\mbox{
\subfigure{\includegraphics[width=0.2\textwidth]{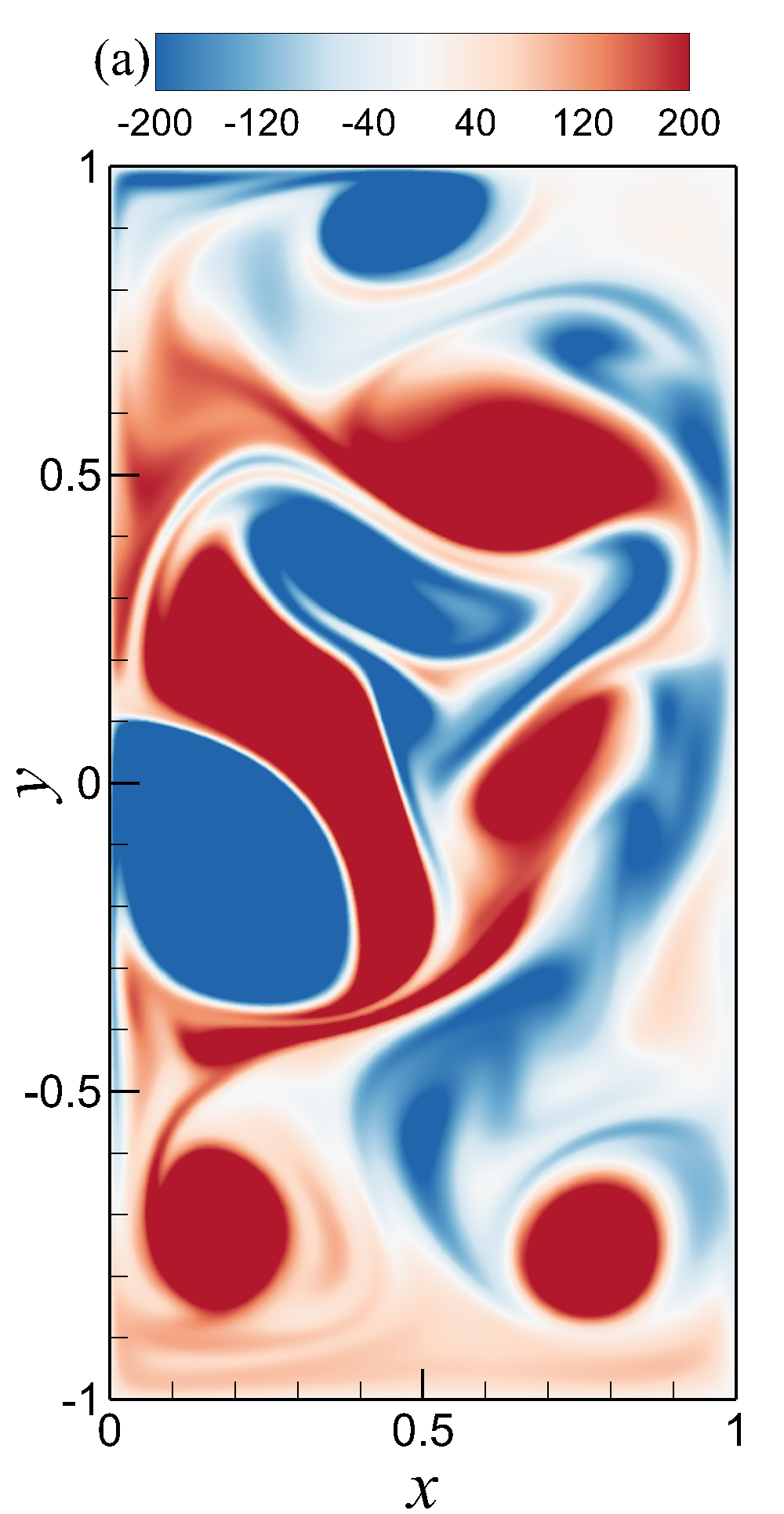}}
\subfigure{\includegraphics[width=0.2\textwidth]{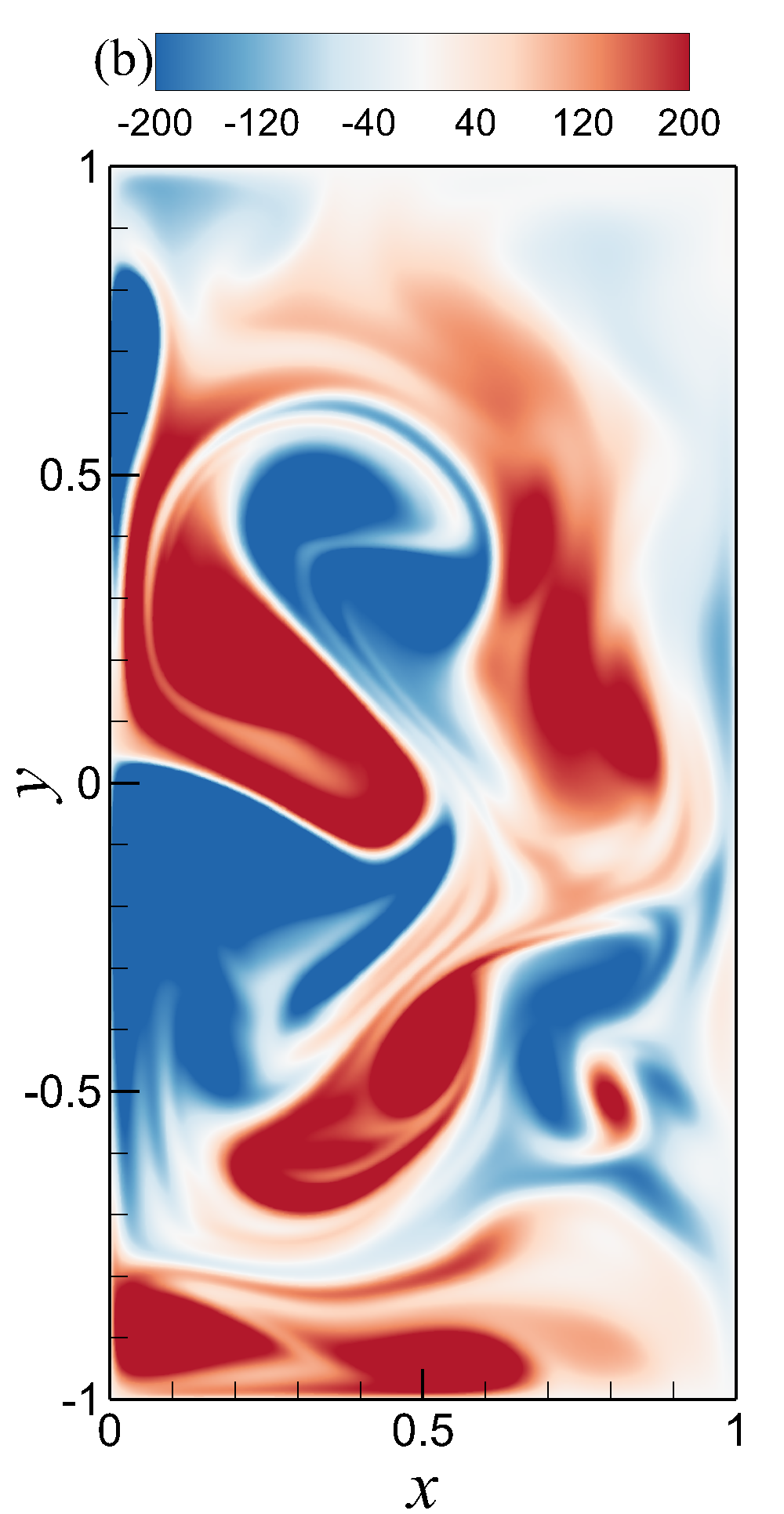}}
\subfigure{\includegraphics[width=0.2\textwidth]{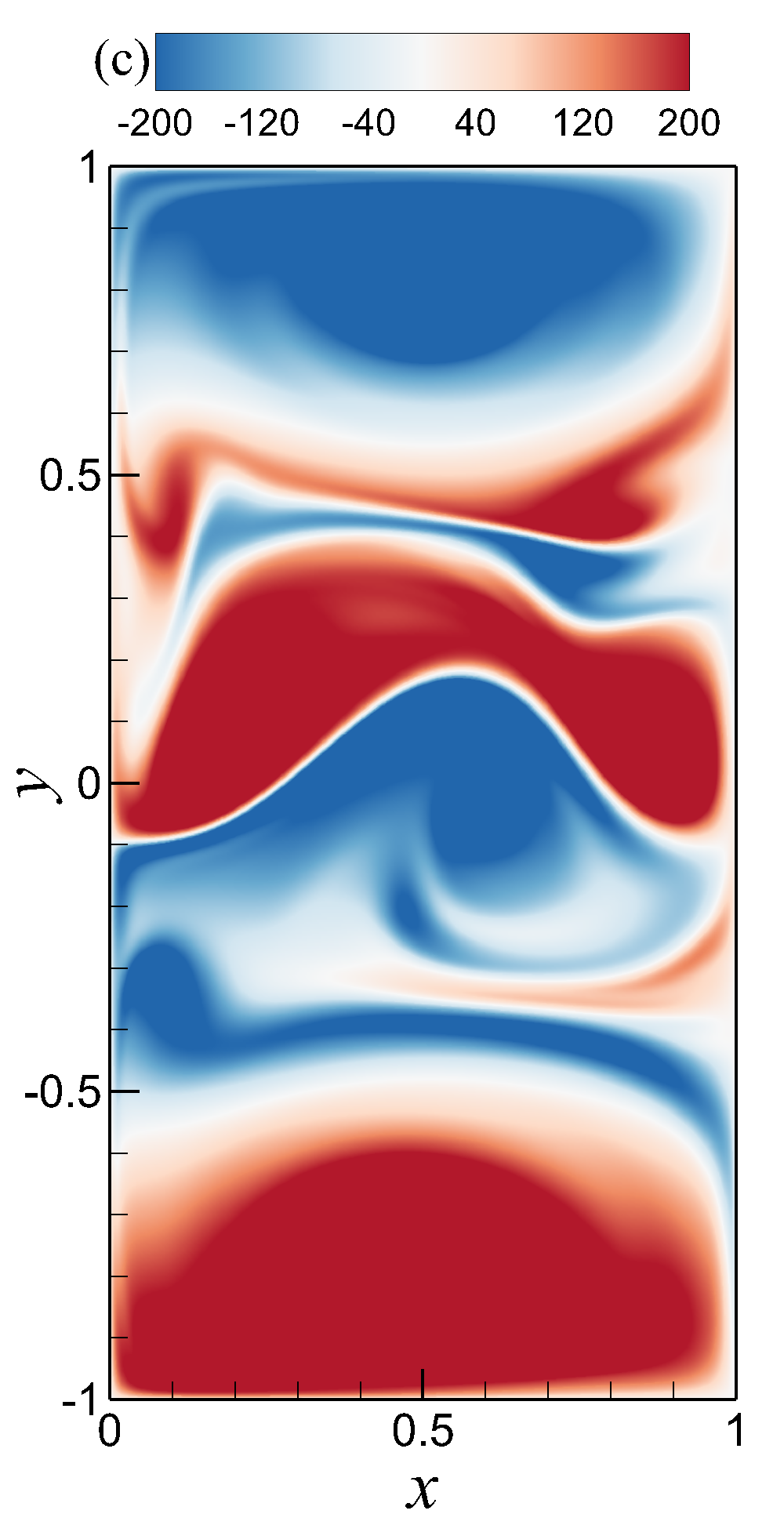}}
\subfigure{\includegraphics[width=0.2\textwidth]{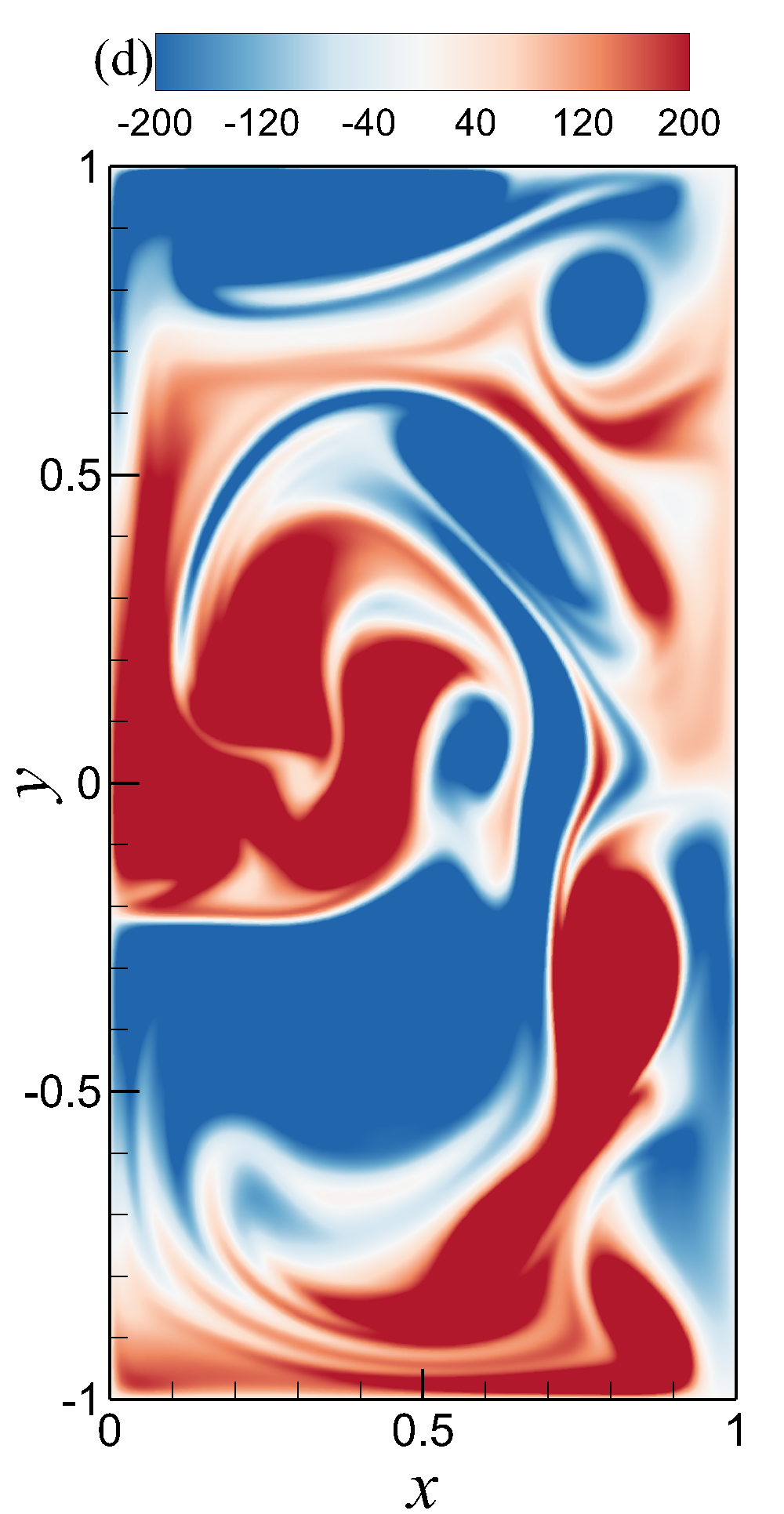}}
}\\
\mbox{
\subfigure{\includegraphics[width=0.2\textwidth]{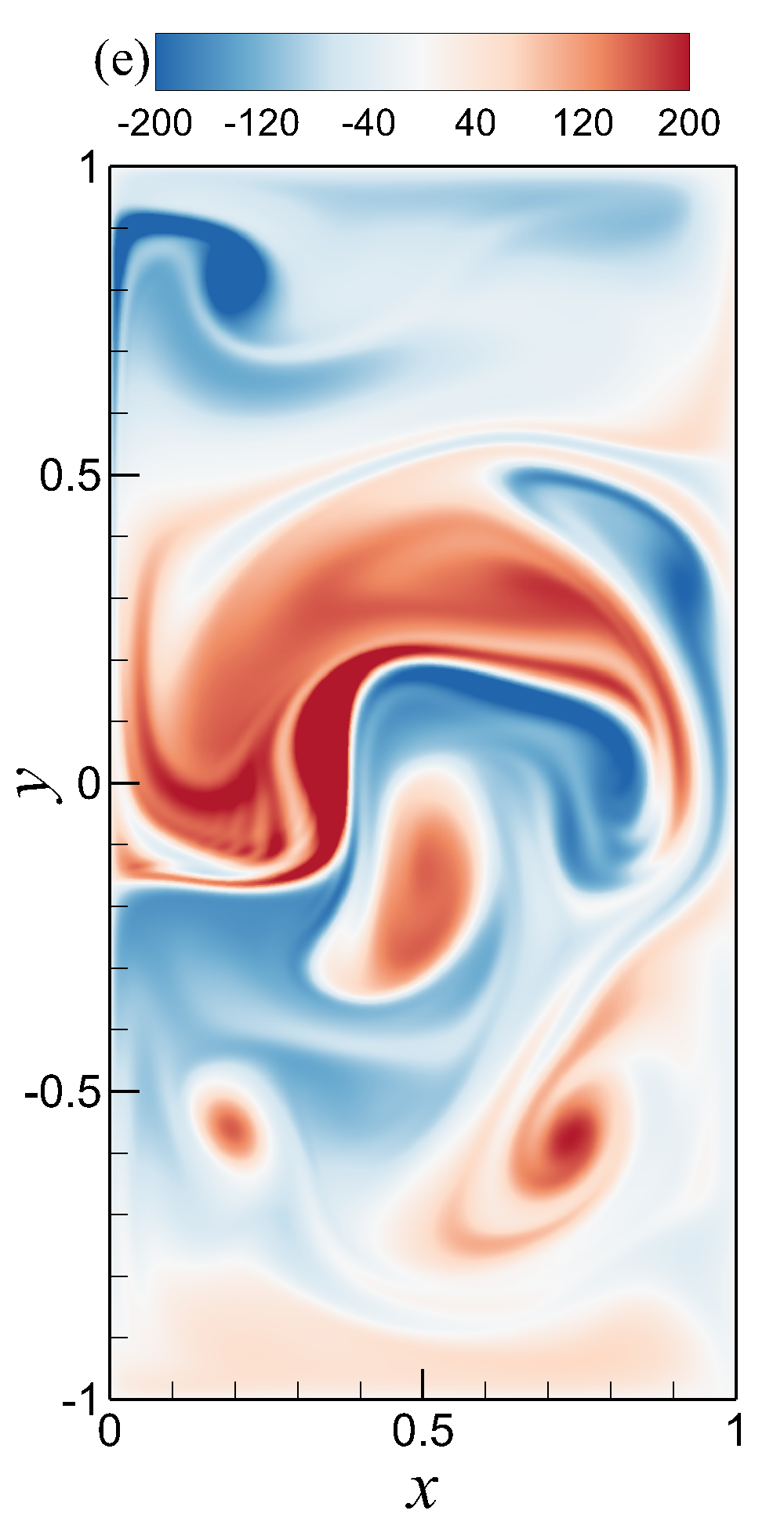}}
\subfigure{\includegraphics[width=0.2\textwidth]{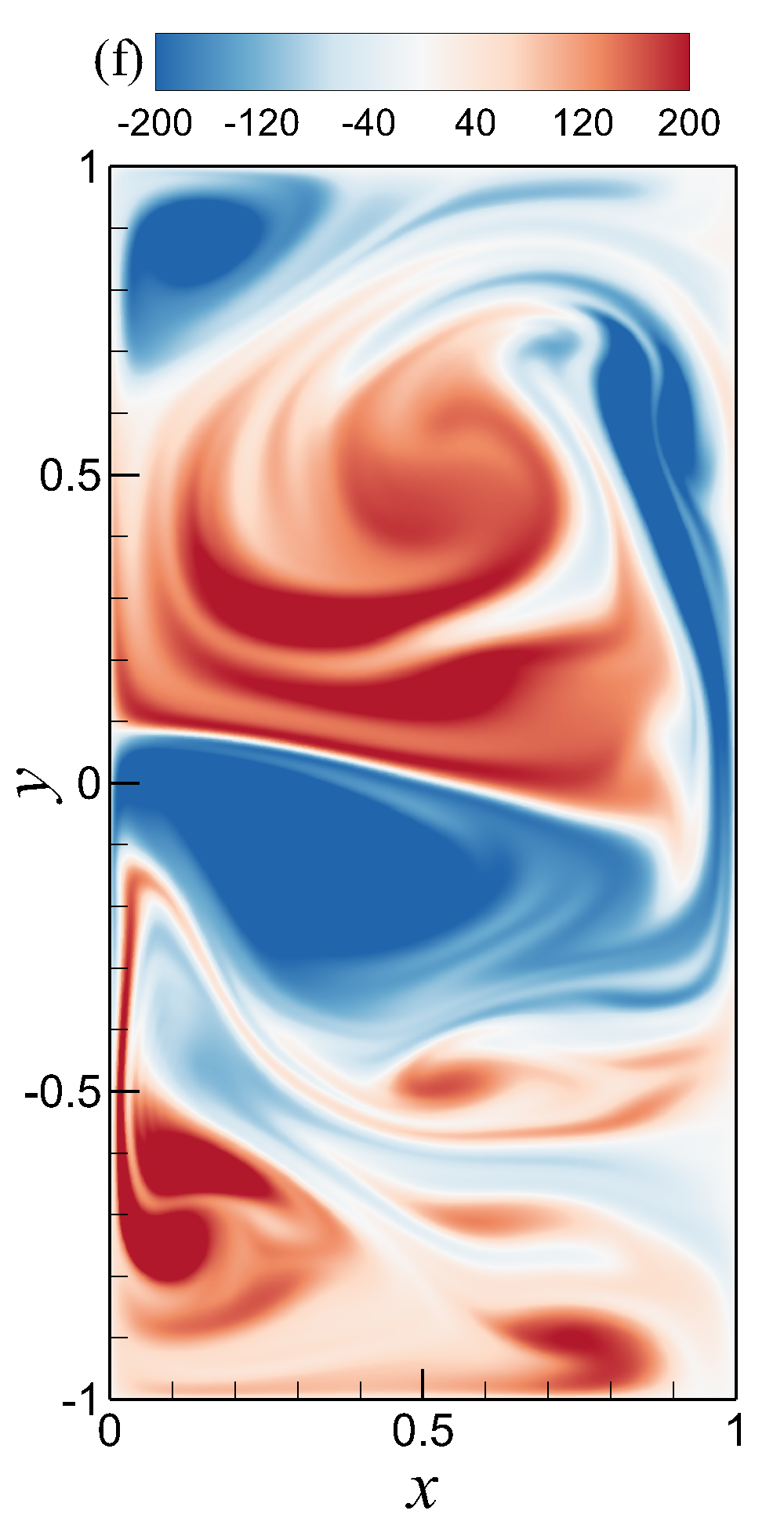}}
\subfigure{\includegraphics[width=0.2\textwidth]{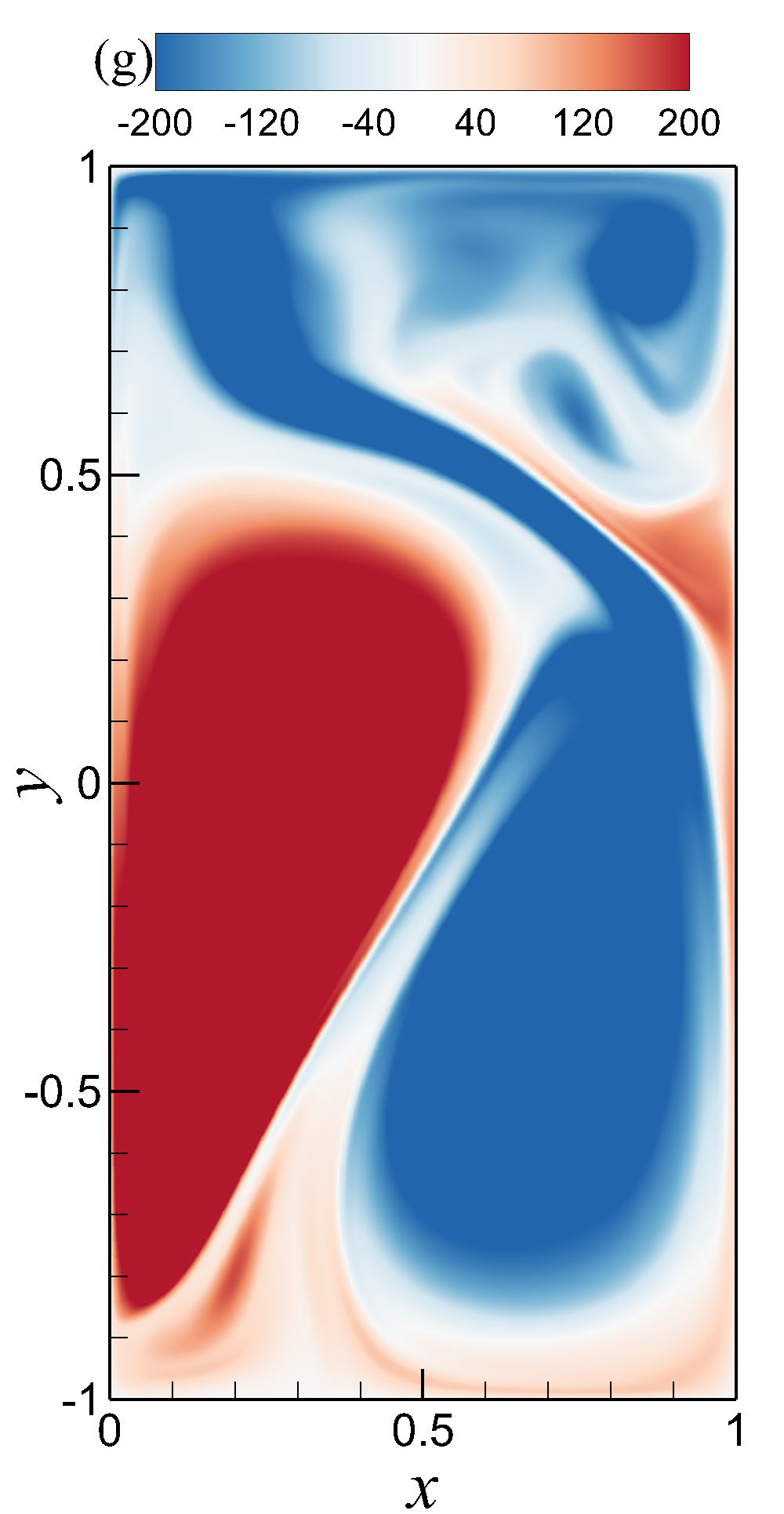}}
\subfigure{\includegraphics[width=0.2\textwidth]{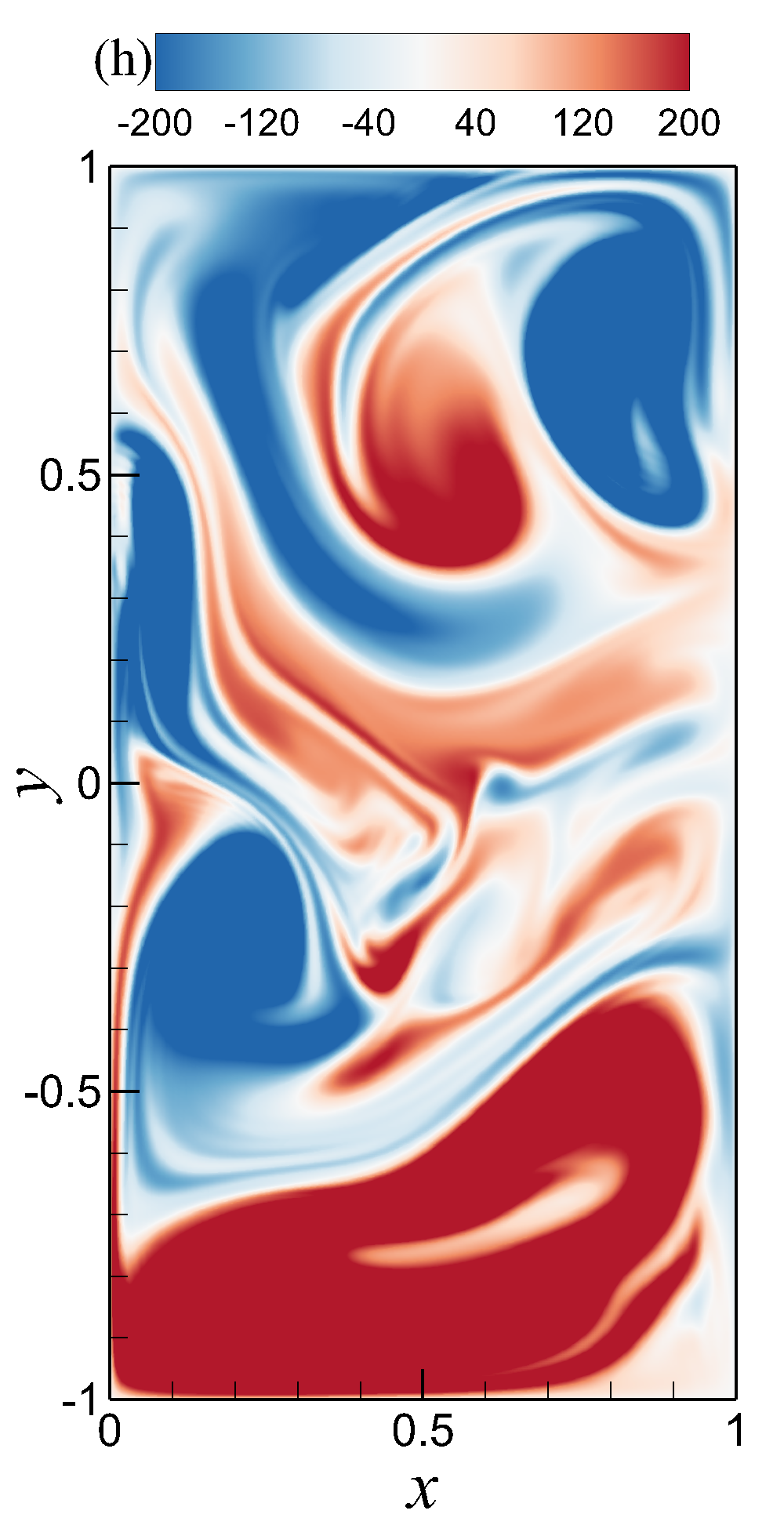}}
}
\caption{Some snapshots from the training data set showing the instantaneous vorticity fields. (a) Experiment I ($t=10$); (b) Experiment I ($t=20$); (c) Experiment I ($t=30$); (d) Experiment I ($t=40$); (e) Experiment II ($t=10$); (f) Experiment II ($t=20$); (g) Experiment II ($t=30$); (h) Experiment II ($t=40$).}
\label{fig:1a}
\end{figure*}

In general, the time evolution of the kinetic energy for QG model shows an initial short transient interval followed by the statistically steady state. As we can see from the time series plots, both (Re, Ro) combinations show a similar trend with the higher (Re, Ro) combination showing comparatively steady state fluctuations (with larger amplitude) and the lower (Re, Ro) combination showing comparatively more unsteady fluctuations. As shown in FIG.~\ref{fig:1}, we store $400$ snapshots from $t = 10$ to $t=50$ to collect data snapshots at statistically steady state after the initial transient period. To get an idea of the energy captured by the POD modes for different flow conditions, we also present the percentage of energy accumulation with respect to the POD modes in FIG.~\ref{fig:1}. We can compute the percentage modal energy by:
\begin{align}
	P(k) = \left(\frac{\sum_{j = 1}^{k}\lambda_j}{\sum_{j=1}^{N}\lambda_j}\right) \times 100,
\end{align}
where the number of snapshots is set to $N = 400$ in our study. It is apparent in the percentage modal energy plot that $50$ POD modes capture around $80 \%$ of total energy of the system for lower (Re, Ro) combination whereas capture around $85 \%$ of total energy of the system for higher (Re, Ro) combination. Since we have seen a comparatively more steady fluctuations in the time series plot of the total energy for higher (Re, Ro) combination, it is expected to accumulate more energy in less POD modes in higher (Re, Ro) combination case. Surely, if we increase the number of modes, the captured percentage modal energy will increase for both cases. Regardless of that, we represent most of our simulation results up to $80$ modes in this study. FIG.~\ref{fig:1a} shows the instantaneous vorticity fields for Experiment I and II, and gives a visualization of the vigorous eddying nature of this test problem. The mixing between outer and inner-gyre in the instantaneous fields demonstrates that the flow is in the turbulent state and a strong distortion of the vorticity contours can be noticed in both experiments. 

\subsection{Experiment I: Both data collection and prediction at Re $\mathbf{=200}$, Ro $\mathbf{=0.0016}$}

\begin{figure*}[htbp]
\centering
\mbox{
\subfigure{\includegraphics[width=0.2\textwidth]{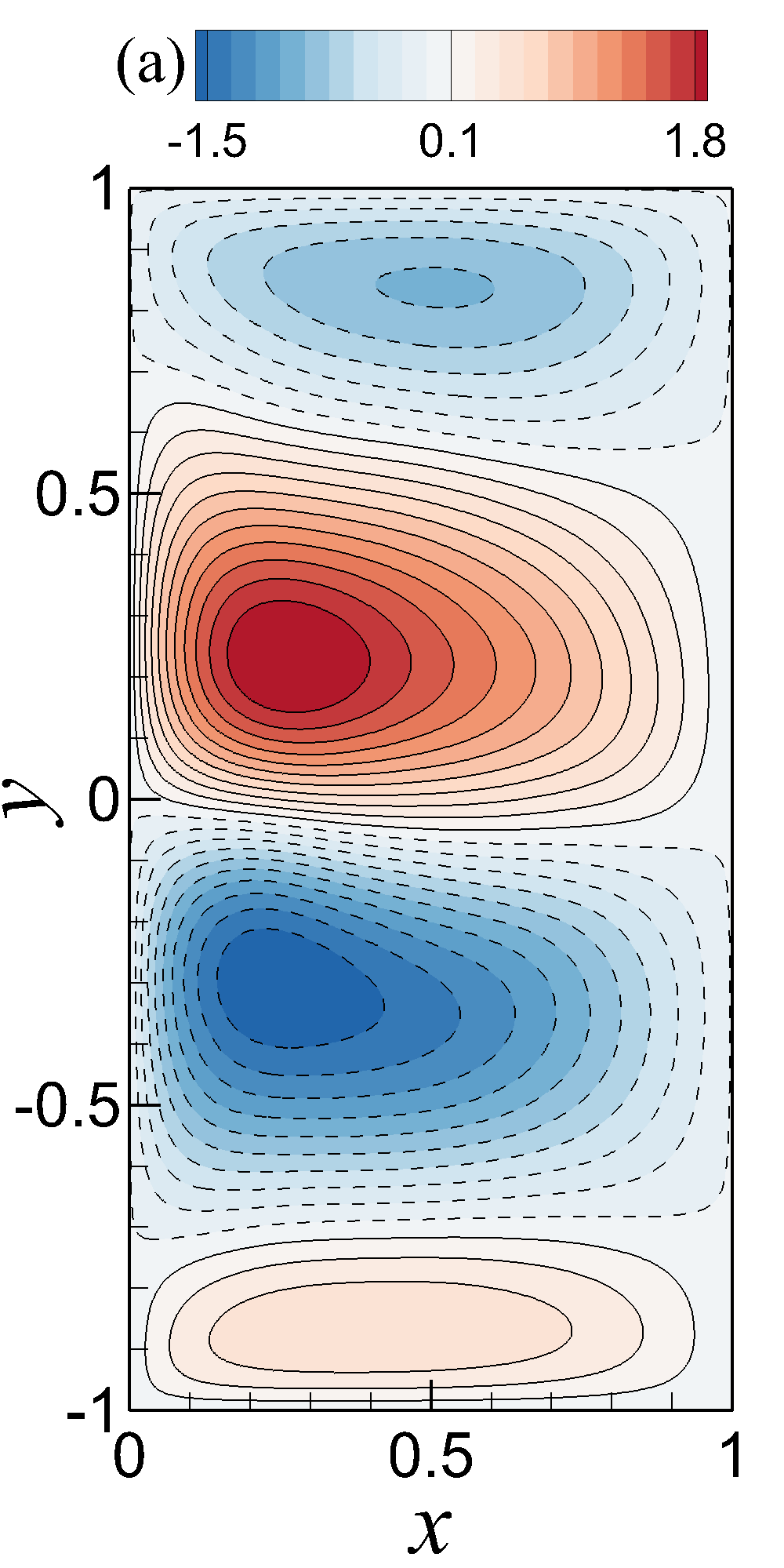}}
\subfigure{\includegraphics[width=0.2\textwidth]{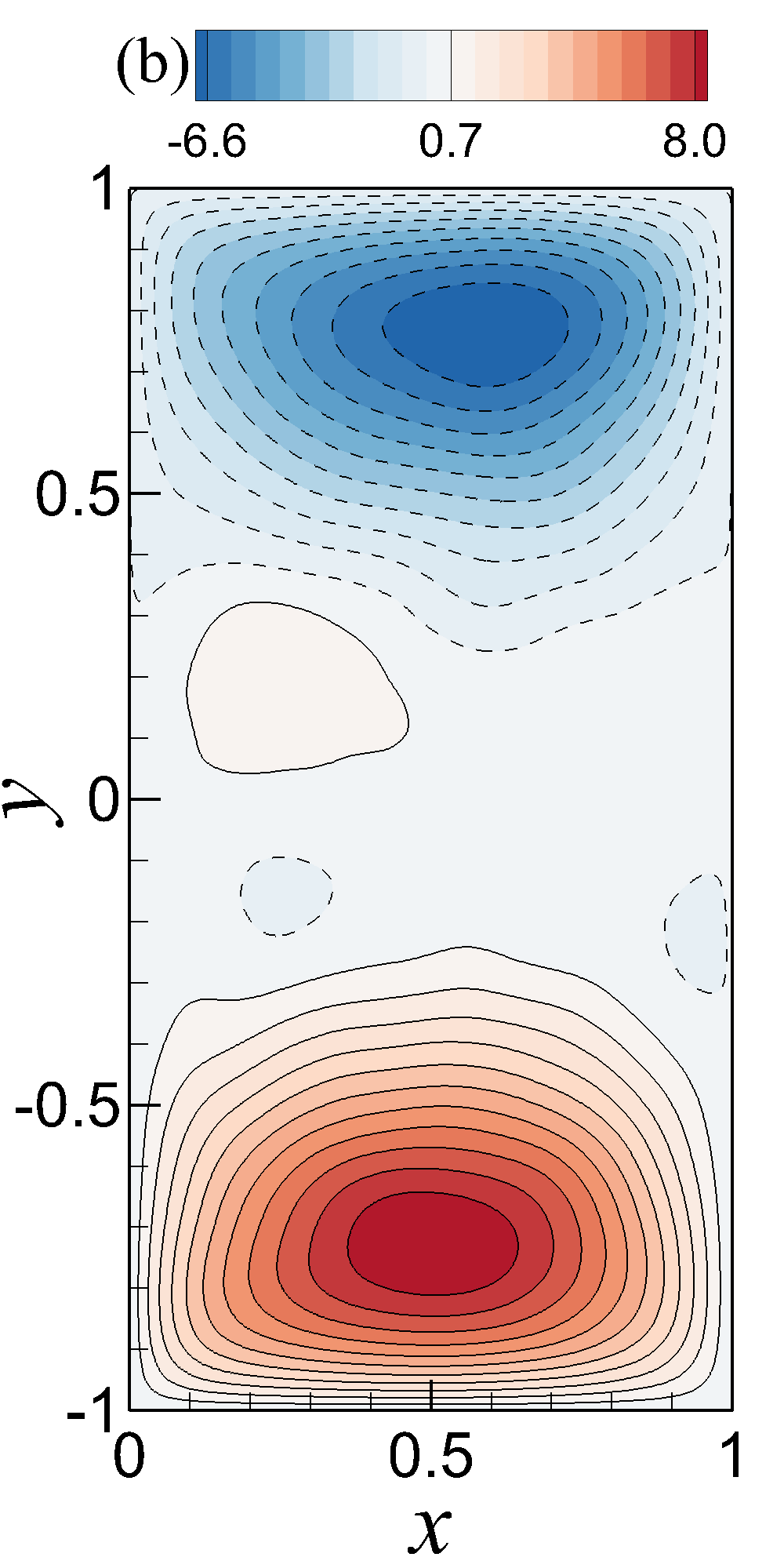}}
\subfigure{\includegraphics[width=0.2\textwidth]{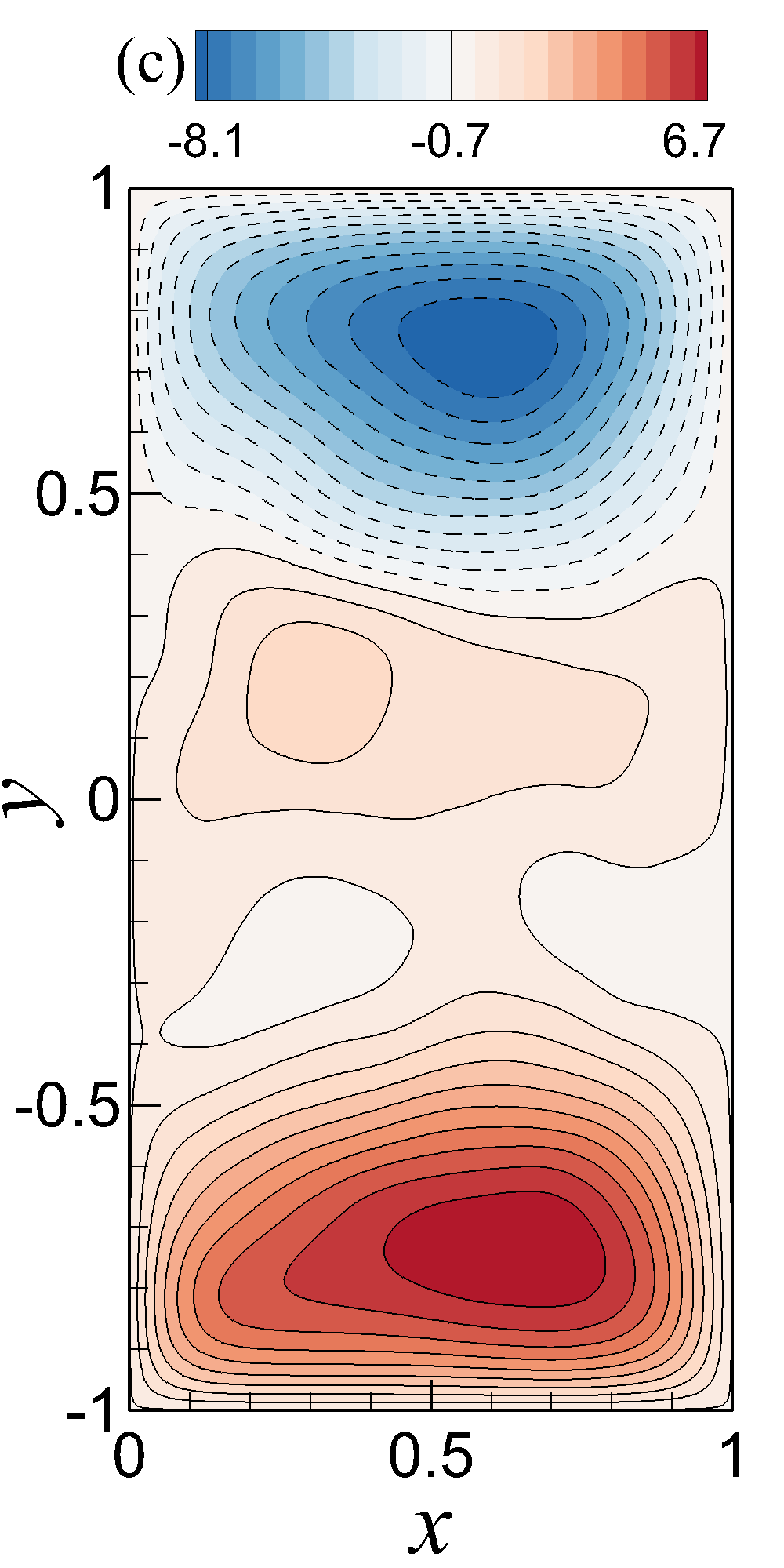}}
\subfigure{\includegraphics[width=0.2\textwidth]{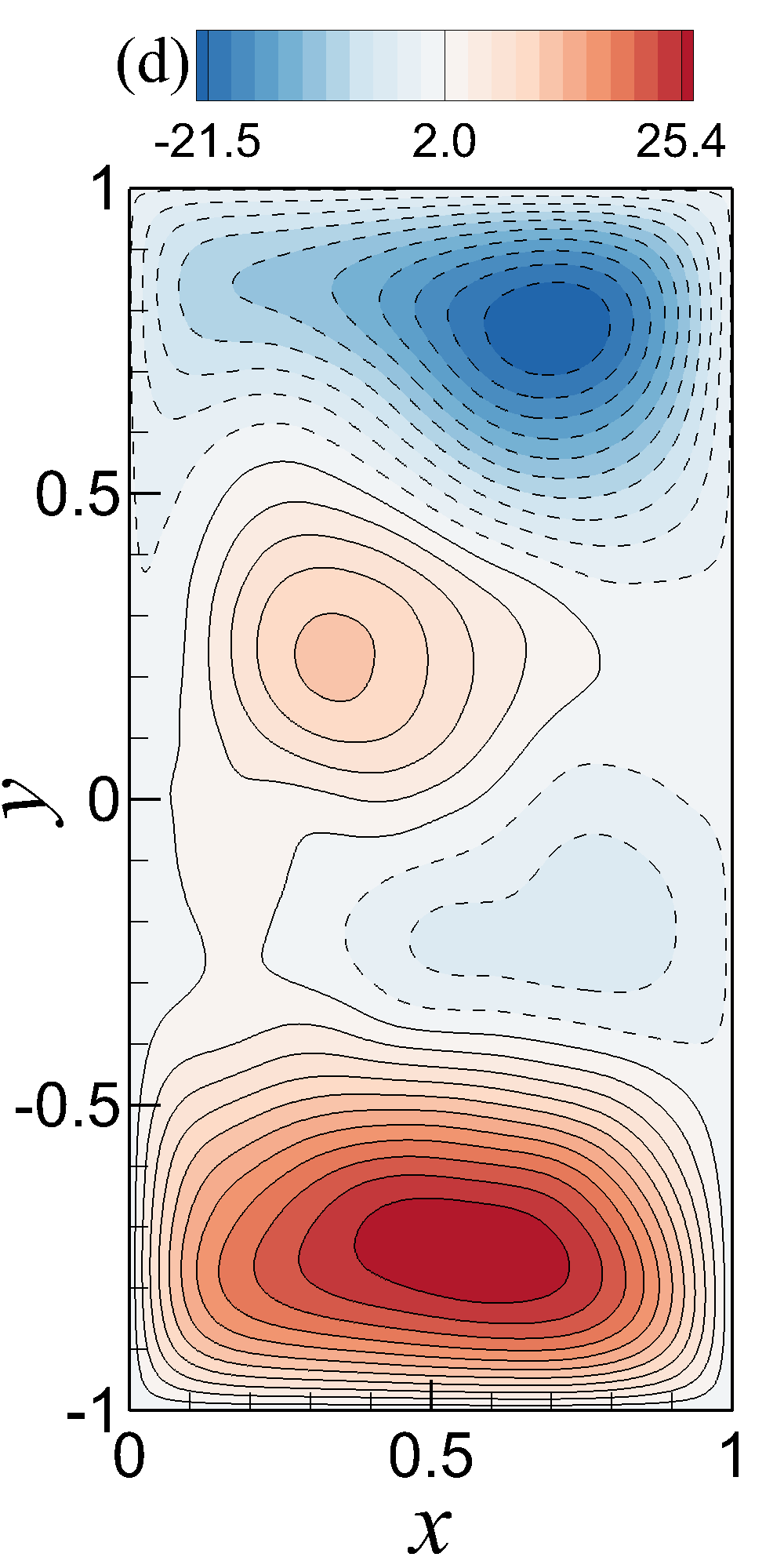}}
}\\
\mbox{
\subfigure{\includegraphics[width=0.2\textwidth]{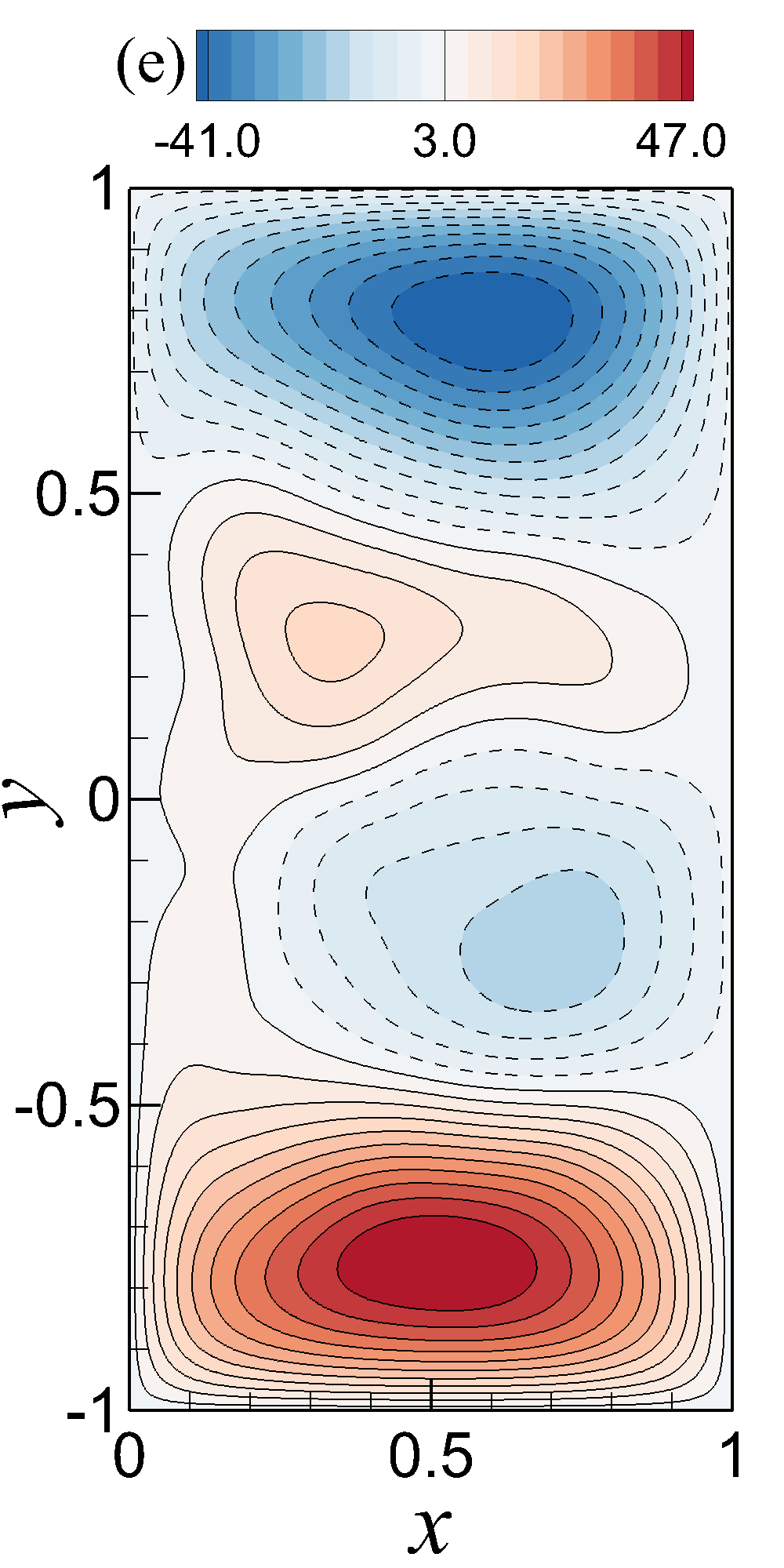}}
\subfigure{\includegraphics[width=0.2\textwidth]{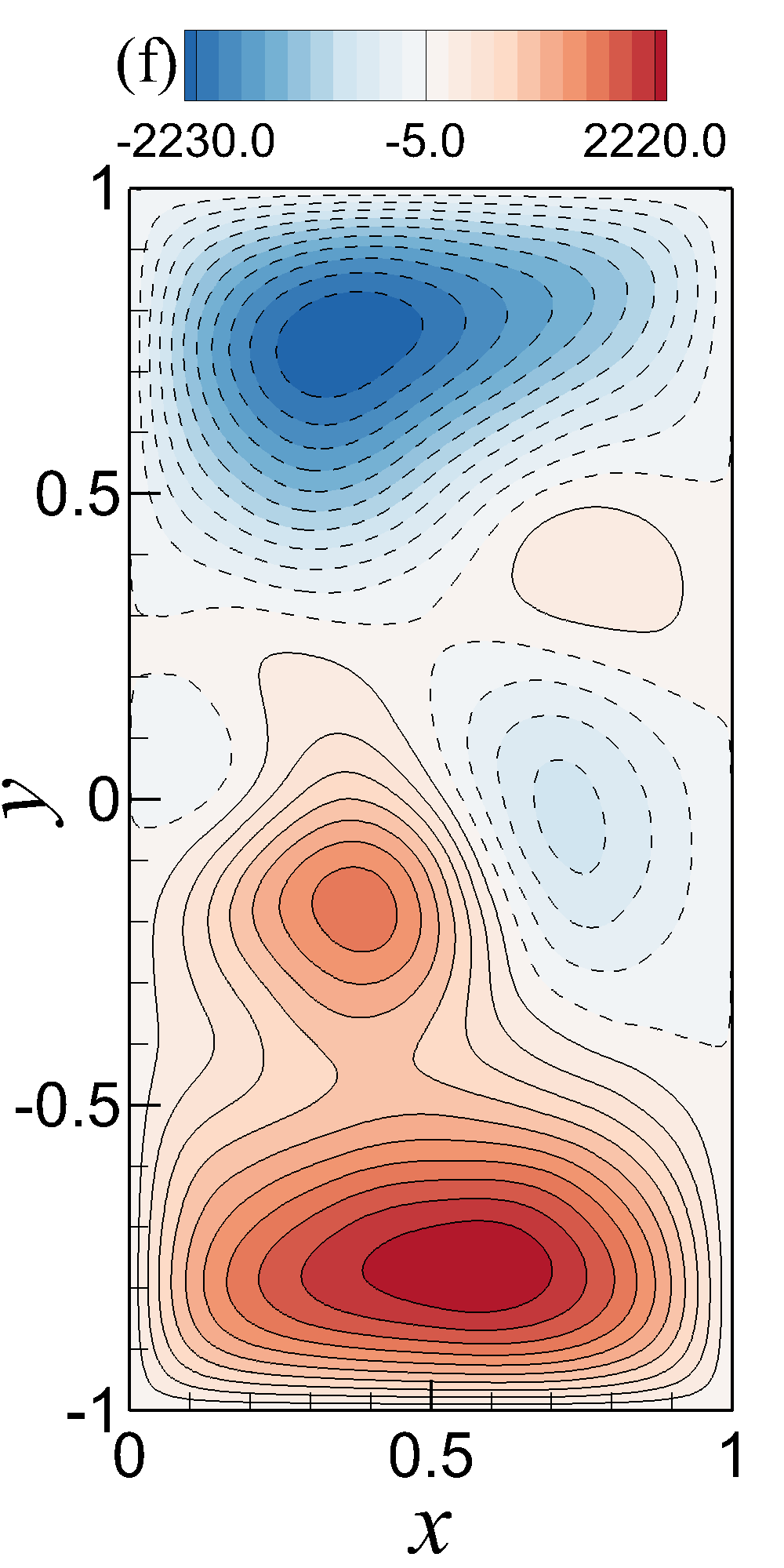}}
\subfigure{\includegraphics[width=0.2\textwidth]{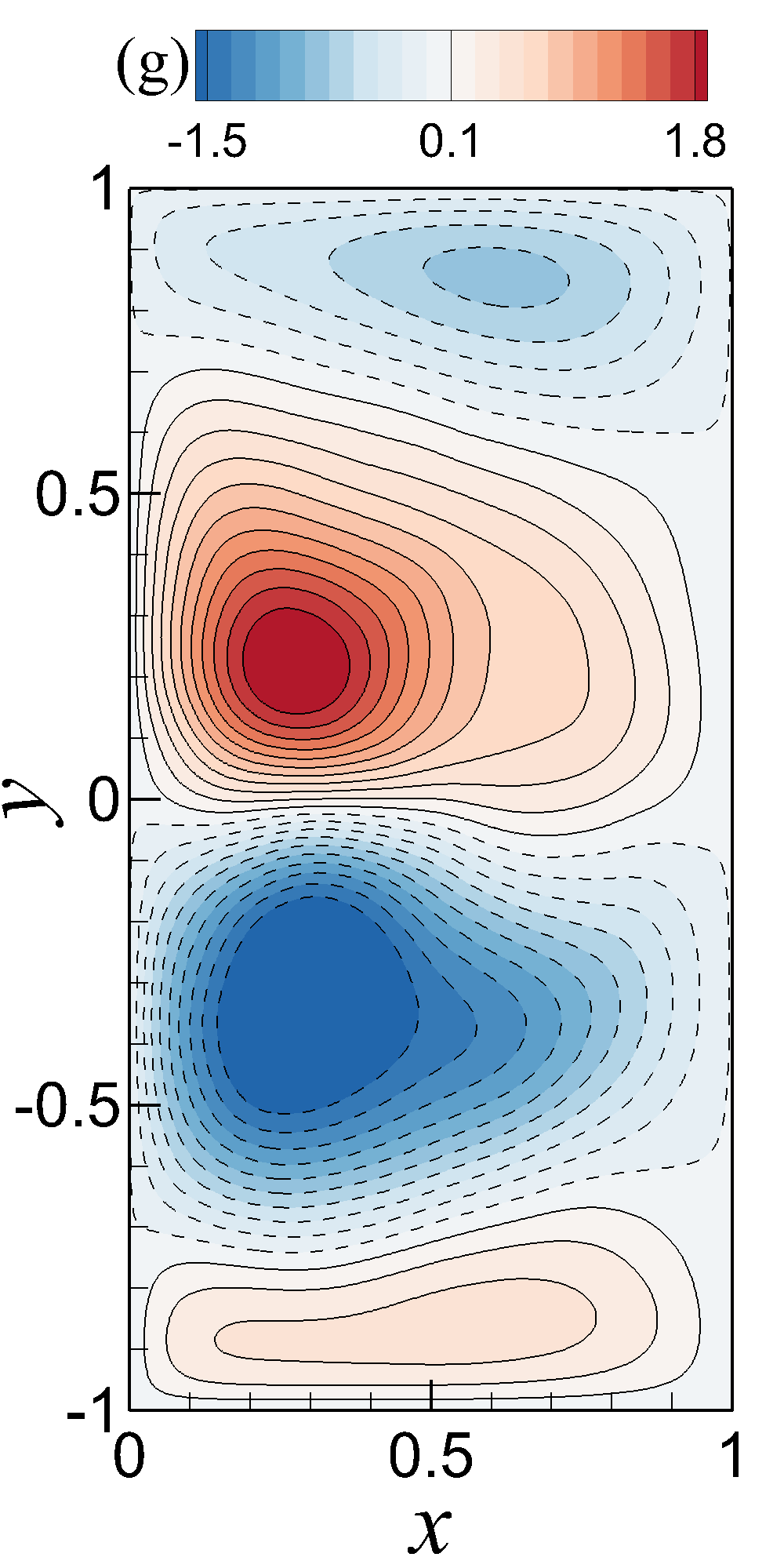}}
\subfigure{\includegraphics[width=0.2\textwidth]{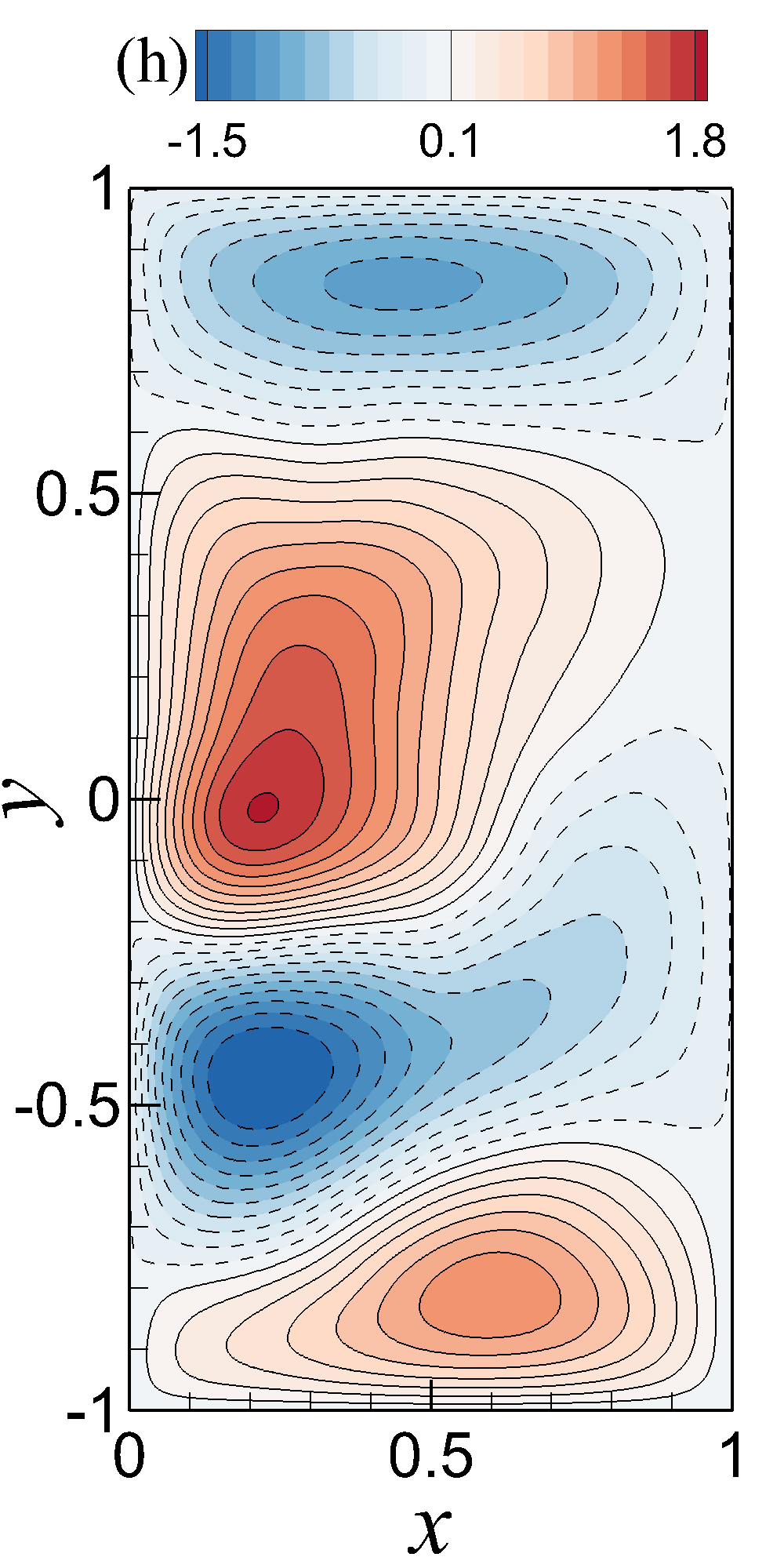}}
}
\caption{Mean streamfunction contours for Experiment I (for training snapshots between $t = 10$ and $t = 50$). (a) FOM at a resolution of $256 \times 512$; (b) ROM-G with $R=50$ modes; (c) ROM-G with $R=40$ modes; (d) ROM-G with $R=30$ modes; (e) ROM-G with $R=20$ modes; (f) ROM-G with $R=10$ modes; (g) proposed ROM-D with $R=20$ modes and $\Delta R=3$; (h) proposed ROM-D with $R=10$ modes and $\Delta R=3$. Note that $\Delta R = R - \tilde{R}$.}
\label{fig:2}
\end{figure*}

In Experiment I, we collect our training snapshots data from a $256 \times 512$ resolution FOM simulation at Re $=200$ and Ro $=0.0016$. FIG.~\ref{fig:2} shows the streamfunction contours, averaged in time, obtained by FOM, ROM-G (with $R = 10$, $20$, $30$, $40$ and $50$) and ROM-D (with $R = 10$ and $20$), respectively. Although the dynamics of instantaneous flows is chaotic and fluctuates in time, the time-averaged contour plot shows the four-gyre circulation pattern for FOM simulation as displayed in FIG.~\ref{fig:2}. This four-gyre pattern indicates that the model reaches a state of turbulent equilibrium where the two inner gyres circulation are similar as the the wind stress curl forcing while the outer gyres correspond to the northern and southern gyres found in geostrophic turbulence experiments (i.e., time-averaged field data is obtained by averaging between $t=10$ and $t=100$). The circulations of the two gyres in both inner and outer regions can be seen in the figure where the red color represents the circulation in positive direction (i.e., counter-clock wise) and the blue color indicates the circulation in opposite direction.

Since we conduct the experiment for this test problem in highly turbulent regime, i.e., turbulence with weak dissipation, the four-gyre circulation pattern is expected when the flow reaches statistically steady state and the simulation result is consistent with the existing literature \cite{greatbatch2000four}. The physical interpretation of this four-gyre circulation can be the wind stress curl represented by the two inner-gyres is equilibriated by the two outer-gyres driven by the eddy flux of potential vorticity. In FIG.~\ref{fig:2}, we can also observe the nonphysical flow prediction in the contour plots for ROM-G solutions with lower $R$ (Note the range in the legend for $R = 30$, $R = 20$ and $R = 10$). For $R = 40$, the ROM-G gives a good prediction of the FOM solution and the $R = 50$ result is comparatively even better. The proposed ROM-D solutions, however, show a better estimation of the FOM solution for only $R = 10$ and $R = 20$ than the ROM-G solution with $R = 50$. We can clearly visualize the presence of four-gyre circulation in reasonable range for only $R = 20$ which indicates the balance between the wind stress curl and the divergence of the eddy potential vorticity flux. A point should be noted here that we utilize $\Delta t = 2.5\times10^{-4}$ for all of our ROM-G and ROM-D computations since ROMs are free of the stability constraint (even though time step is set to $\Delta t = 2.5\times10^{-5}$ for the FOM simulation). 

\begin{figure*}[htbp]
\centering
\mbox{
\subfigure{\includegraphics[width=0.9\textwidth]{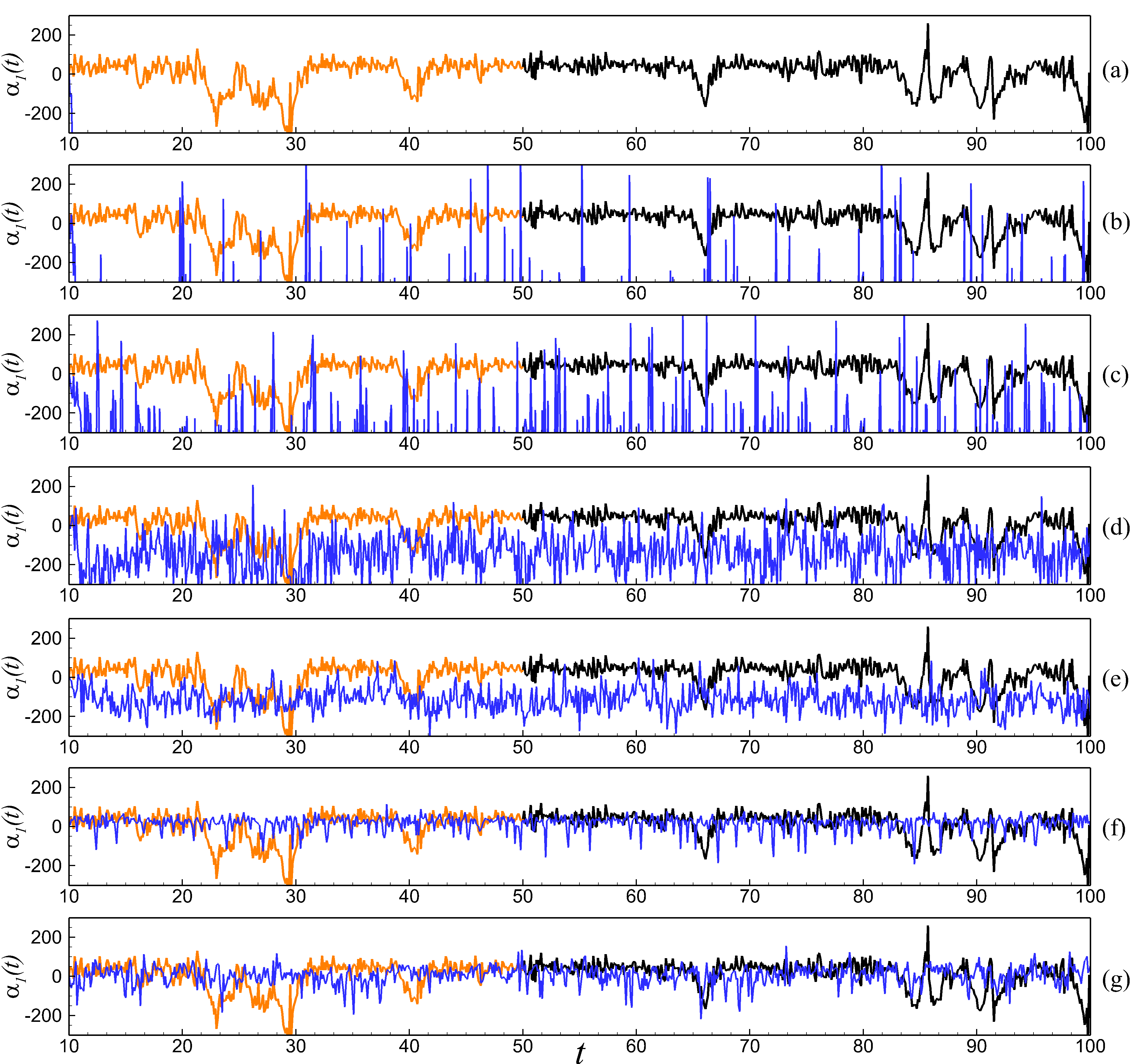}}
}
\caption{Time series of the first modal coefficient for Experiment I. (a) ROM-G with $R=10$ modes; (b) ROM-G with $R=20$ modes; (c) ROM-G with $R=30$ modes; (d) ROM-G with $R=40$ modes; (e) ROM-G with $R=50$ modes; (f) proposed ROM-D with $R=10$ modes and $\Delta R=3$; (g) proposed ROM-D with $R=20$ modes and $\Delta R=3$. Note that $\Delta R = R - \tilde{R}$. True projection data is underlined in each figure with orange (training zone) and black (extended zone).}
\label{fig:3}
\end{figure*}

In another analysis, we plot the time series evolution of the first modal coefficient for Experiment I in FIG.~\ref{fig:3} which show the true data projection of the FOM simulation in each row, the ROM-G solutions with $R = 10$, $R = 20$, $R = 30$, $R = 40$ and $R = 50$ in consequent rows starting from the first row, and the ROM-D solution with $R = 10$ and $R = 20$ in the last two rows, respectively. We extend the time series plot to $100$ to show the range of prediction capability of different models. In the plots, the orange colored part of the true projection data represents the training zone, the black colored part represents the extended zone, and the blue colored line in each figure indicates the solution obtained by the ROMs. Similar to the findings from mean contour plots, we can observe that the ROM-G model with $R = 50$ produces the result closest to the FOM solution although a clear gap is present between the true projection and ROM-G solutions for all the modes up to $R = 50$. In contrast, the ROM-D solutions for $R = 10$ and $R = 20$, undoubtedly, show better performance than ROM-G solutions and an almost overlap can be noticed between ROM-D with $R = 20$ and the true projection. Similar statistical observations can be seen for other modal coefficients (not shown here due to clarity and space limitation reasons). To investigate the robustness of the proposed ROM-D model, we perform a sensitivity test with respect to the dynamic model parameter, $\tilde{R}$ for different test truncation, $\Delta{R}$ values. In our ROM-D framework formulation, we define $\Delta R = R-\tilde{R}$ as a modeling parameter analogous to the test filter strength in LES. We must mention here that we use $\Delta R = 3$ value in all of our experiments (except the sensitivity analyses where we vary the $\Delta{R}$ value) in this study. 

\begin{figure*}[htbp]
\centering
\mbox{
\subfigure{\includegraphics[width=0.9\textwidth]{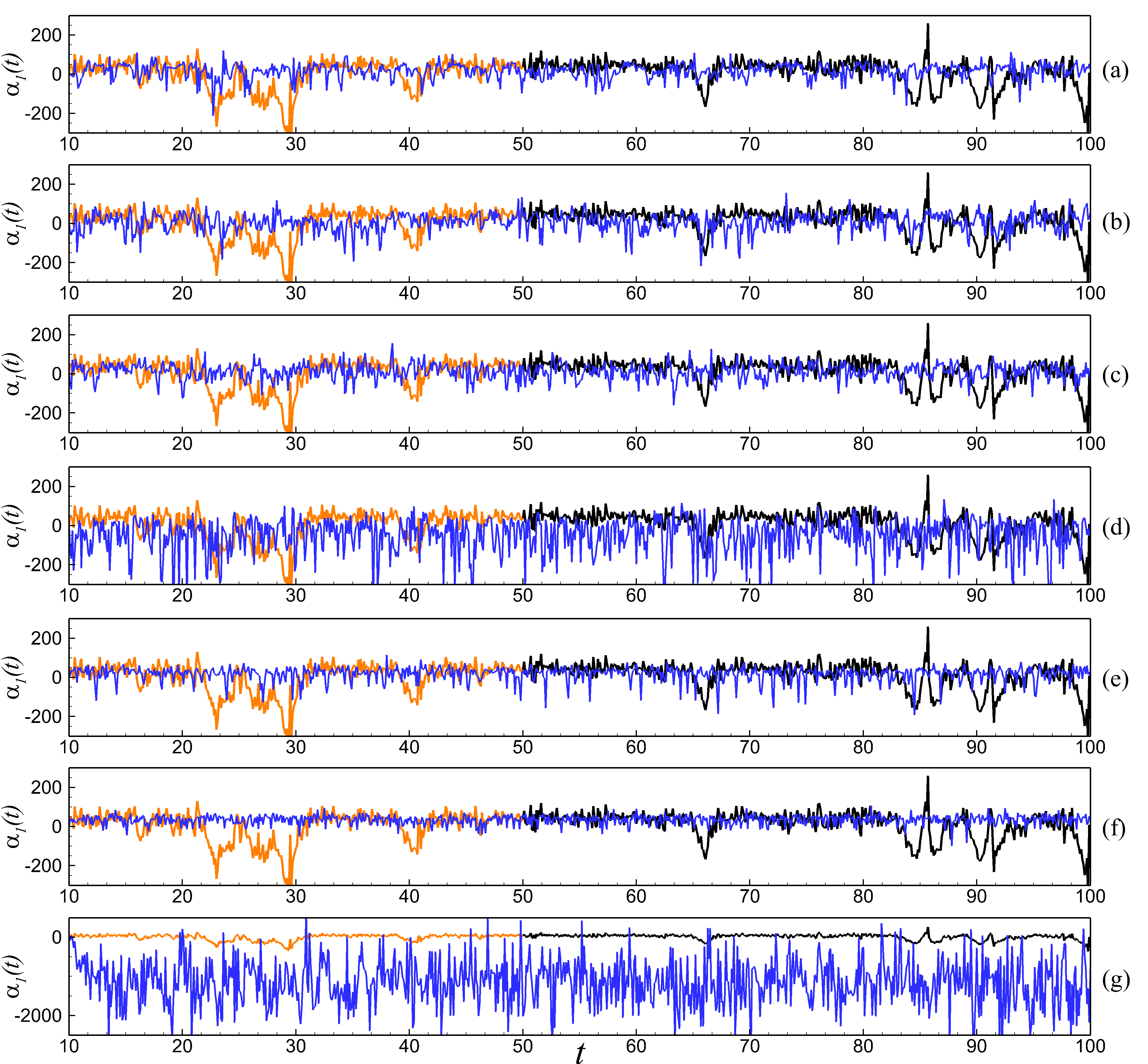}}
}
\caption{A sensitivity test with respect to the dynamic model parameter $\tilde{R}$ showing the time series of the first modal coefficient for Experiment I. (a) ROM-D ($R=20$) with $\Delta R=2$; (b) ROM-D ($R=20$) with $\Delta R=3$; (c) ROM-D ($R=20$) with $\Delta R=4$; (d) ROM-D ($R=10$) with $\Delta R=2$; (e) ROM-D ($R=10$) with $\Delta R=3$; (f) ROM-D ($R=10$) with $\Delta R=4$; (g) ROM-G with $R=20$ modes. Note that $\Delta R = R - \tilde{R}$. True projection data is presented in each figure with orange (training zone) and black (extended zone).}
\label{fig:4}
\end{figure*}

For sensitivity analysis, we first present the time series of the first modal coefficient for Experiment I with different ($\Delta R$, $R$) combinations of the proposed ROM-D model in FIG.~\ref{fig:4}. As we can see, the results do not vary much statistically for any $R$ with different $\Delta R$ combinations. However, it seems that there are some correlations between $\Delta R$ and the amplitude of the fluctuations for ROM-D with $R = 10$. On the other hand, the time series of ROM-G with $R = 20$ on the bottom last row reveals that the both $R = 10$ and $R = 20$ combined with different $\Delta{R}$ values ROM-D models exhibit a better estimation of the true projection than the ROM-G solution. Since we take into account the truncated modes in ROM-D models, the results improve significantly with respect to the ROM-G model with lower $R$.

\begin{figure*}[htbp]
\centering
\mbox{
\subfigure{\includegraphics[width=0.2\textwidth]{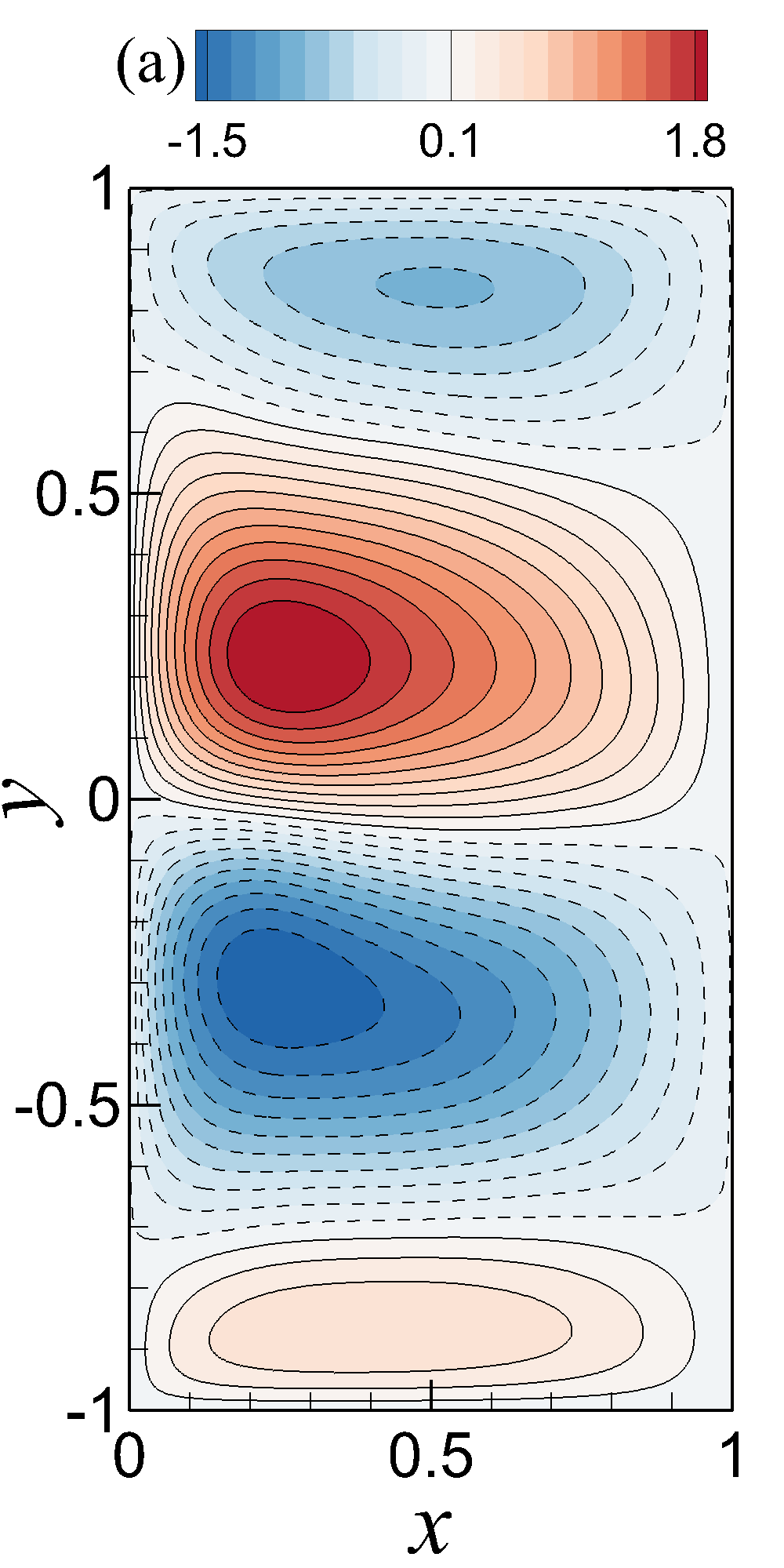}}
\subfigure{\includegraphics[width=0.2\textwidth]{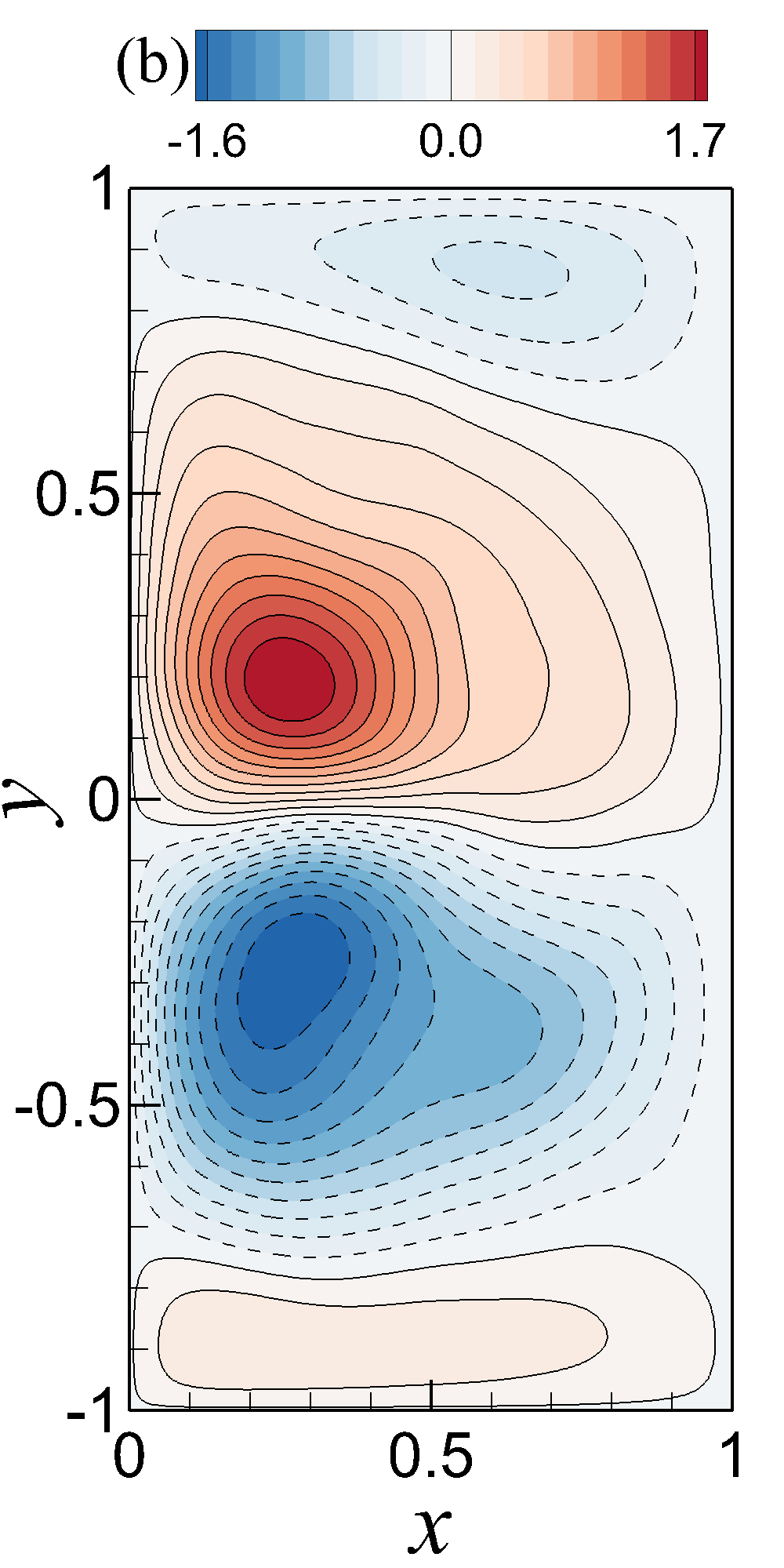}}
\subfigure{\includegraphics[width=0.2\textwidth]{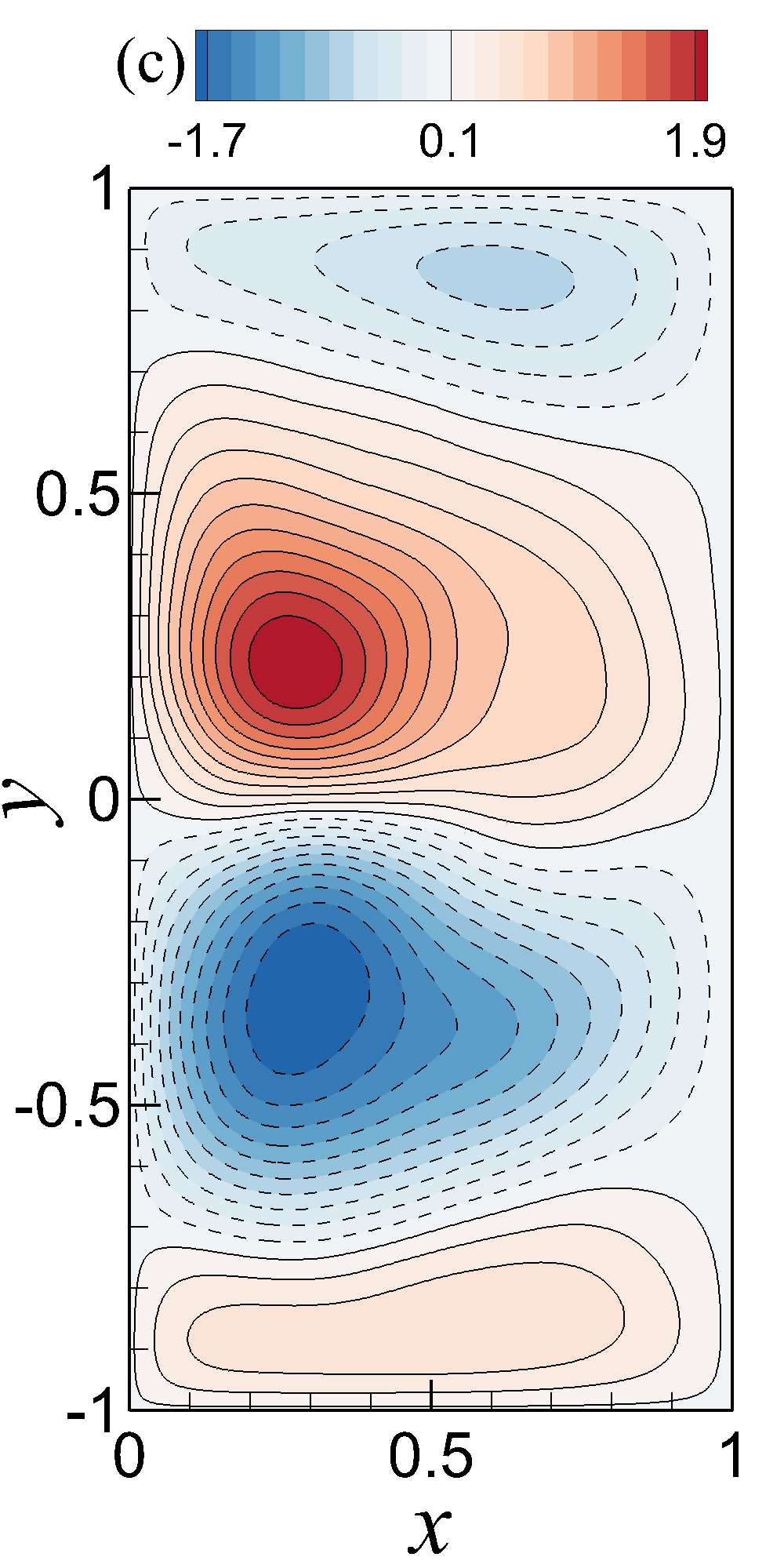}}
\subfigure{\includegraphics[width=0.2\textwidth]{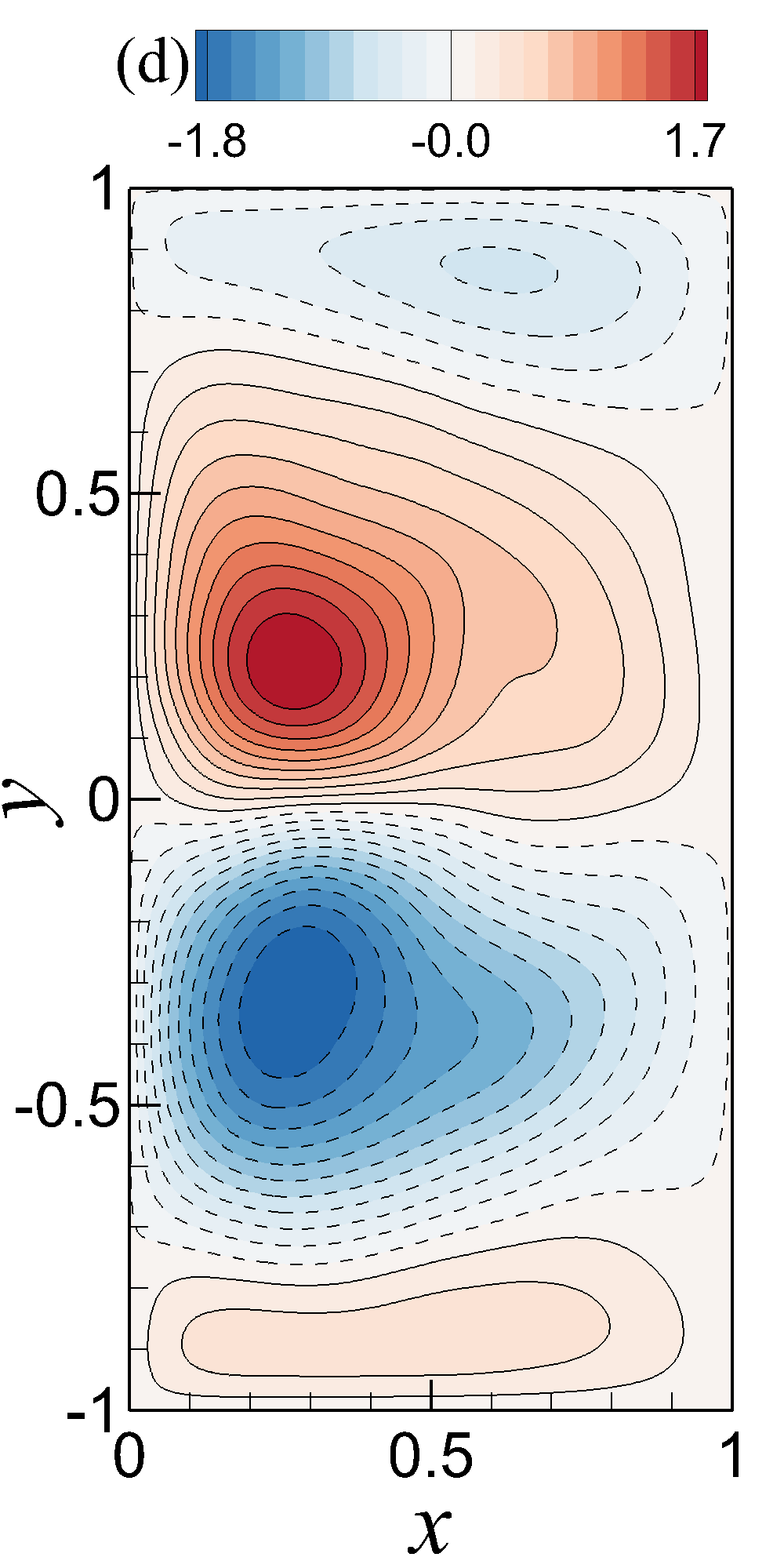}}
}\\
\mbox{
\subfigure{\includegraphics[width=0.2\textwidth]{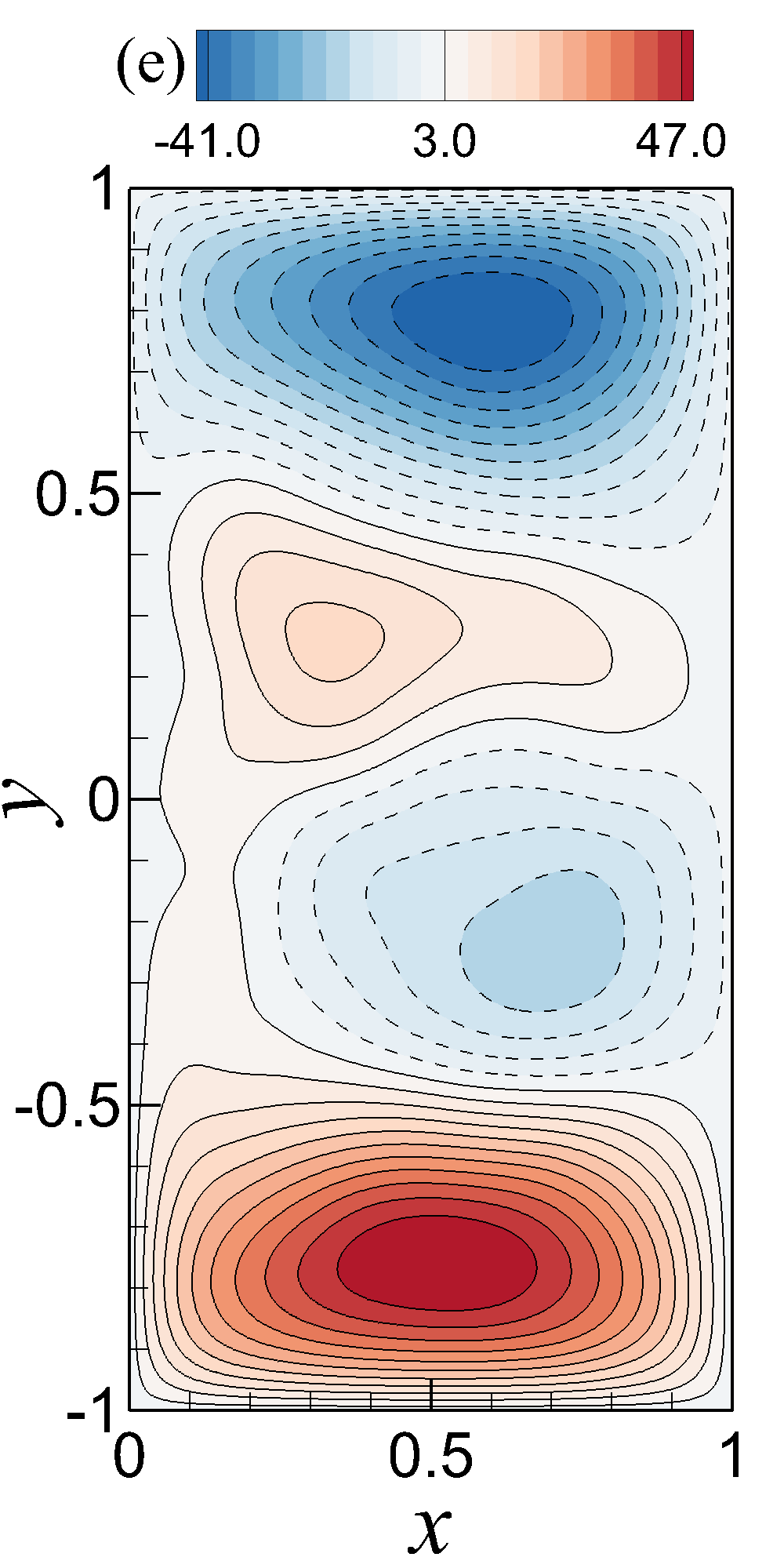}}
\subfigure{\includegraphics[width=0.2\textwidth]{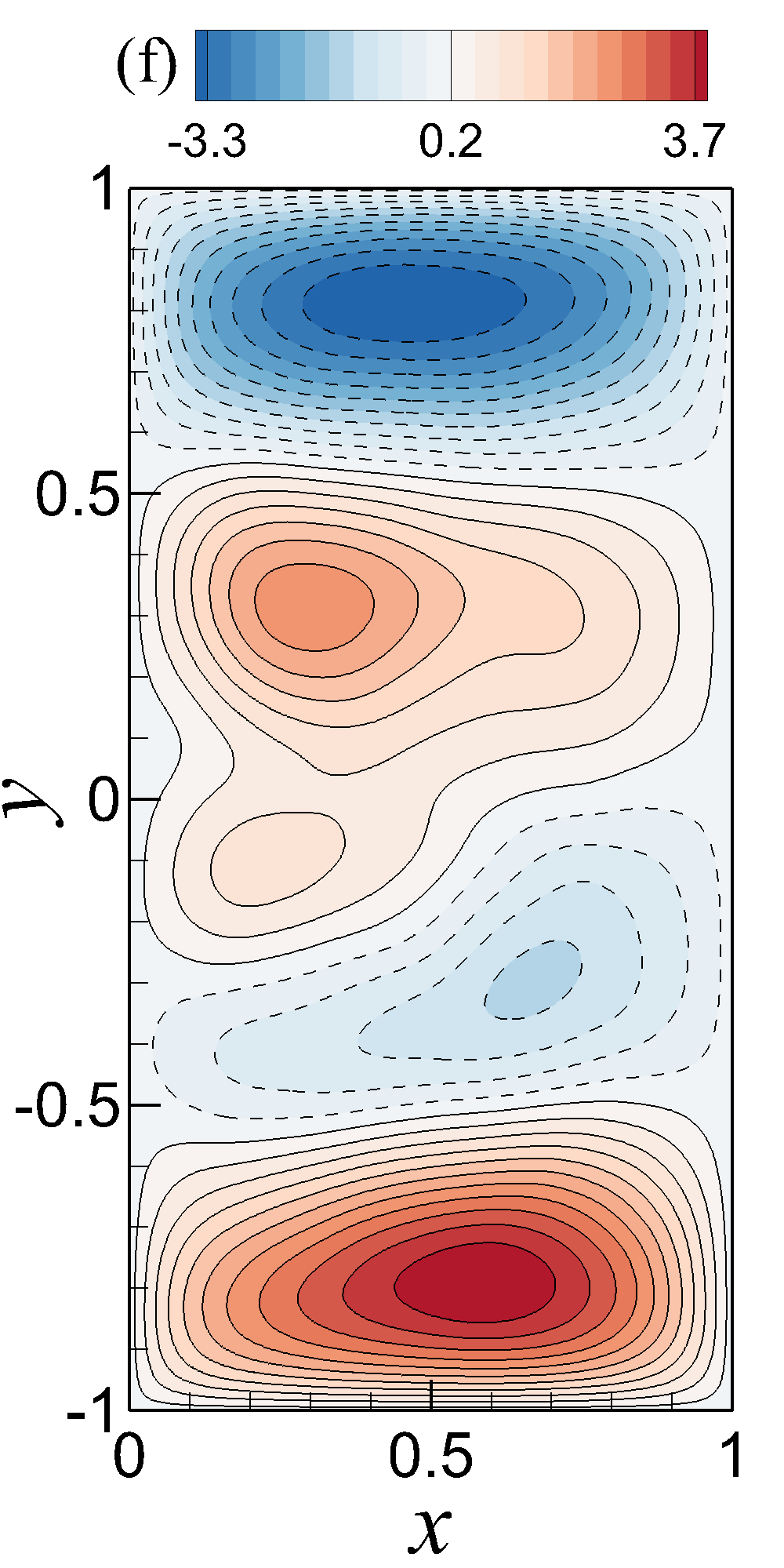}}
\subfigure{\includegraphics[width=0.2\textwidth]{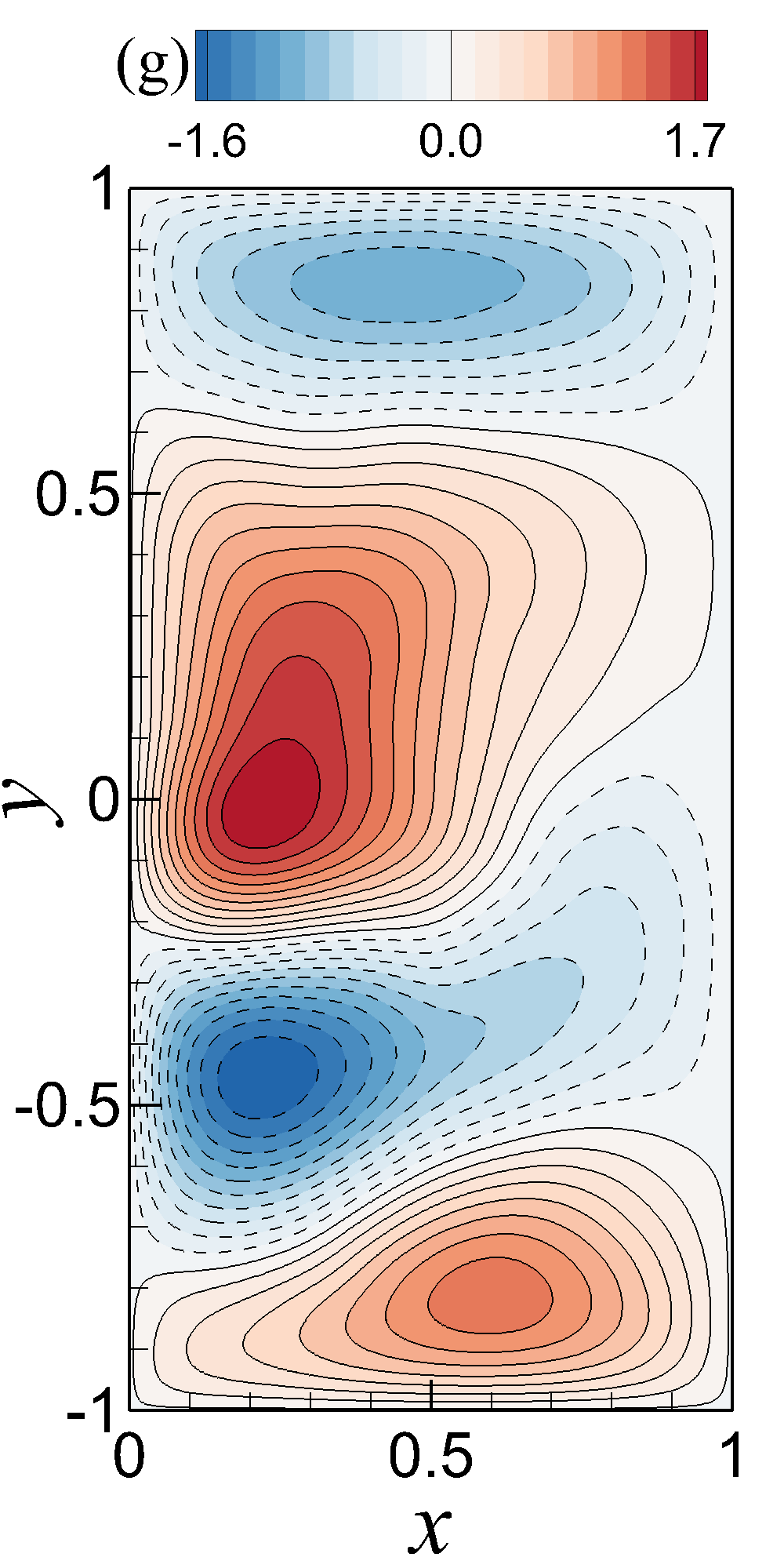}}
\subfigure{\includegraphics[width=0.2\textwidth]{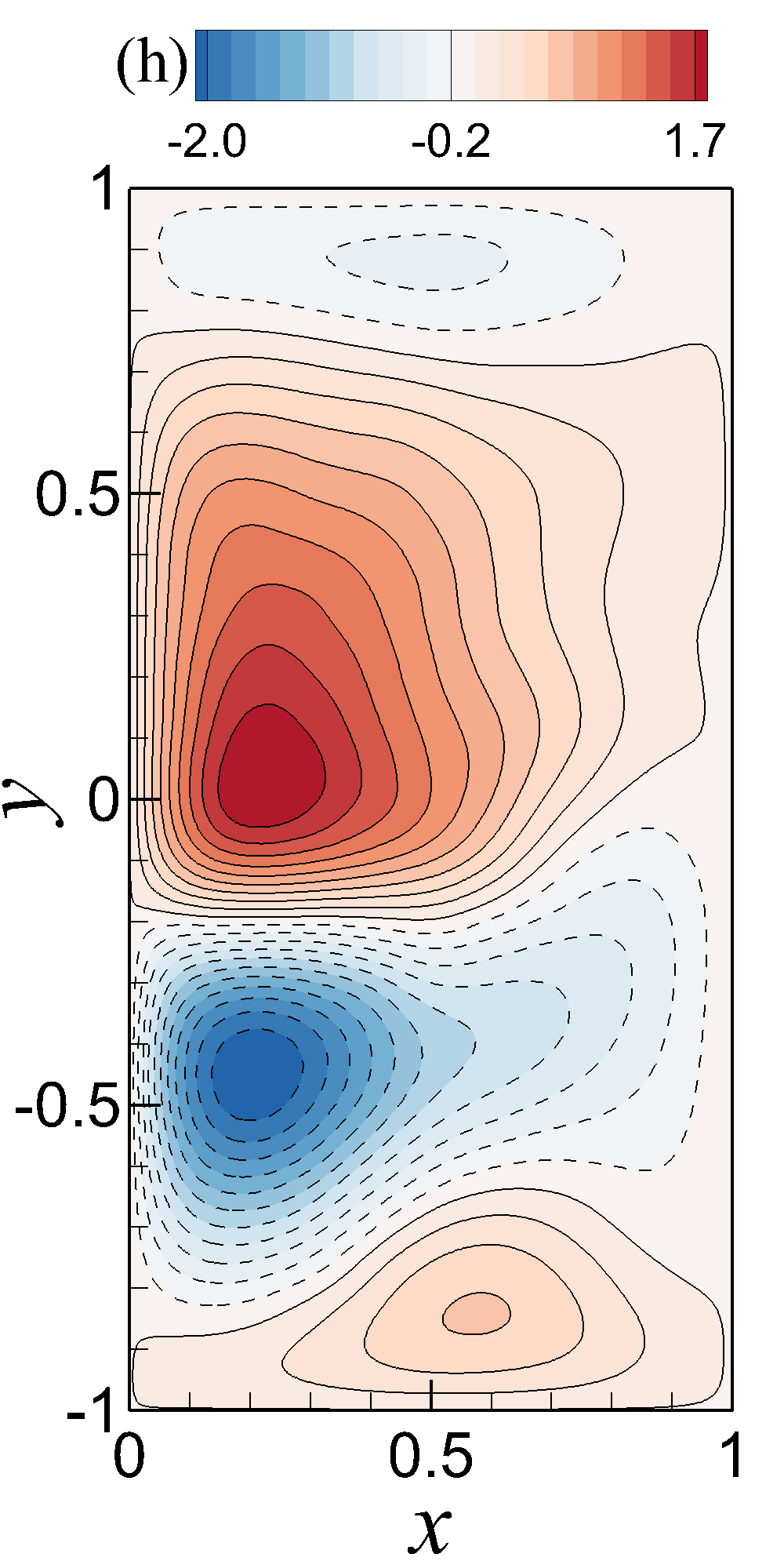}}
}
\caption{A sensitivity test with respect to the dynamic model parameter $\tilde{R}$ showing the mean streamfunction contours for Experiment I. (a) FOM at a resolution of $256 \times 512$; (b) ROM-D ($R=20$) with $\Delta R=2$; (c) ROM-D ($R=20$) with $\Delta R=3$; (d) ROM-D ($R=20$) with $\Delta R=4$; (e) ROM-G with $R=20$ modes; (f) ROM-D ($R=10$) with $\Delta R=2$; (g) ROM-D ($R=10$) with $\Delta R=3$; (h) ROM-D ($R=10$) with $\Delta R=4$. Note that $\Delta R = R - \tilde{R}$.}
\label{fig:5}
\end{figure*}

We further illustrate the sensitivity analysis based on the mean streamfunction contour plots in FIG.~\ref{fig:5} which clearly show that the $R = 20$ for different $\Delta R$ give a very promising prediction of the FOM solution. Furthermore, results do not vary much qualitatively with respect to different $\Delta R$. These results indicate the ROM-D model for $R = 20$ is robust in predicting the true solution for this experiment. Moreover, we provided our quantitative assessments of the ROM-G and ROM-D models for Experiment I in Table~\ref{t1} showing the computational overhead and L\textsubscript{2}-norm error for the mean streamfunction field. It is evident that we can achieve more accuracy for ROM-G model with the increment of $R$ and computational time (for example, ROM-G with $R=80$ gives $1.18 \times 10^{-1}$ of accuracy in $1741.60$ seconds). However, the similar order of accuracy can be obtained by the ROM-D with $R=10$ and $\Delta R =4$ in around $224$ times speedup. Additionally, $\Delta R$ in ROM-D model gives us a freedom (combining with higher $R$) to increase the accuracy of the solution with a little increase in computational time.

\begin{table*}[htbp]
\centering
\caption{Quantitative assessments for Experiment I demonstrating the CPU time in seconds for ROM simulations (using computational time step $\Delta t=2.5\times 10^{-4}$), and L\textsubscript{2}-norm error for the mean streamfunction field (with respect to FOM). Note that the CPU time for the FOM simulation is about 135 hours (between $t=0$ and $t=100$), where computational time step is set $\Delta t=2.5\times 10^{-5}$ due to the CFL restriction of numerical stability for our explicit forward model on the resolution of $256 \times 512$. Offline computing time for solving the eigensystem to find POD modes is about 21 minutes (including about 8 seconds (per 10 modes) for performing numerical integration to calculate the predetermined coefficients). Note that $\Delta R = R-\tilde{R}$.}
\label{t1}
\begin{tabular}{p{0.34\textwidth}p{0.18\textwidth}p{0.24\textwidth}}
\hline\noalign{\smallskip}
& CPU (s) & $|| \psi_{\mbox{\tiny ROM}} - \psi_{\mbox{\tiny FOM}}  ||^2$ \\
\noalign{\smallskip}\hline\noalign{\smallskip}
\multicolumn{2}{l}{\textsl{\underline{Galerkin ROM}}} \\
ROM-G ($R=80$)  & 1741.60 & $1.18 \times 10^{-1}$  \\
ROM-G ($R=60$) & 723.88   & $1.35 \times 10^{0}$  \\
ROM-G ($R=50$) & 432.35 & $1.17 \times 10^{1}$  \\
ROM-G ($R=40$) & 287.59 & $1.19 \times 10^{1}$  \\
ROM-G ($R=30$) & 94.54 & $1.11 \times 10^{2}$  \\
ROM-G ($R=20$) & 29.71 & $3.74 \times 10^{2}$  \\
ROM-G ($R=10$) & 4.43 & $1.07 \times 10^{6}$  \\
\multicolumn{2}{l}{\textsl{\underline{Dynamic ROM}}} \\
ROM-D ($R=10$, $\Delta R =4$)  & 7.75 & $2.39\times 10^{-1}$  \\
ROM-D ($R=10$, $\Delta R =3$)  & 8.58 & $2.67\times 10^{-1}$  \\
ROM-D ($R=10$, $\Delta R =2$)  & 9.75 & $1.69\times 10^{0}$  \\
ROM-D ($R=20$, $\Delta R =4$)  & 68.45 & $5.25\times 10^{-2}$  \\
ROM-D ($R=20$, $\Delta R =3$)  & 72.72 & $3.15\times 10^{-2}$  \\
ROM-D ($R=20$, $\Delta R =2$)  & 77.26 & $7.72\times 10^{-2}$  \\
\noalign{\smallskip}\hline
\end{tabular}
\end{table*}

\subsection{Experiment II: Both data collection and prediction at Re $\mathbf{=450}$, Ro $\mathbf{=0.0036}$}

\begin{figure*}[htbp]
\centering
\mbox{
\subfigure{\includegraphics[width=0.2\textwidth]{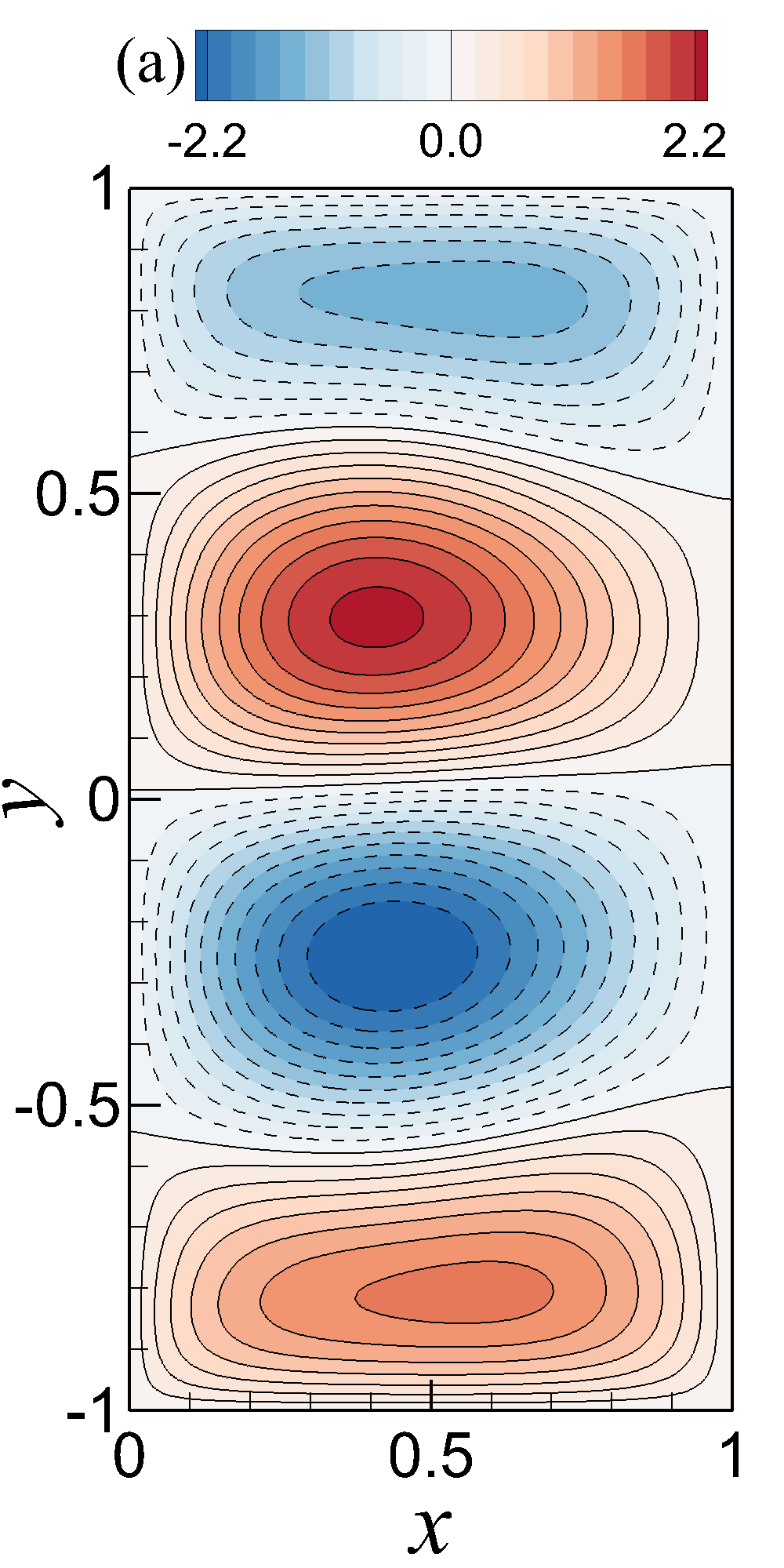}}
\subfigure{\includegraphics[width=0.2\textwidth]{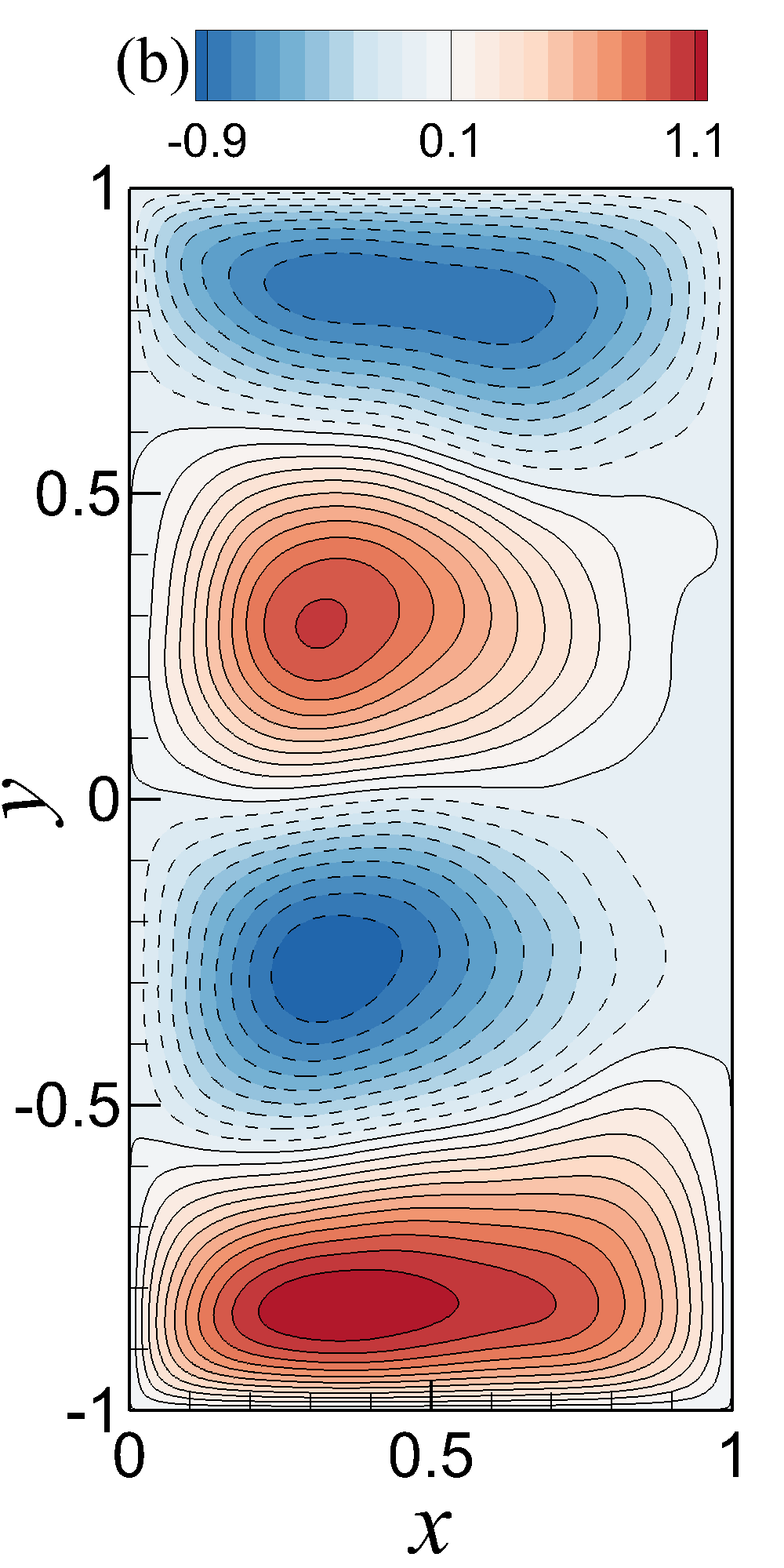}}
\subfigure{\includegraphics[width=0.2\textwidth]{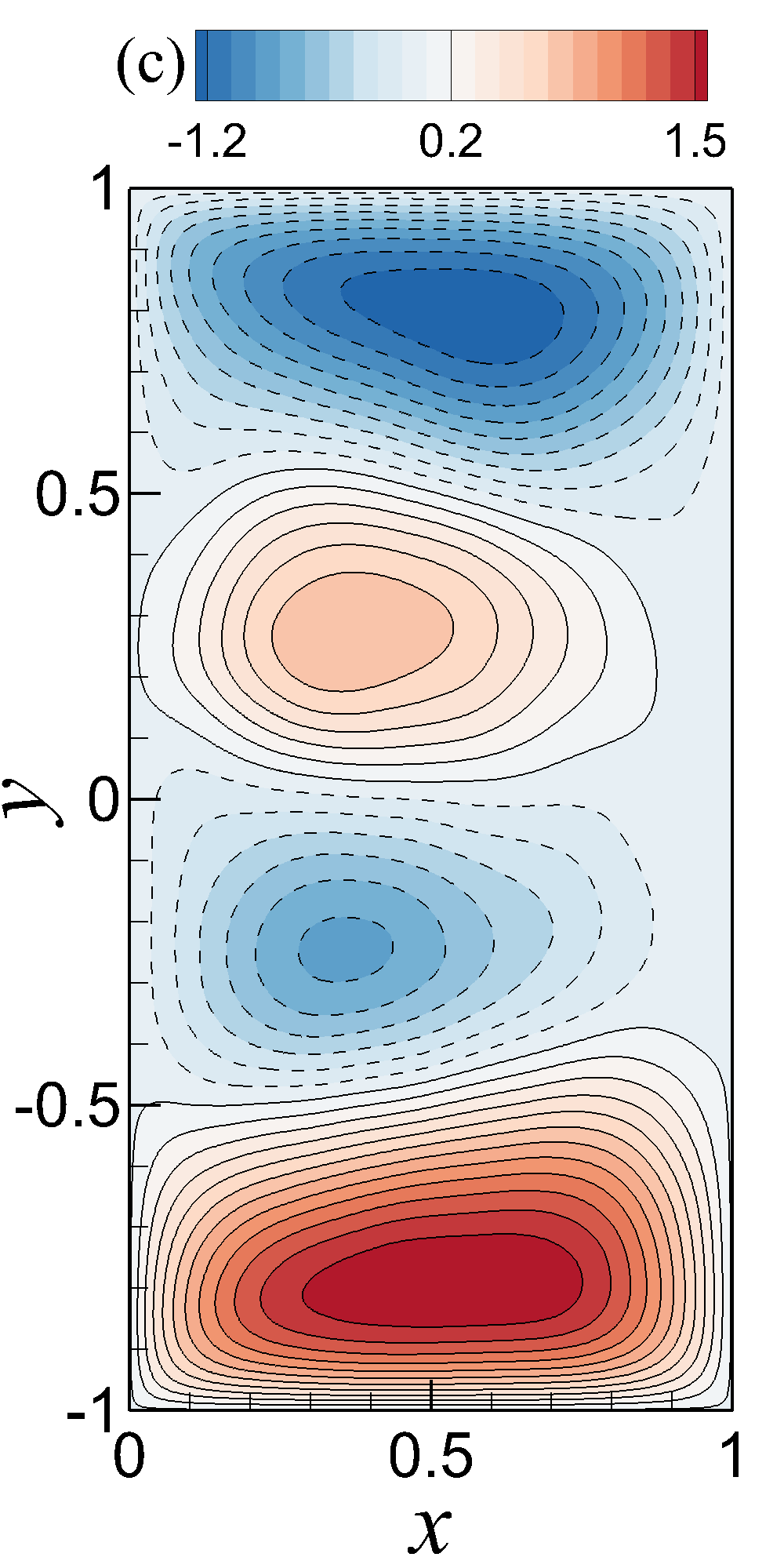}}
\subfigure{\includegraphics[width=0.2\textwidth]{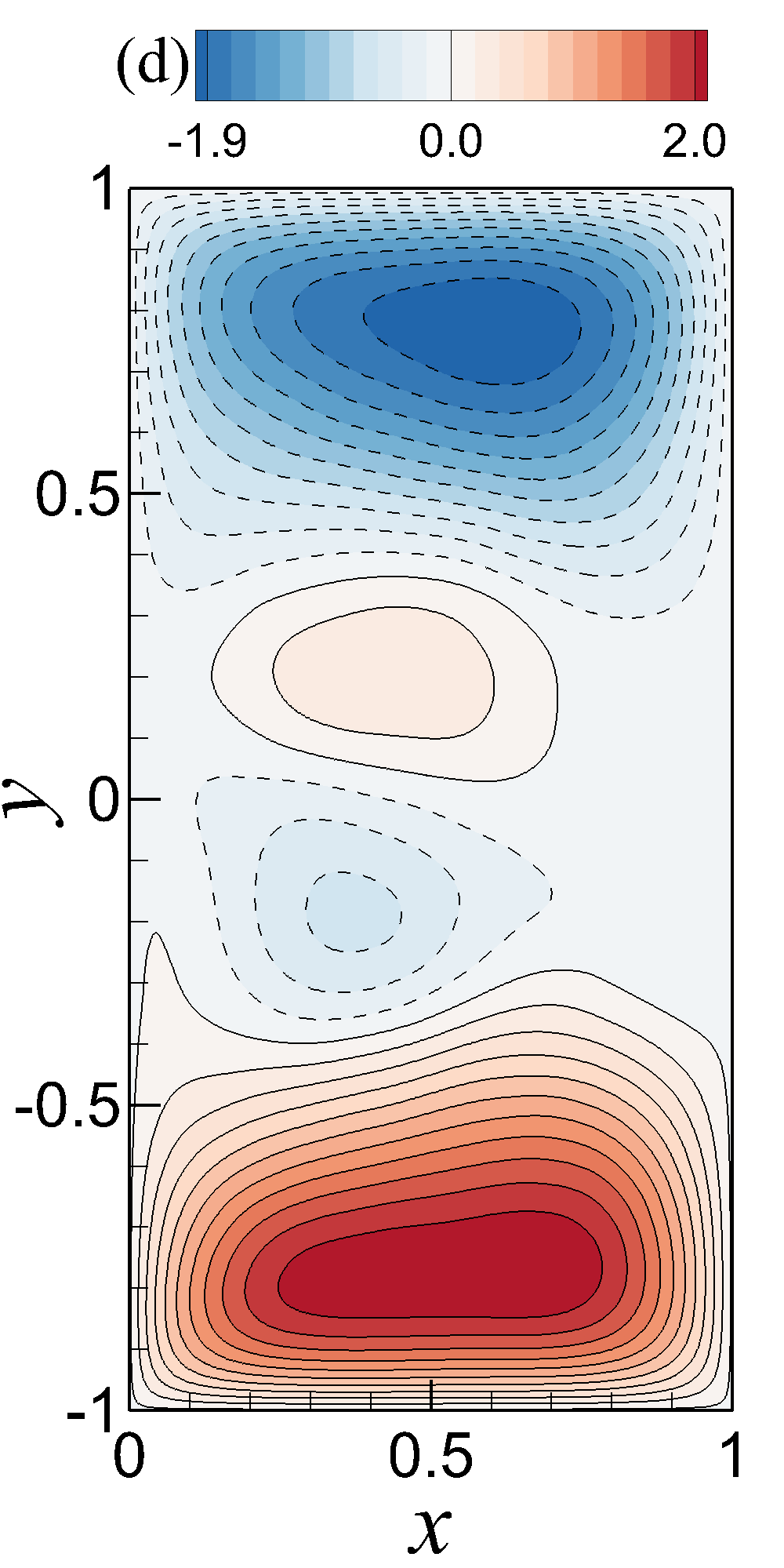}}
}\\
\mbox{
\subfigure{\includegraphics[width=0.2\textwidth]{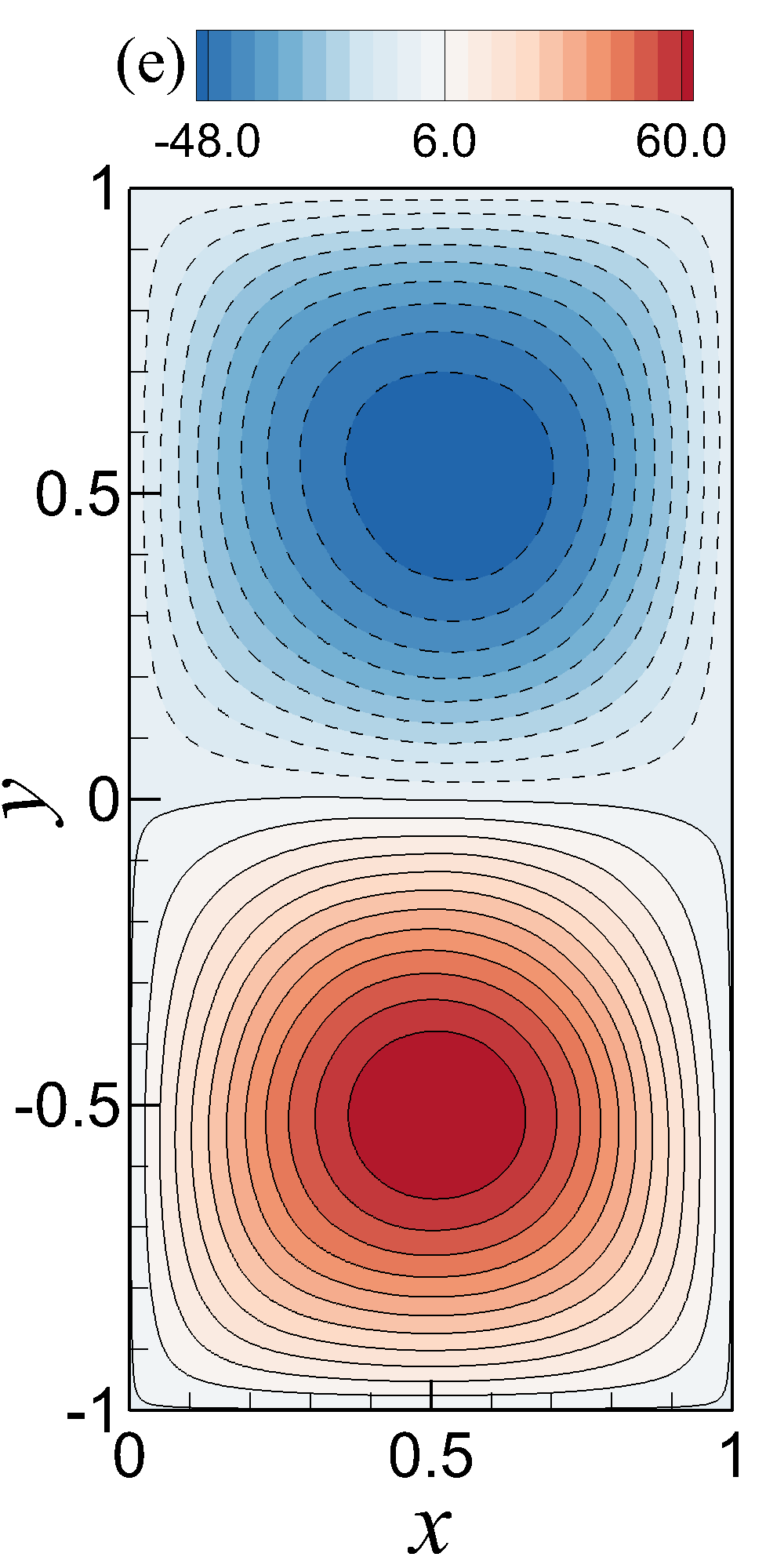}}
\subfigure{\includegraphics[width=0.2\textwidth]{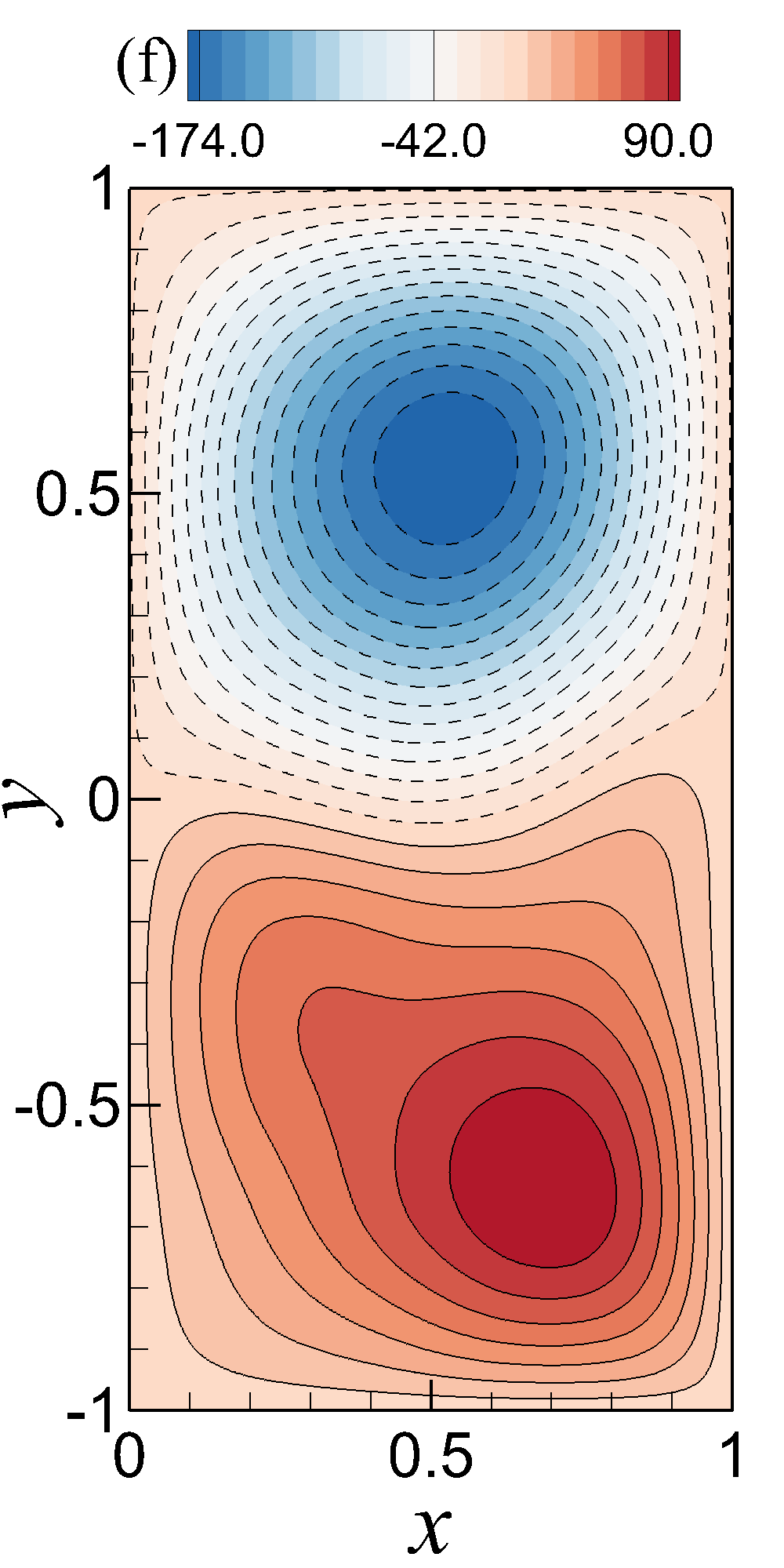}}
\subfigure{\includegraphics[width=0.2\textwidth]{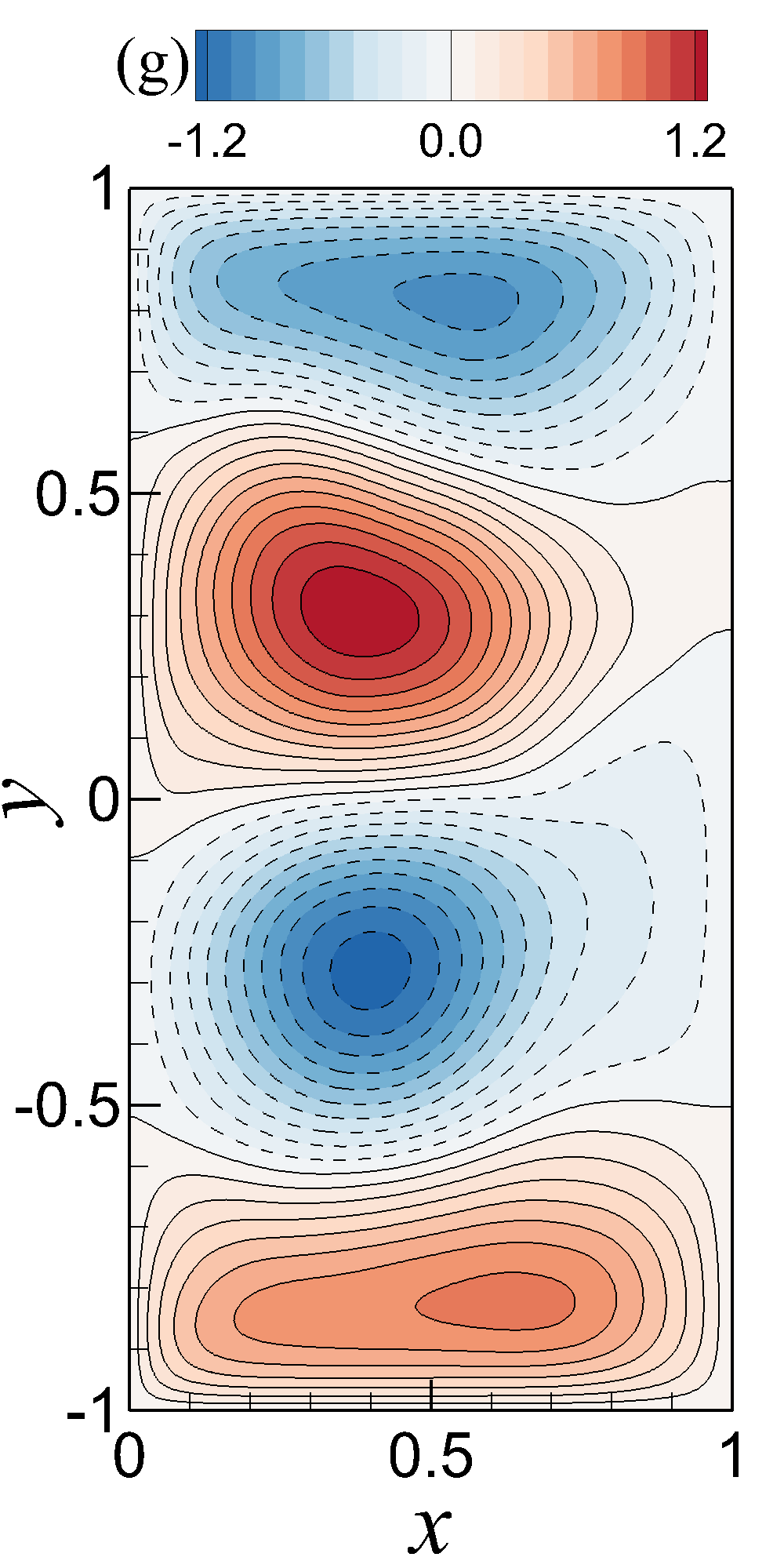}}
\subfigure{\includegraphics[width=0.2\textwidth]{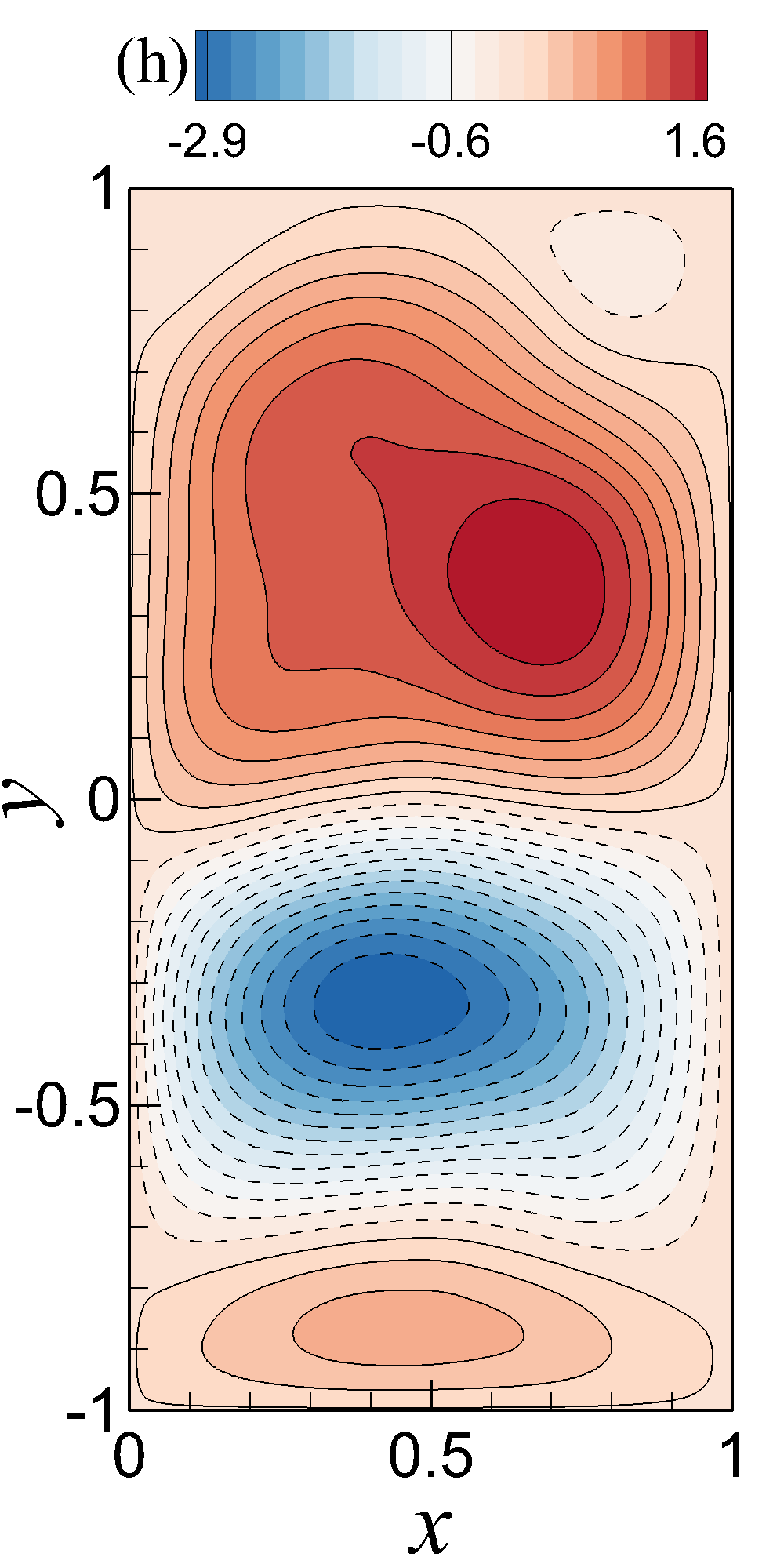}}
}
\caption{Mean streamfunction contours for Experiment II (for training snapshots between $t = 10$ and $t = 50$). (a) FOM at a resolution of $256 \times 512$; (b) ROM-G with $R=50$ modes; (c) ROM-G with $R=40$ modes; (d) ROM-G with $R=30$ modes; (e) ROM-G with $R=20$ modes; (f) ROM-G with $R=10$ modes; (g) proposed ROM-D with $R=20$ modes and $\Delta R=3$; (h) proposed ROM-D with $R=10$ modes and $\Delta R=3$. Note that $\Delta R = R - \tilde{R}$.}
\label{fig:6}
\end{figure*}

\begin{figure*}[htbp]
\centering
\mbox{
\subfigure{\includegraphics[width=0.9\textwidth]{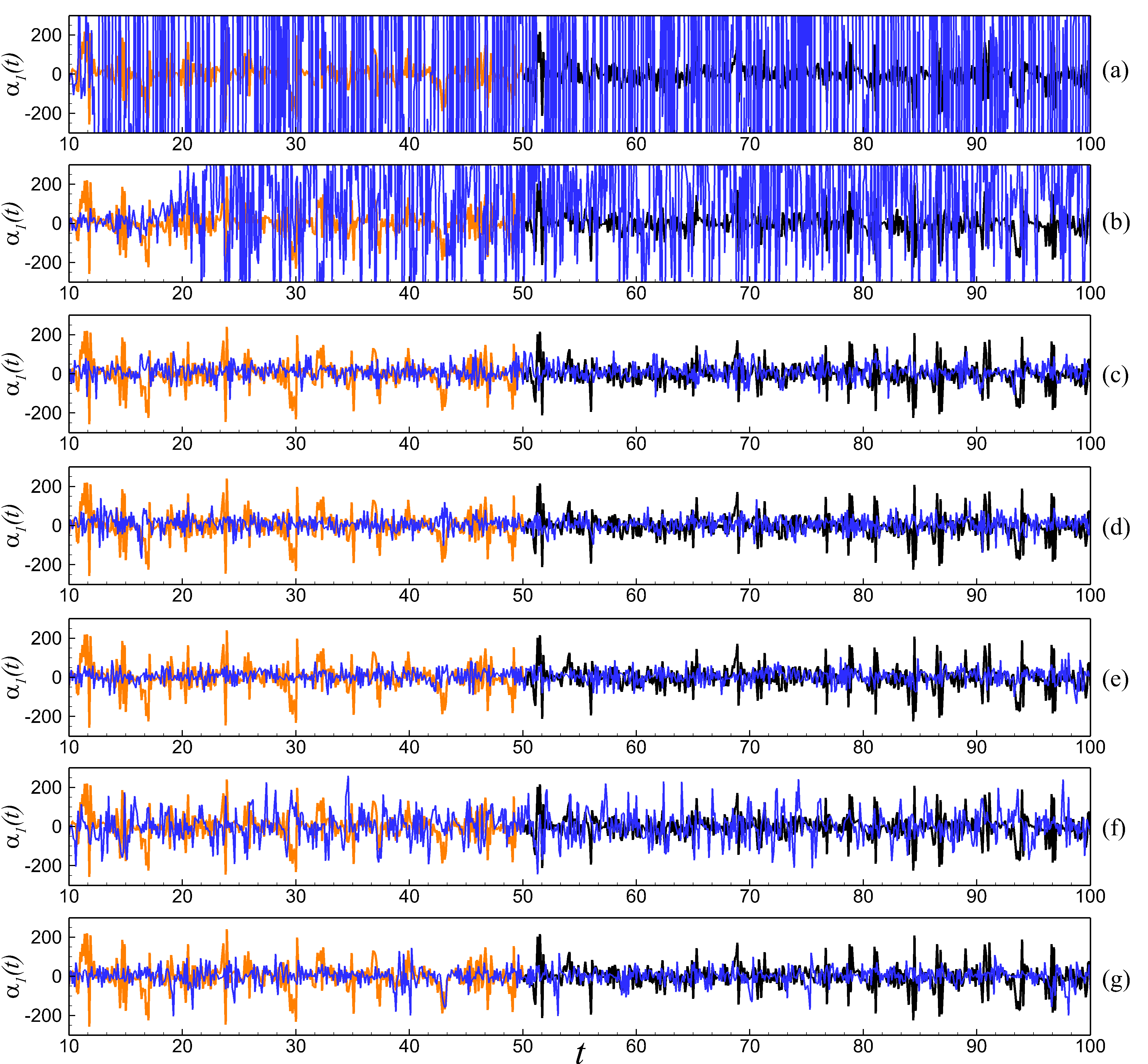}}
}
\caption{Time series of the first modal coefficient for Experiment II. (a) ROM-G with $R=10$ modes; (b) ROM-G with $R=20$ modes; (c) ROM-G with $R=30$ modes; (d) ROM-G with $R=40$ modes; (e) ROM-G with $R=50$ modes; (f) proposed ROM-D with $R=10$ modes and $\Delta R=3$; (g) proposed ROM-D with $R=20$ modes and $\Delta R=3$. Note that $\Delta R = R - \tilde{R}$. True projection data is underlined in each figure with orange (training zone) and black (extended zone).}
\label{fig:7}
\end{figure*}

For Experiment II, we carry out similar analyses as Experiment I using higher (Re, Ro) combination. Since the flow is in highly turbulent regime for this experiment, we expect to observe the four-gyre circulation again in the mean field contours because the eddy flux of potential vorticity balances the vorticity input from the wind stress forcing. Based on the visualizations of the mean streamfunction contour plots in FIG.~\ref{fig:6}, Experiment II reveals that the ROM-G gives a physical result from $R = 30$ although $R = 20$ and $R = 10$ show a nonphysical two-gyre circulation instead of four-gyre circulation (i.e., due to the instability of the POD-G approach with such number of retained modes). In contrast, the ROM-D models with $R = 10$ and $R = 20$ exhibit a good estimation of the FOM solution whereas the $R = 20$ shows the visible four-gyre circulation. The time series evolution for the first modal coefficient plots in FIG.~\ref{fig:7} gives us a better statistical view on the comparative performance of both models. As we can see in the first two rows of the figure that the solutions obtained by ROM-G with $R = 10$ and $R = 20$ modes become nonphysical after a while whereas the true physics suggests a statistically steady flow field after $t = 10$. At the same time range, the $R = 30$, $R = 40$ and $R = 50$  modes for ROM-G give a statistically steady and satisfactory prediction of the true solution which is consistent with the findings of the meanstream function contour plots. On the other hand, the ROM-D model with both $R = 10$ and $R = 20$ modes give an excellent prediction of the truth compared to the ROM-G solutions in lower modes. As we have seen in the POD analysis in FIG.~\ref{fig:1}, Experiment II displays a statistically more stable time series evolution (compared to Experiment I), and it is expected that comparatively more energy will be accumulated in lower $R$ for this case. 

\begin{figure*}[htbp]
\centering
\mbox{
\subfigure{\includegraphics[width=0.9\textwidth]{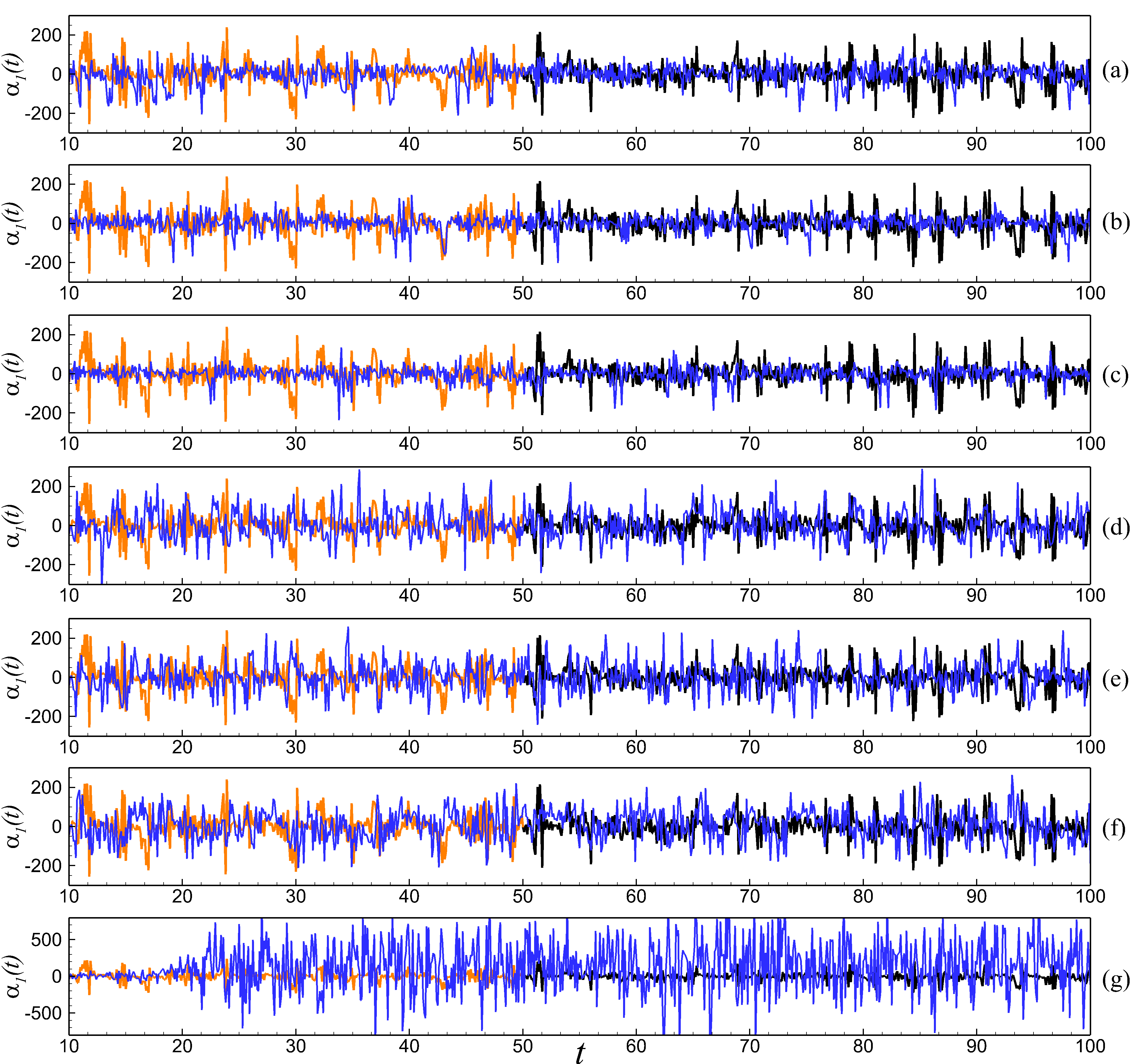}}
}
\caption{A sensitivity test with respect to the dynamic model parameter $\tilde{R}$ showing the time series of the first modal coefficient for Experiment II. (a) ROM-D ($R=20$) with $\Delta R=2$; (b) ROM-D ($R=20$) with $\Delta R=3$; (c) ROM-D ($R=20$) with $\Delta R=4$; (d) ROM-D ($R=10$) with $\Delta R=2$; (e) ROM-D ($R=10$) with $\Delta R=3$; (f) ROM-D ($R=10$) with $\Delta R=4$; (g) ROM-G with $R=20$ modes. Note that $\Delta R = R - \tilde{R}$. True projection data is presented in each figure with orange (training zone) and black (extended zone).}
\label{fig:8}
\end{figure*}

\begin{figure*}[htbp]
\centering
\mbox{
\subfigure{\includegraphics[width=0.2\textwidth]{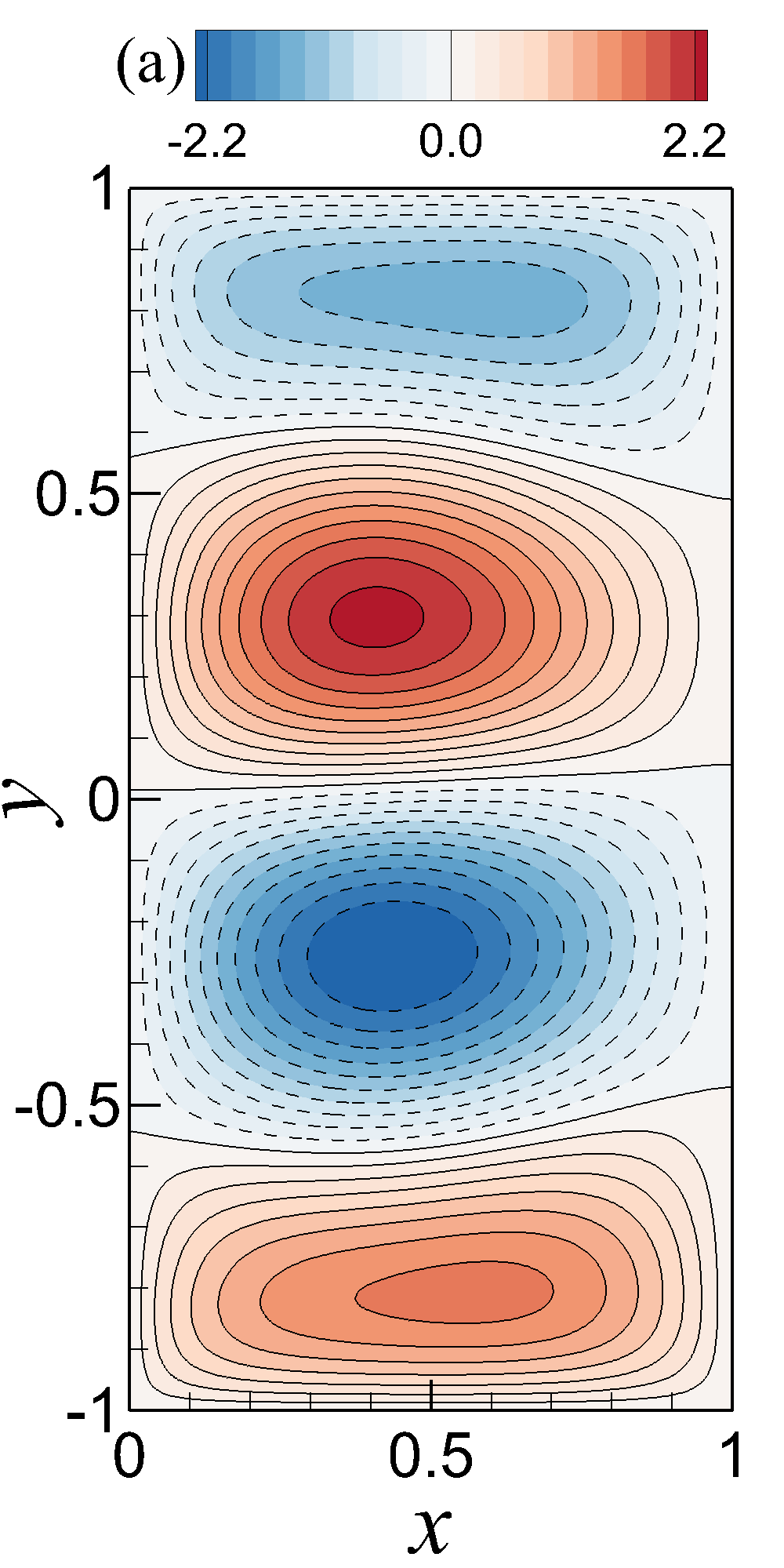}}
\subfigure{\includegraphics[width=0.2\textwidth]{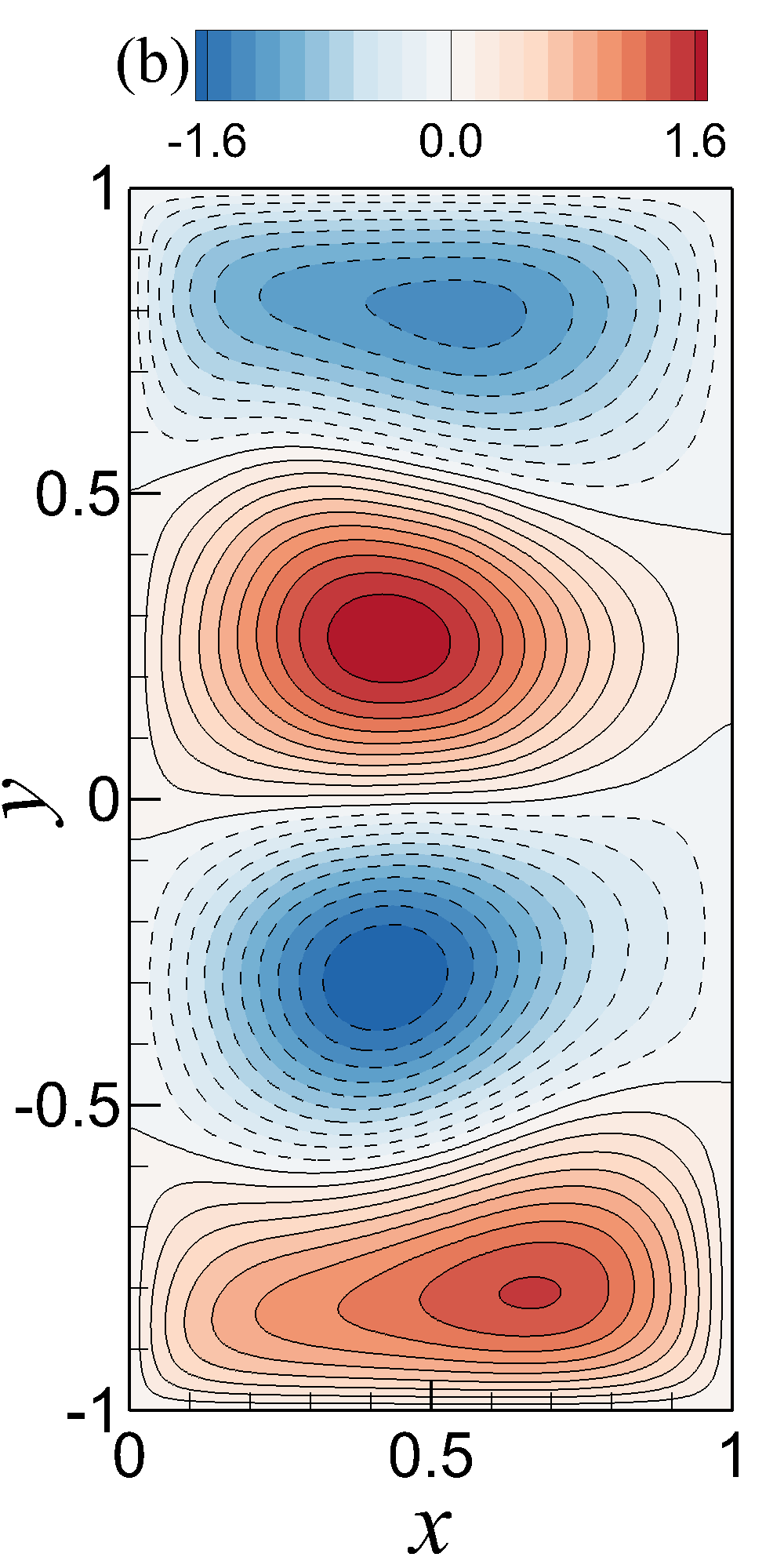}}
\subfigure{\includegraphics[width=0.2\textwidth]{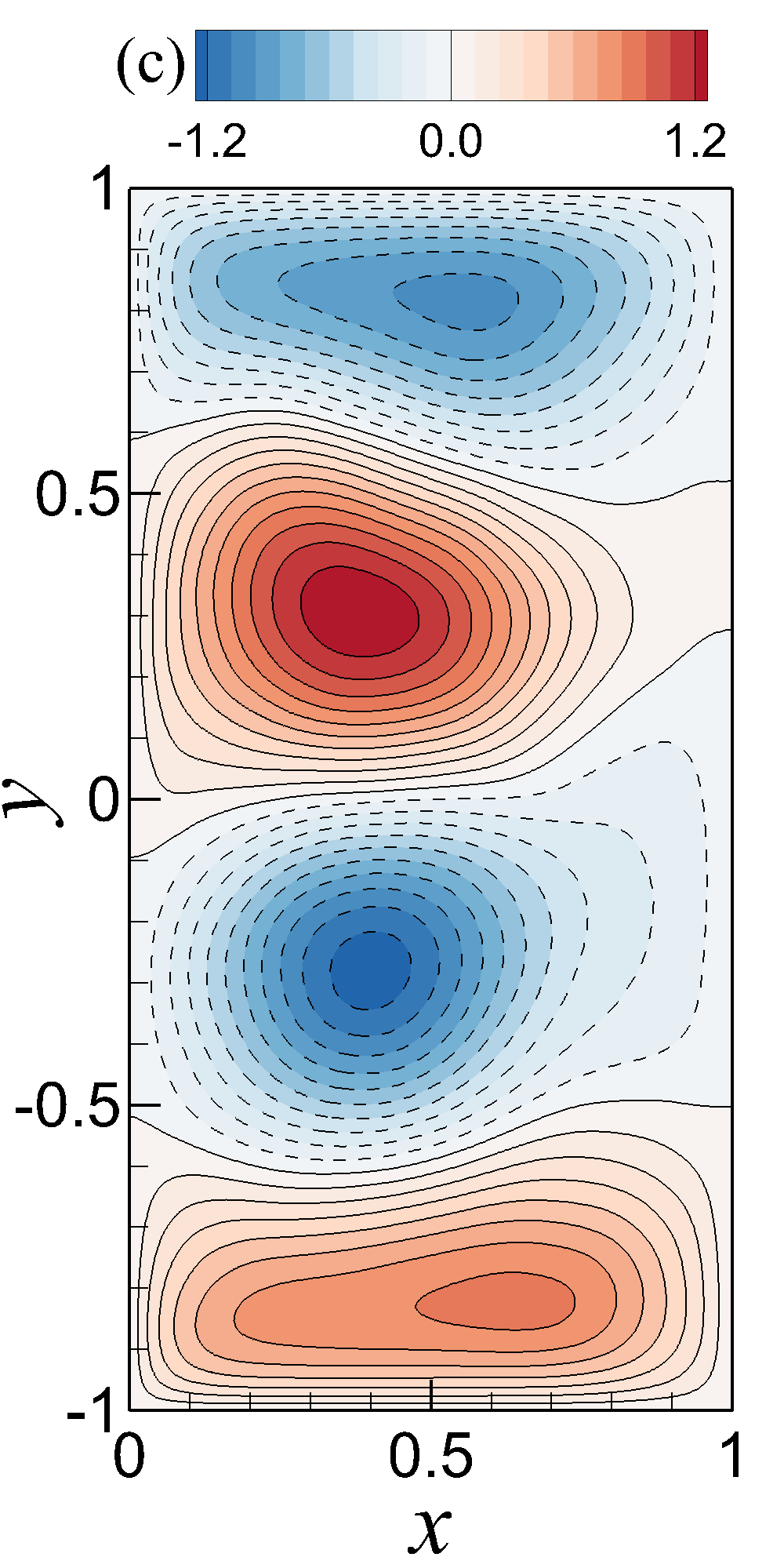}}
\subfigure{\includegraphics[width=0.2\textwidth]{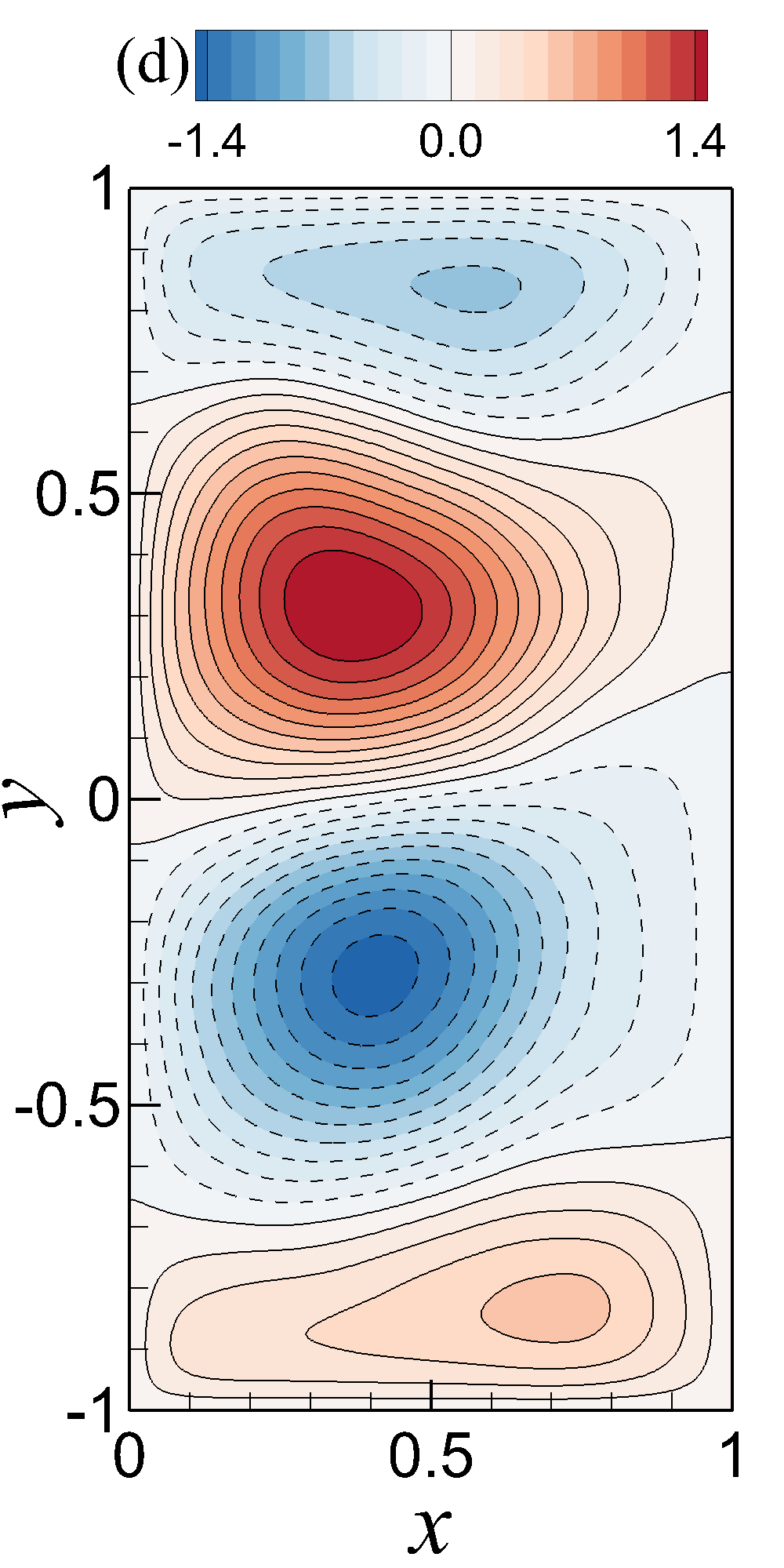}}
}\\
\mbox{
\subfigure{\includegraphics[width=0.2\textwidth]{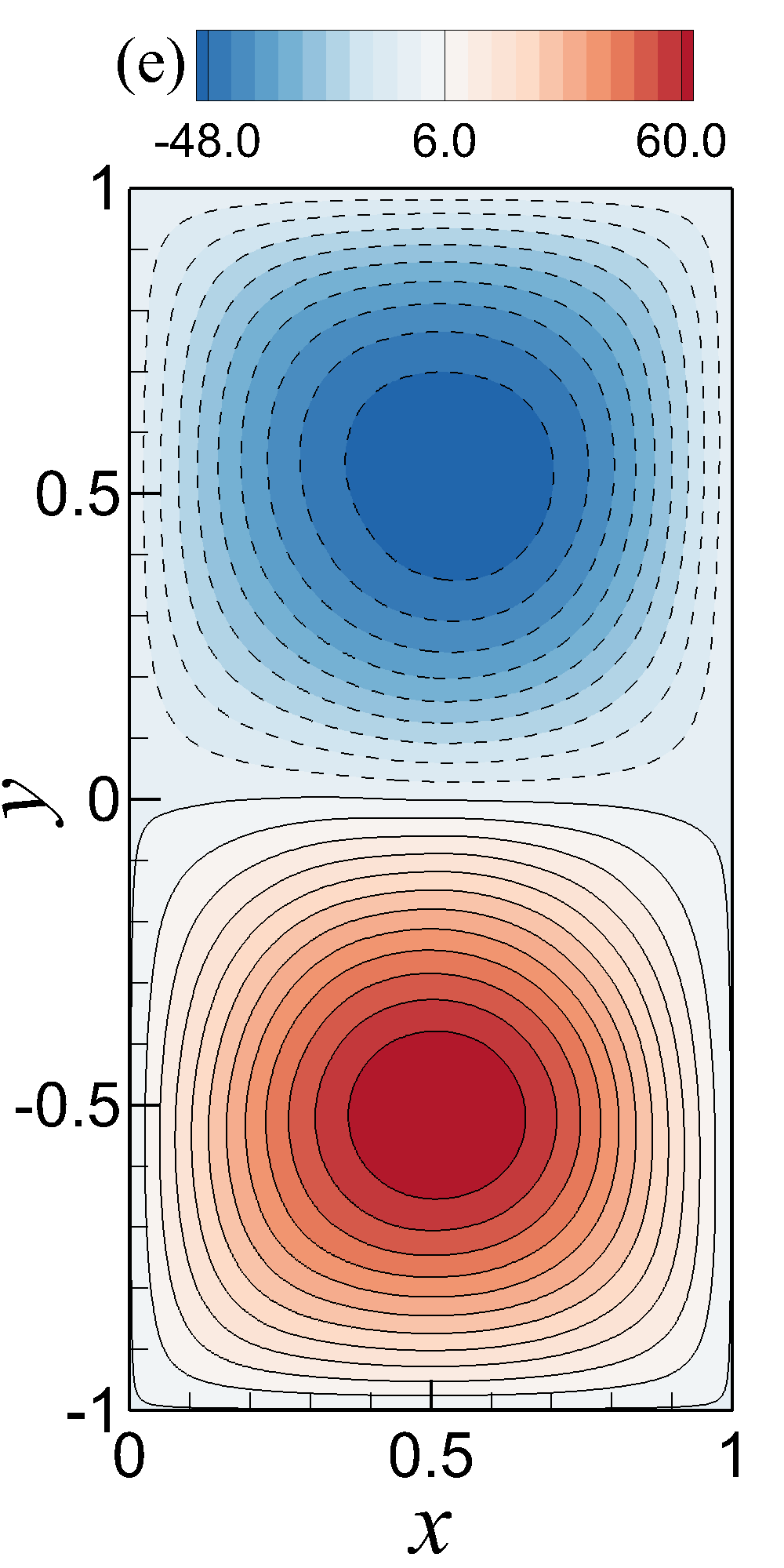}}
\subfigure{\includegraphics[width=0.2\textwidth]{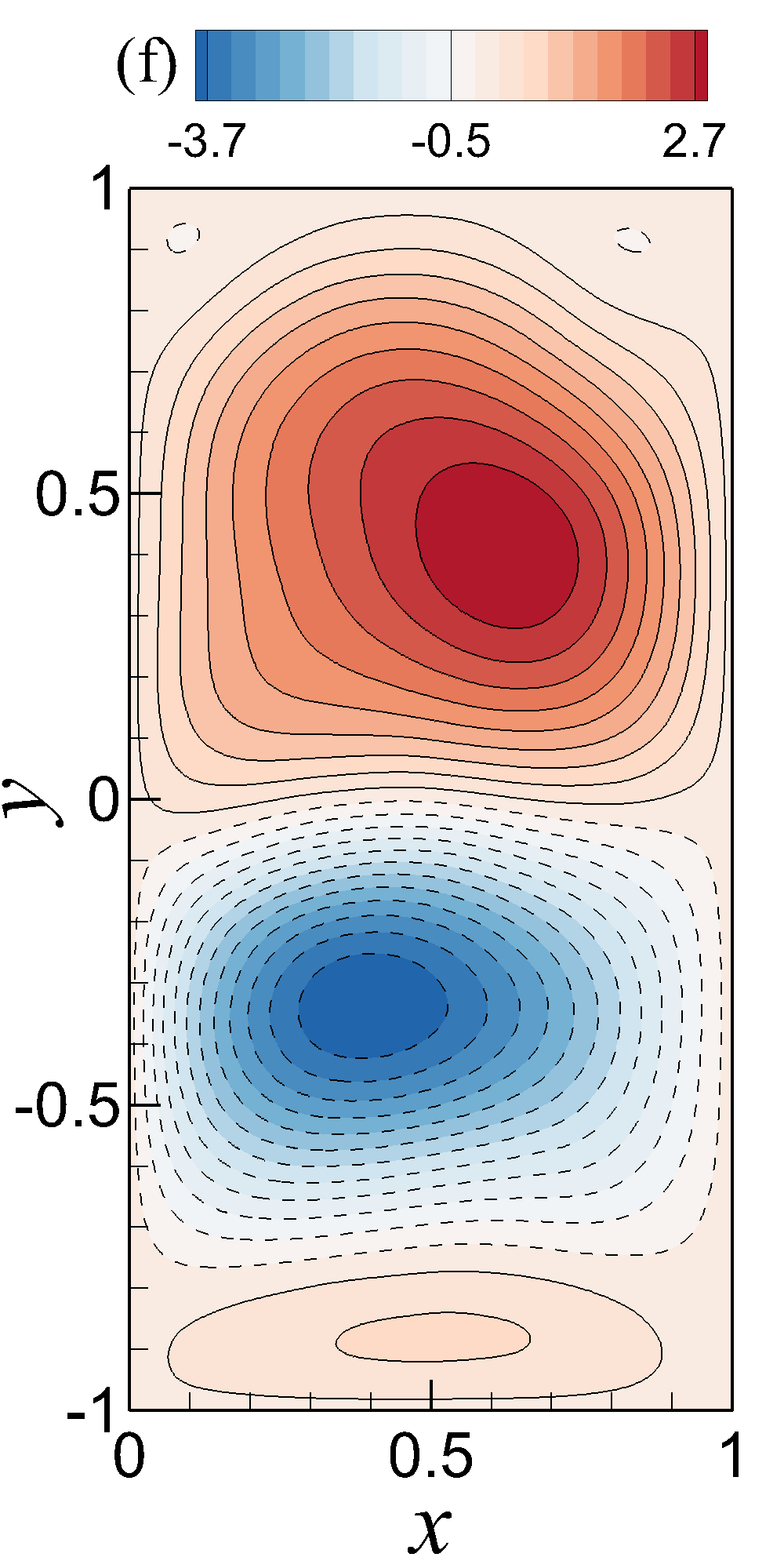}}
\subfigure{\includegraphics[width=0.2\textwidth]{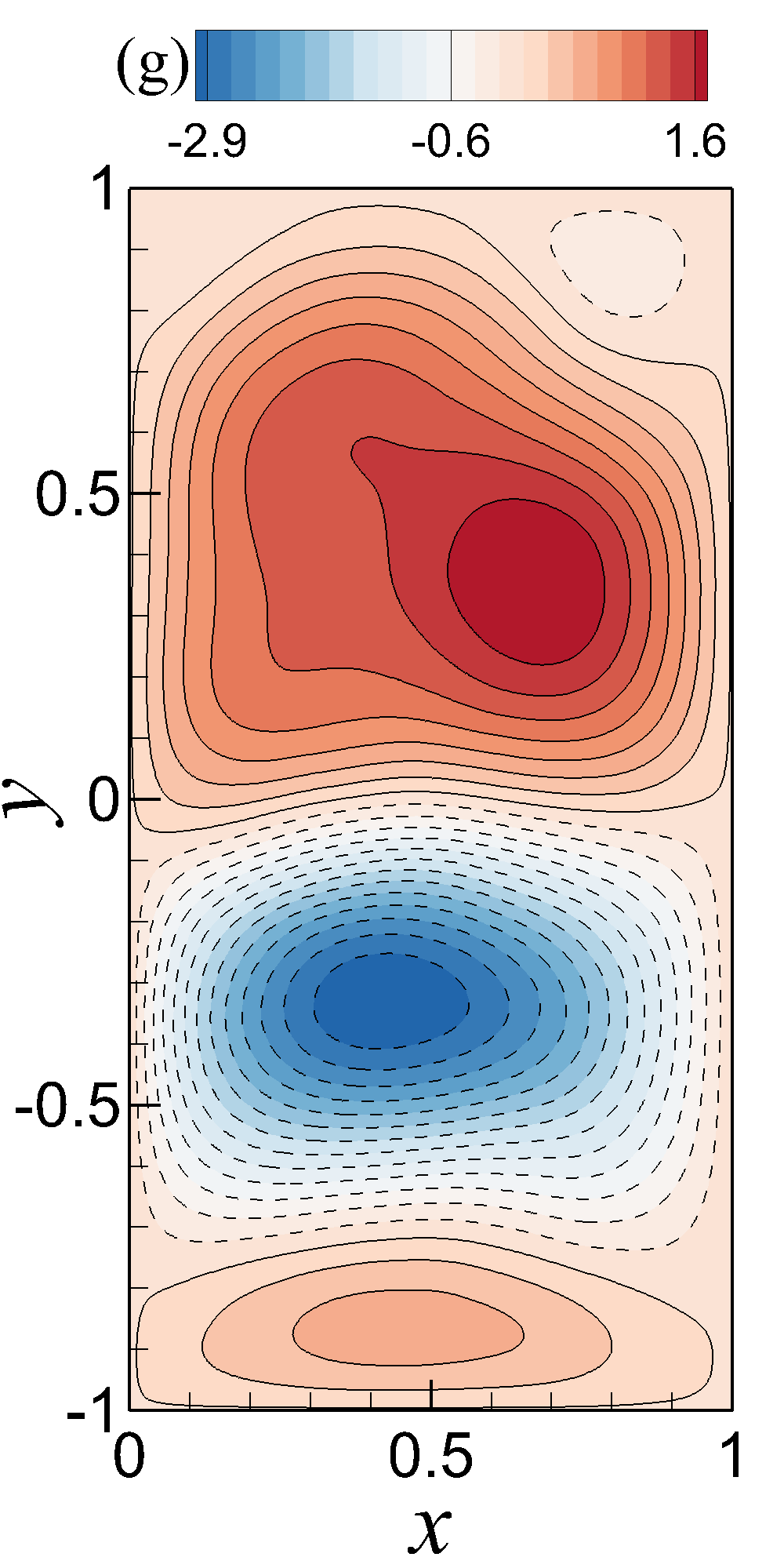}}
\subfigure{\includegraphics[width=0.2\textwidth]{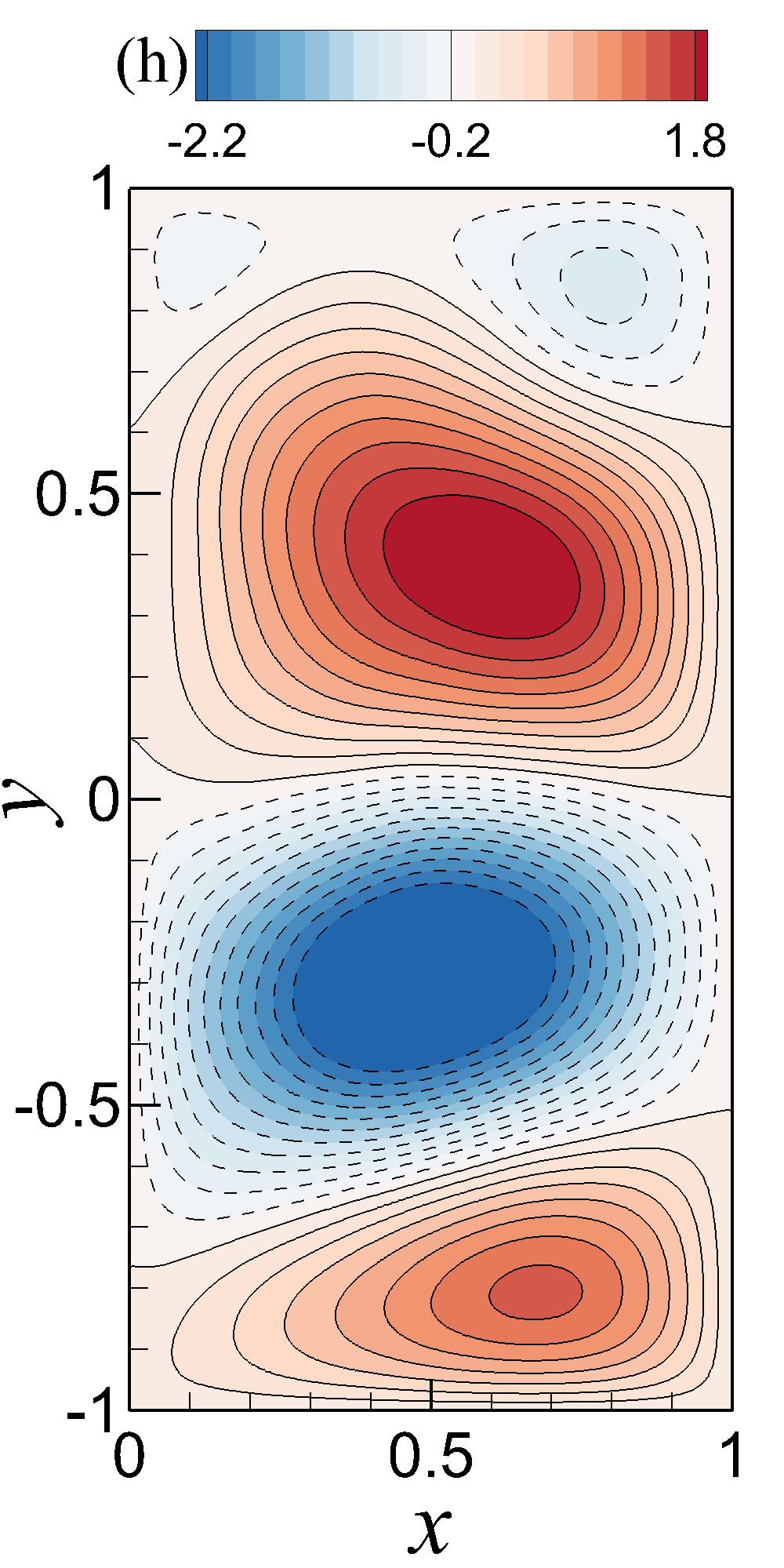}}
}
\caption{A sensitivity test with respect to the dynamic model parameter $\tilde{R}$ showing the mean streamfunction contours for Experiment II. (a) FOM at a resolution of $256 \times 512$; (b) ROM-D ($R=20$) with $\Delta R=2$; (c) ROM-D ($R=20$) with $\Delta R=3$; (d) ROM-D ($R=20$) with $\Delta R=4$; (e) ROM-G with $R=20$ modes; (f) ROM-D ($R=10$) with $\Delta R=2$; (g) ROM-D ($R=10$) with $\Delta R=3$; (h) ROM-D ($R=10$) with $\Delta R=4$. Note that $\Delta R = R - \tilde{R}$.}
\label{fig:9}
\end{figure*}

Next, we perform the sensitivity tests for Experiment II and it can be observed in FIG.~\ref{fig:8} that the ROM-D model is showing consistent predictions for even $R = 10$ invariant to the change in the value of $\Delta R$. Nonetheless, the ROM-G with $R = 20$ prediction is nonphysical yet again after certain time. Similarly, the mean streamfunction contours in FIG.~\ref{fig:9} indicates the ROM-D model is very robust for $R = 20$ and also, showing a good prediction for $R = 10$. In Table~\ref{t2}, we report the computational time and L\textsubscript{2}-norm error of the ROM-G and ROM-D model simulation results for Experiment II. Similar to Experiment I, we can obtain equivalent order of accuracy as ROM-G ($R = 80$) in ROM-D ($R = 10$, $\Delta R = 4$) with around $219$ times reduction in overall computational time. It is also apparent that we can gain more accurate solutions for $R = 20$ with different $\Delta R$ using ROM-D model.

\begin{table*}[htbp]
\centering
\caption{Quantitative assessments for Experiment II demonstrating the CPU time in seconds for ROM simulations (using computational time step $\Delta t=2.5\times 10^{-4}$), and L\textsubscript{2}-norm error for the mean streamfunction field (with respect to FOM). Note that the CPU time for the FOM simulation is about 130 hours (between $t=0$ and $t=100$), where computational time step is set $\Delta t=2.5\times 10^{-5}$ due to the CFL restriction of numerical stability for our explicit forward model on the resolution of $256 \times 512$. Offline computing time for solving the eigensystem to find POD modes is about 22 minutes (including about 8 seconds (per 10 modes) for performing numerical integration to calculate the predetermined coefficients). Note that $\Delta R = R-\tilde{R}$.}
\label{t2}
\begin{tabular}{p{0.34\textwidth}p{0.18\textwidth}p{0.24\textwidth}}
\hline\noalign{\smallskip}
& CPU (s) & $|| \psi_{\mbox{\tiny ROM}} - \psi_{\mbox{\tiny FOM}}  ||^2$ \\
\noalign{\smallskip}\hline\noalign{\smallskip}
\multicolumn{2}{l}{\textsl{\underline{Galerkin ROM}}} \\
ROM-G ($R=80$)  & 1752.79 & $3.84 \times 10^{-1}$  \\
ROM-G ($R=60$) & 722.02   & $3.33 \times 10^{-1}$  \\
ROM-G ($R=50$) & 554.14 & $3.63 \times 10^{-1}$  \\
ROM-G ($R=40$) & 218.06 & $4.33 \times 10^{-1}$  \\
ROM-G ($R=30$) & 96.00 & $9.99 \times 10^{-1}$  \\
ROM-G ($R=20$) & 29.32 & $9.87 \times 10^{2}$  \\
ROM-G ($R=10$) & 4.36 & $5.59 \times 10^{3}$  \\
\multicolumn{2}{l}{\textsl{\underline{Dynamic ROM}}} \\
ROM-D ($R=10$, $\Delta R =4$)  & 8.00 & $3.55\times 10^{-1}$  \\
ROM-D ($R=10$, $\Delta R =3$)  & 8.96 & $7.82\times 10^{-1}$  \\
ROM-D ($R=10$, $\Delta R =2$)  & 9.51 & $1.51\times 10^{0}$  \\
ROM-D ($R=20$, $\Delta R =4$)  & 68.32 & $3.02\times 10^{-1}$  \\
ROM-D ($R=20$, $\Delta R =3$)  & 72.70 & $2.88\times 10^{-1}$  \\
ROM-D ($R=20$, $\Delta R =2$)  & 77.27 & $1.22\times 10^{-1}$  \\
\noalign{\smallskip}\hline
\end{tabular}
\end{table*}

\subsection{Experiment III: Data collection at Re $\mathbf{=450}$, Ro $\mathbf{=0.0036}$, prediction at Re $\mathbf{=200}$, Ro $\mathbf{=0.0016}$}

The first two experiments clearly address the improvements we can achieve through the dynamic closure ROM approach for the prediction of the flow field within the range of training data. However, to further investigate the prediction capability of the proosed ROM-D model for out-of-sample flow condition, we perform an extrapolatory predictive performance analysis where we collect the training data for POD basis function generation at higher (Re, Ro) combination, i.e., at (Re $=450$, Ro $=0.0036$) and then, test to predict the flow field at (Re $=200$, Ro $=0.0016$). Indeed, it should be challenging since we have seen the analyses in the previous sections that the (Re $=200$, Ro $=0.0016$) combination introduces comparatively uneven fluctuations than the higher (Re, Ro) flow condition. As a result, the lower (Re, Ro) flow condition requires higher POD modes to capture greater percentage of energy of the system. 

\begin{figure*}[htbp]
\centering
\mbox{
\subfigure{\includegraphics[width=0.2\textwidth]{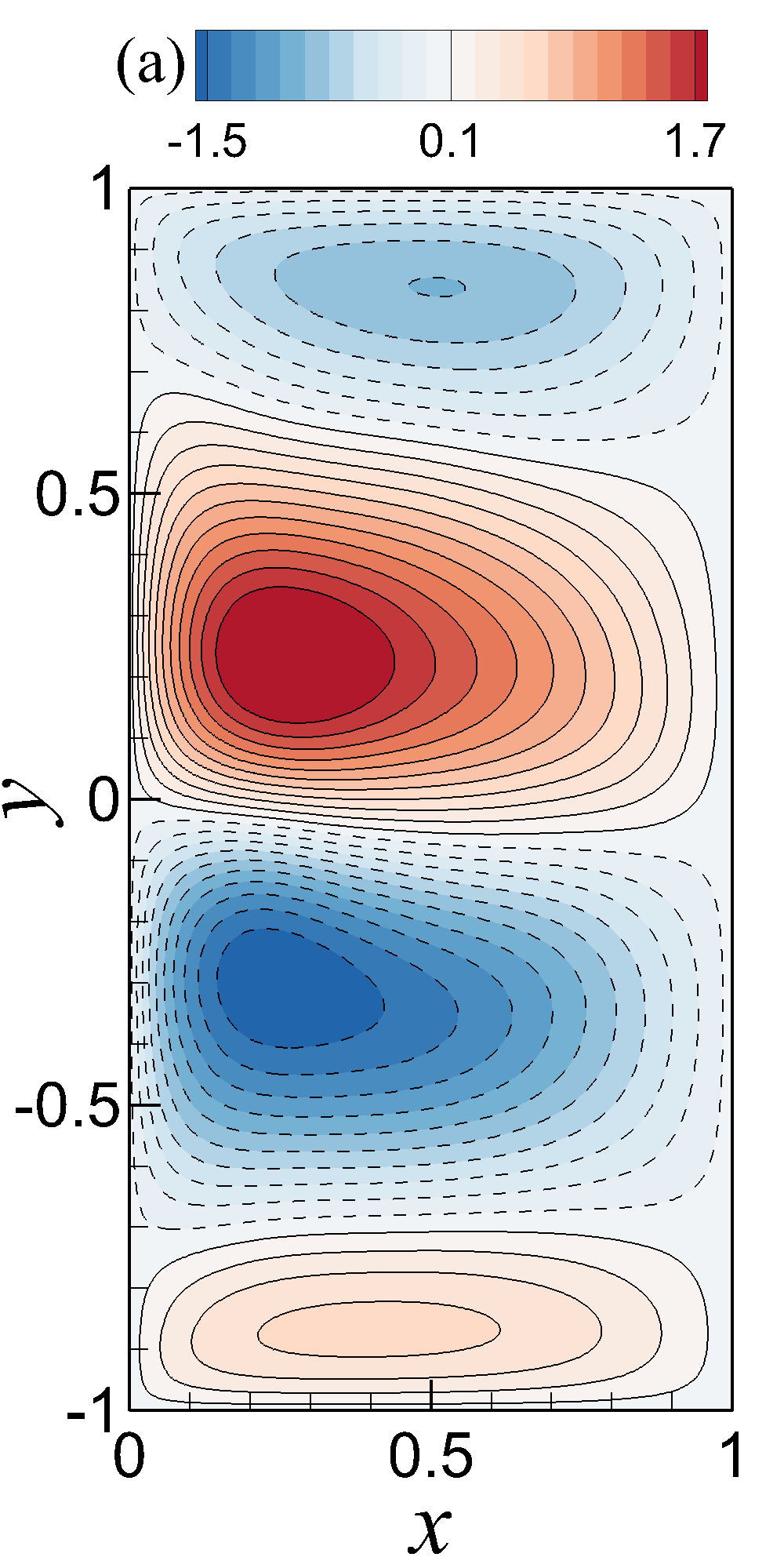}}
\subfigure{\includegraphics[width=0.2\textwidth]{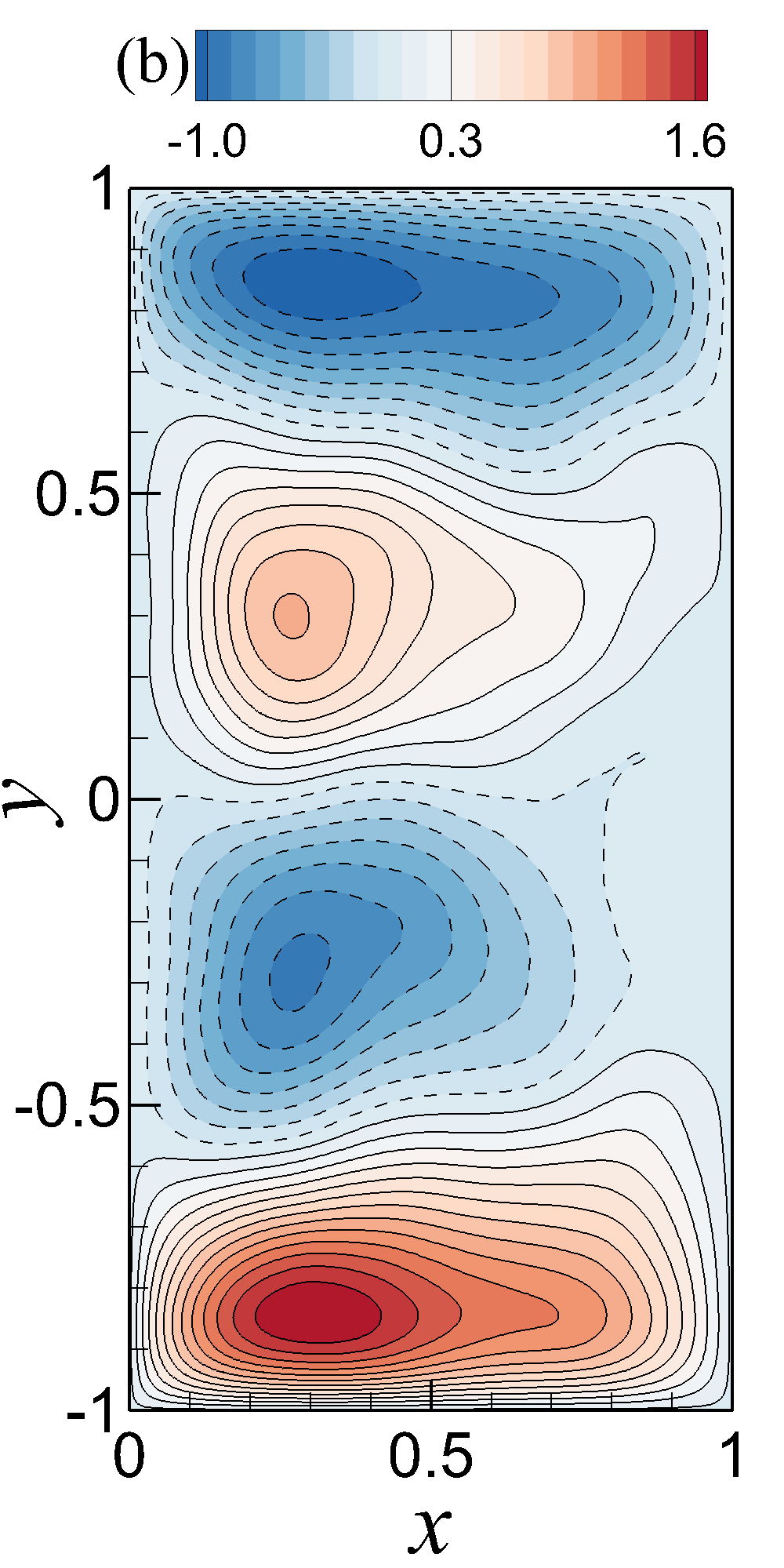}}
\subfigure{\includegraphics[width=0.2\textwidth]{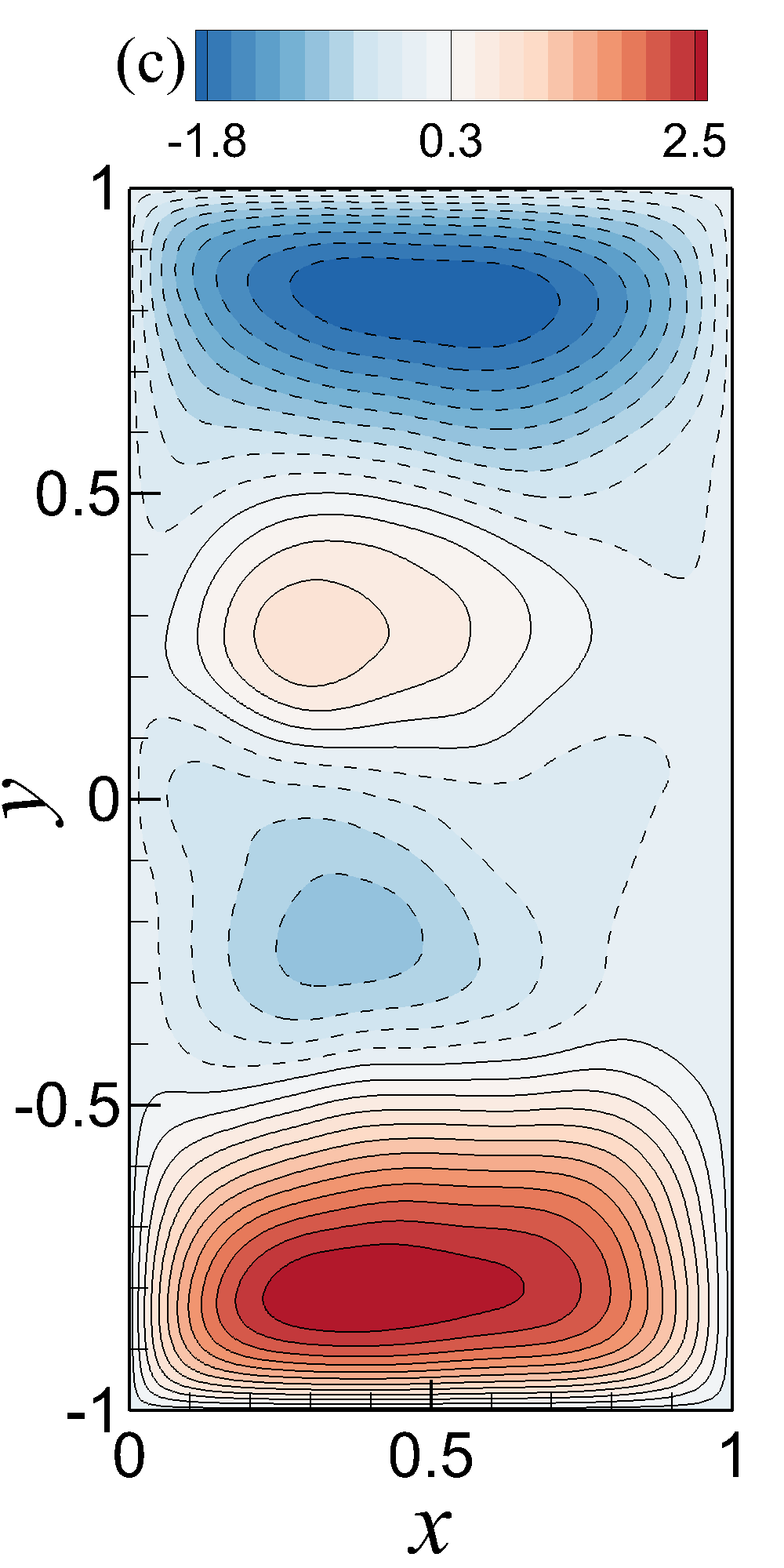}}
\subfigure{\includegraphics[width=0.2\textwidth]{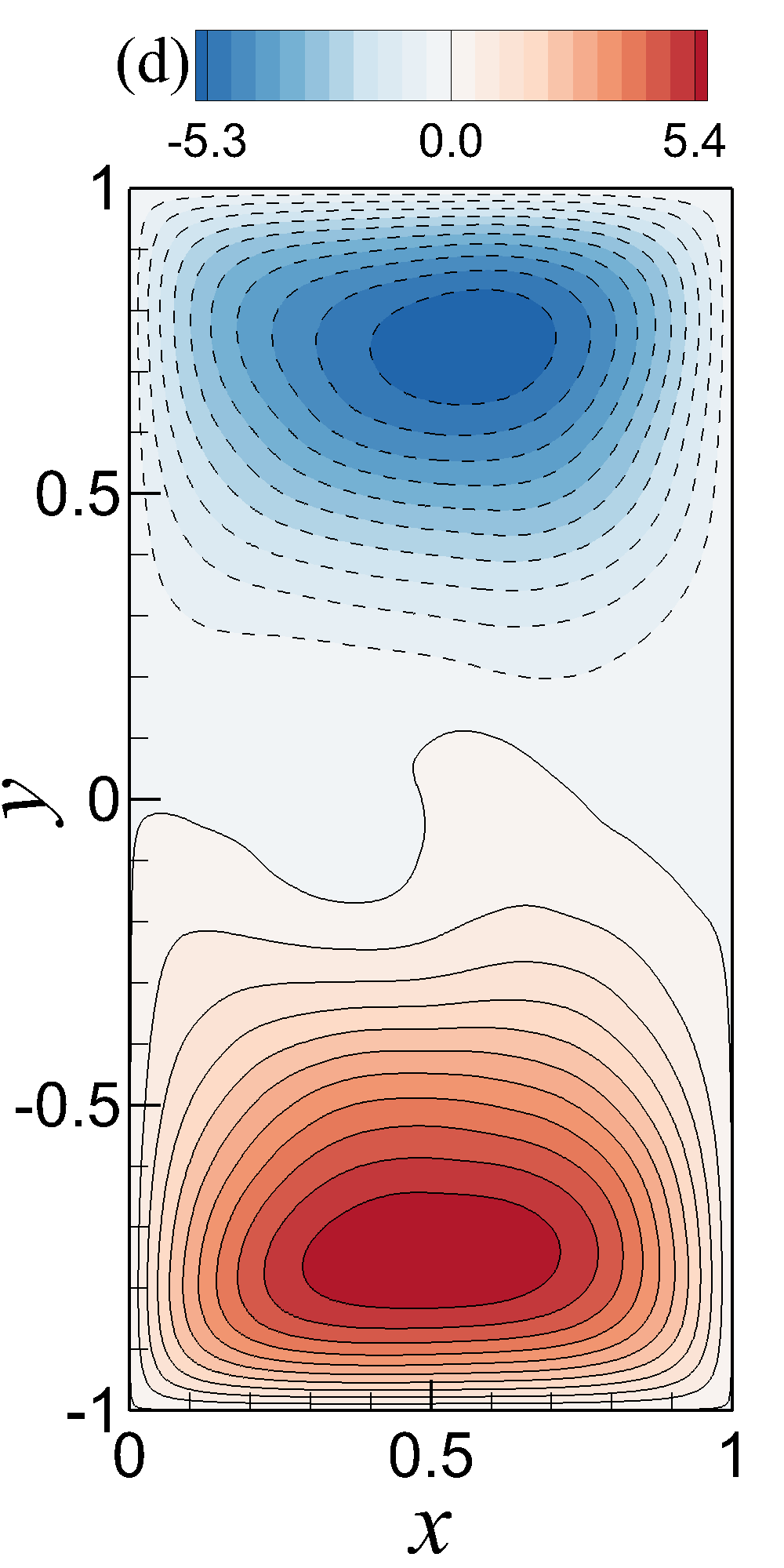}}
}\\
\mbox{
\subfigure{\includegraphics[width=0.2\textwidth]{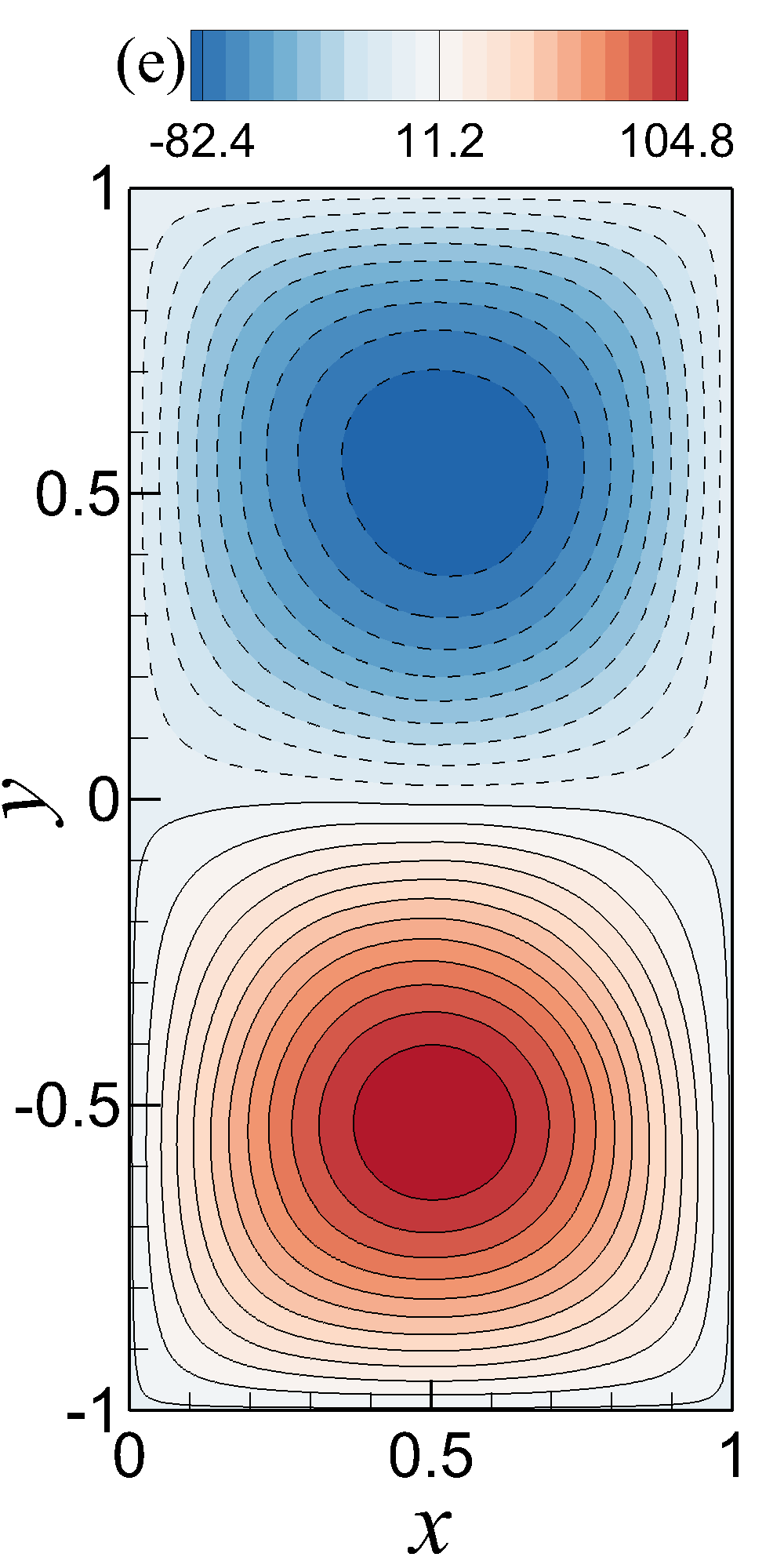}}
\subfigure{\includegraphics[width=0.2\textwidth]{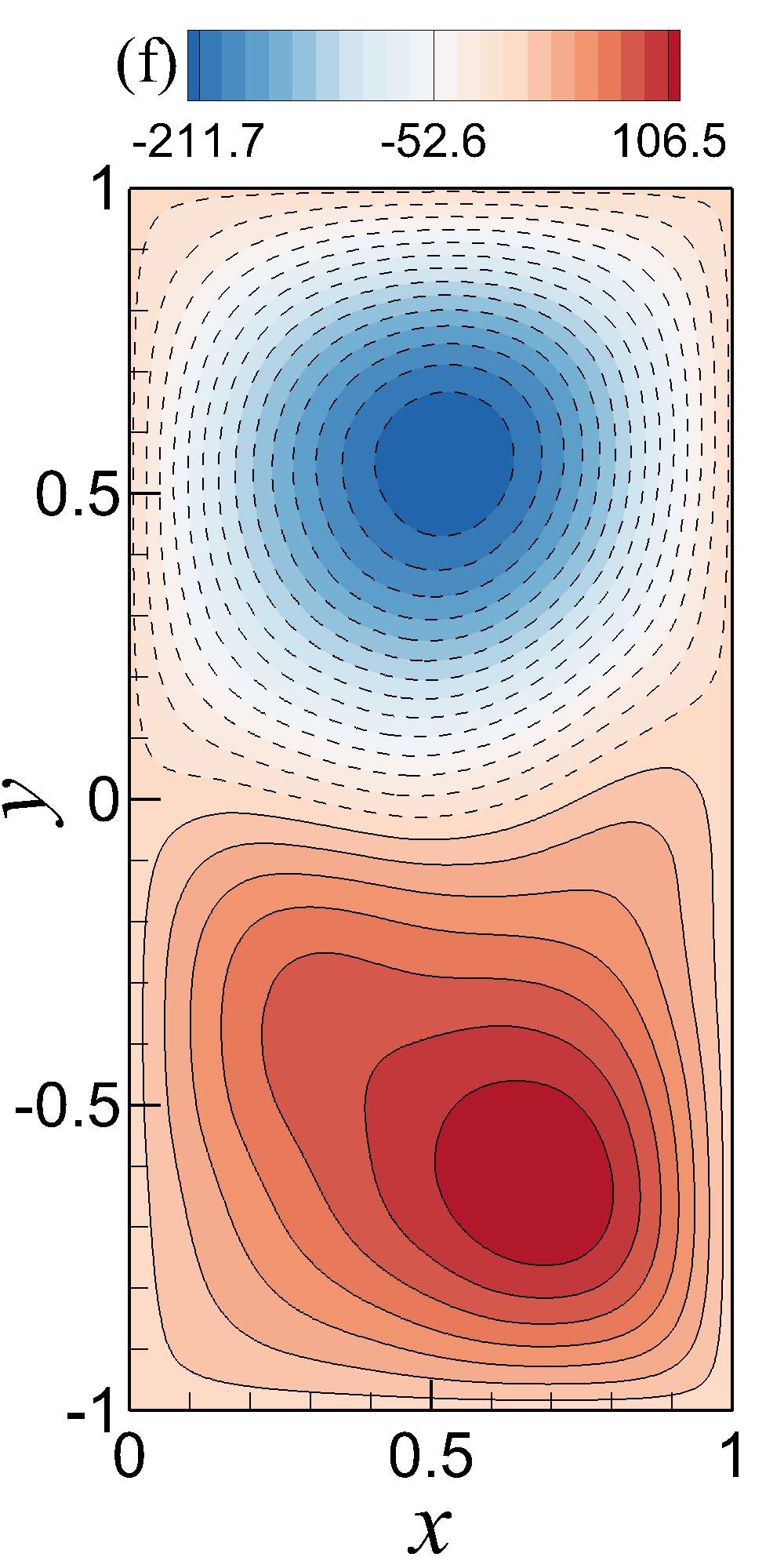}}
\subfigure{\includegraphics[width=0.2\textwidth]{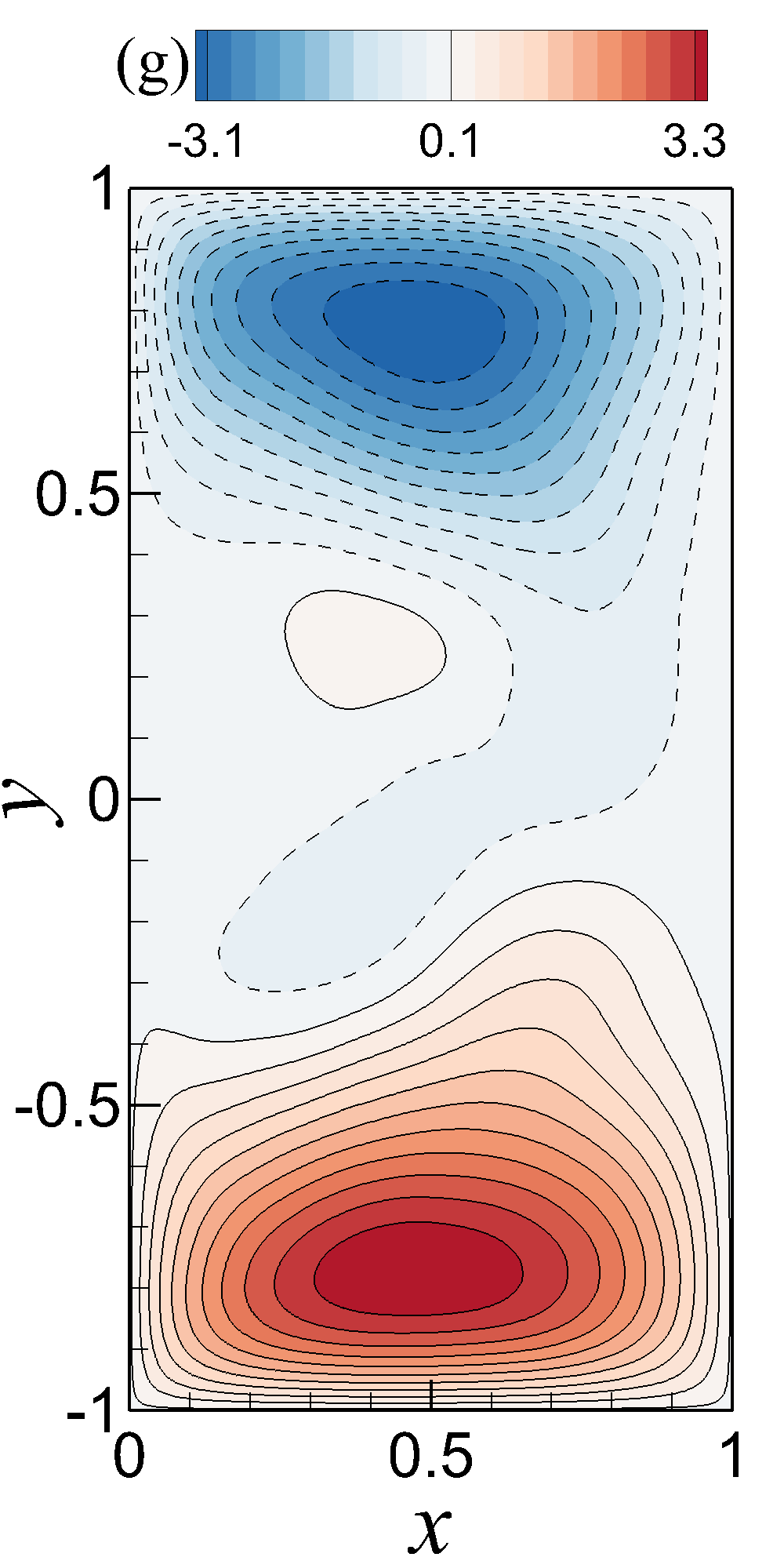}}
\subfigure{\includegraphics[width=0.2\textwidth]{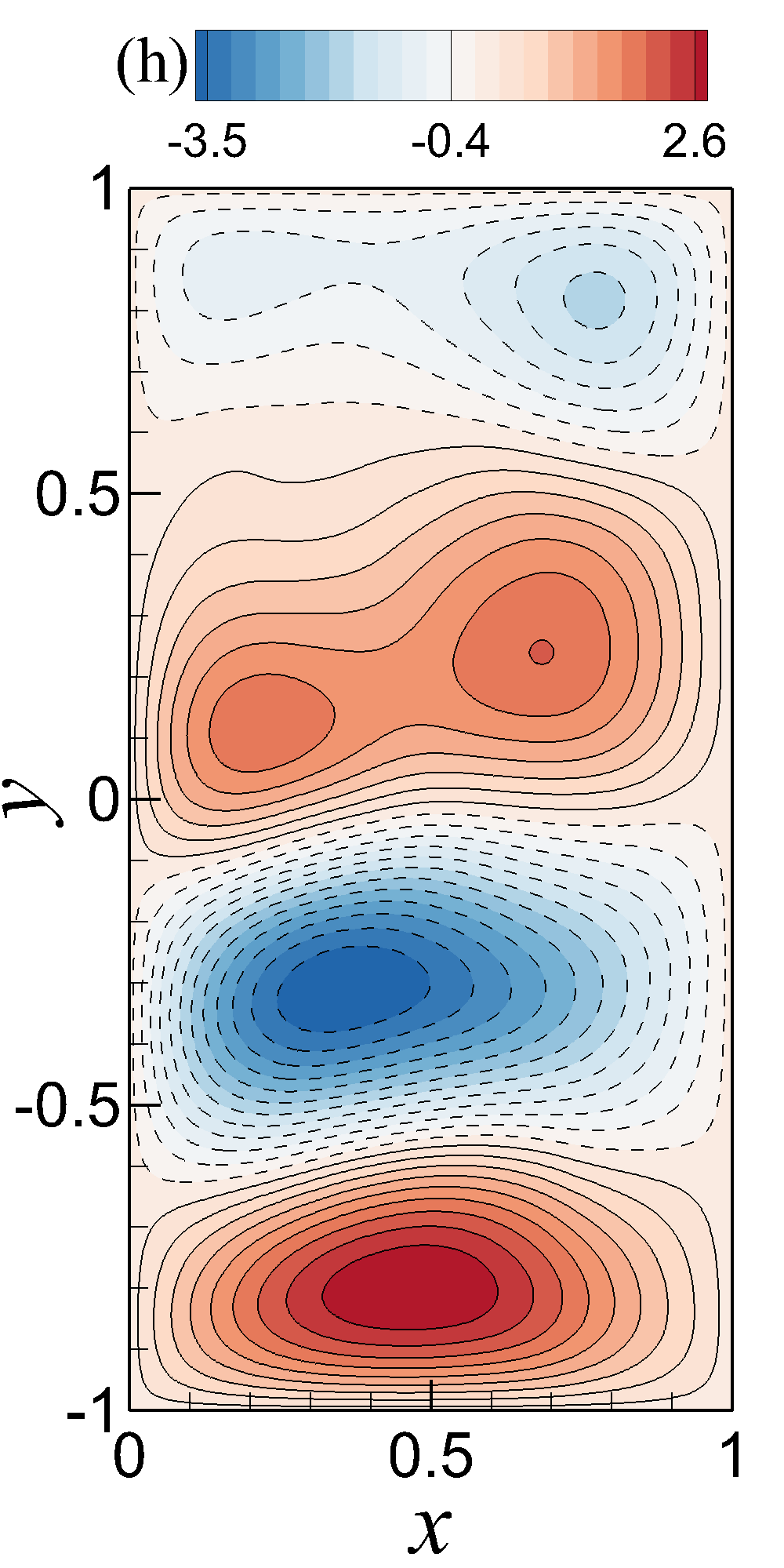}}
}
\caption{Mean streamfunction contours for Experiment III showing the extrapolatory predictive performance at $\mbox{Re}=200$, and $\mbox{Ro}=0.0016$ (i.e, using POD basis functions and mean fields associated with the training data obtained at $\mbox{Re}=450$, and $\mbox{Ro}=0.0036$). (a) FOM at a resolution of $256 \times 512$; (b) ROM-G with $R=50$ modes; (c) ROM-G with $R=40$ modes; (d) ROM-G with $R=30$ modes; (e) ROM-G with $R=20$ modes; (f) ROM-G with $R=10$ modes; (g) proposed ROM-D with $R=20$ modes and $\Delta R=3$; (h) proposed ROM-D with $R=10$ modes and $\Delta R=3$. Note that $\Delta R = R - \tilde{R}$.}
\label{fig:10}
\end{figure*}

\begin{figure*}[htbp]
\centering
\mbox{
\subfigure{\includegraphics[width=0.9\textwidth]{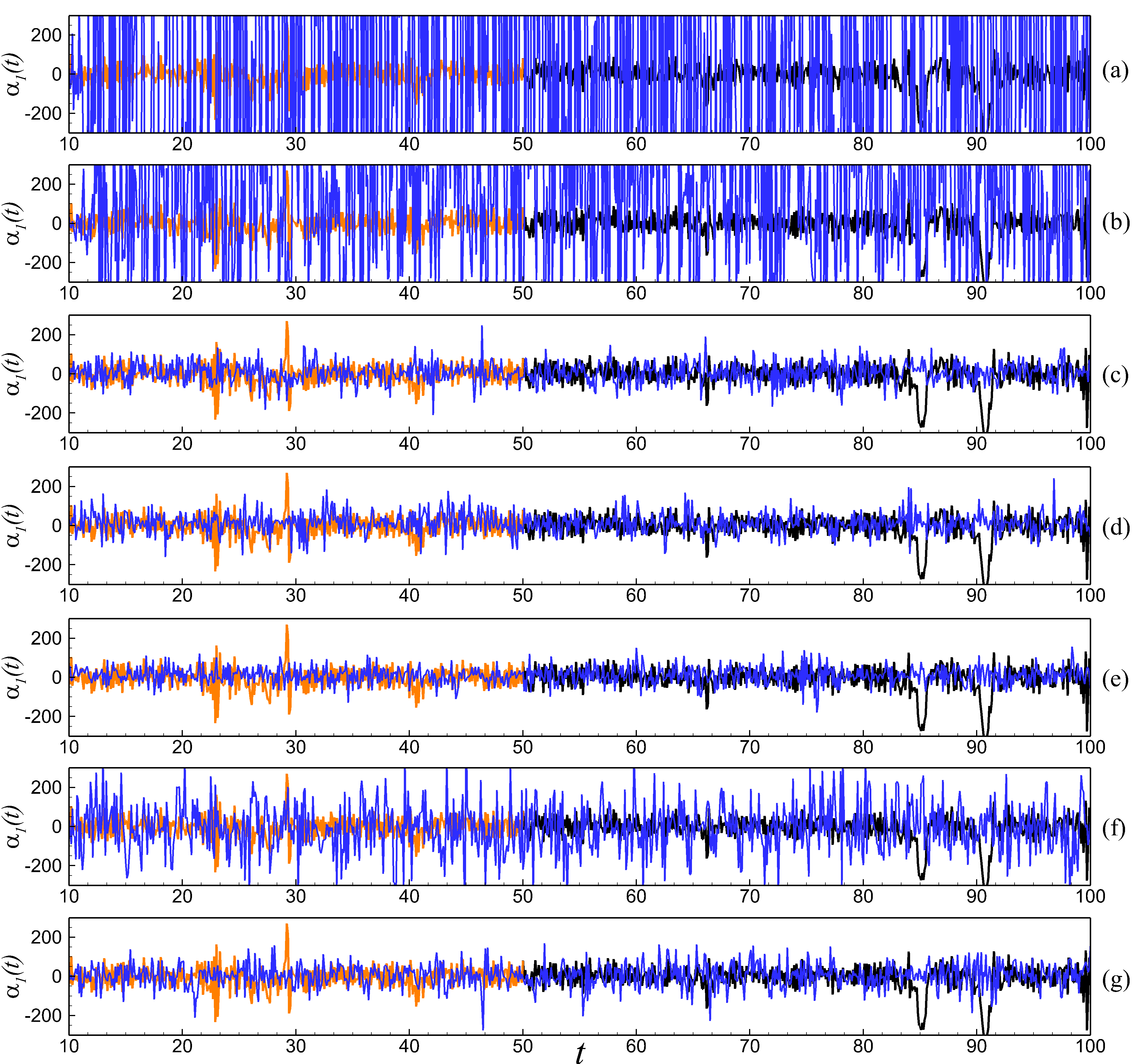}}
}
\caption{Time series of the first modal coefficient for Experiment III showing the extrapolatory predictive performance at $\mbox{Re}=200$, and $\mbox{Ro}=0.0016$ (i.e, using POD basis functions and mean fields associated with the training data obtained at $\mbox{Re}=450$, and $\mbox{Ro}=0.0036$). (a) ROM-G with $R=10$ modes; (b) ROM-G with $R=20$ modes; (c) ROM-G with $R=30$ modes; (d) ROM-G with $R=40$ modes; (e) ROM-G with $R=50$ modes; (f) proposed ROM-D with $R=10$ modes and $\Delta R=3$; (g) proposed ROM-D with $R=20$ modes and $\Delta R=3$. Note that $\Delta R = R - \tilde{R}$. True projection data is underlined in each figure with orange (training zone) and black (extended zone).}
\label{fig:11}
\end{figure*}

The mean streamfunction plots for this set up is presented in FIG.~\ref{fig:10} which show the FOM simulation result at (Re $=200$, Ro $=0.0016$) on the top left corner of the figure. In the same figure, we can also observe that the physical four-gyre pattern is captured by the standard ROM-G with $R = 50$ and $R = 40$. However, retaining the lower $R$ number of modes might fail to capture the true physics. On the contrary, the ROM-D model shows a hint of capturing the four-gyre for $R = 20$. It is evident from the figure that the ROM-D with lower $R$ values clearly showing a better prediction than the ROM-G with higher $R$ values. Eventually, we present the time series evolution of the first modal coefficient plots for Experiment III in FIG.~\ref{fig:11} to get a lucid idea of the underlying physics. It can be observed that the ROM-D with $R = 10$ exhibits the fluctuations with larger amplitude, but the $R = 20$ yields a comparatively better estimation. Also, we can see that the estimations of ROM-G models are showing a good prediction for $R = 40$ and $R = 50$. Even so, if we only compare the performance of both ROM-G and ROM-D for the same retained number of modes (e.g., $R = 20$), we can clearly identify the difference in performance.

\subsection{Experiment IV: Data collection at Re $\mathbf{=200}$, Ro $\mathbf{=0.0016}$, prediction at Re $\mathbf{=450}$, Ro $\mathbf{=0.0036}$}

\begin{figure*}[htbp]
\centering
\mbox{
\subfigure{\includegraphics[width=0.2\textwidth]{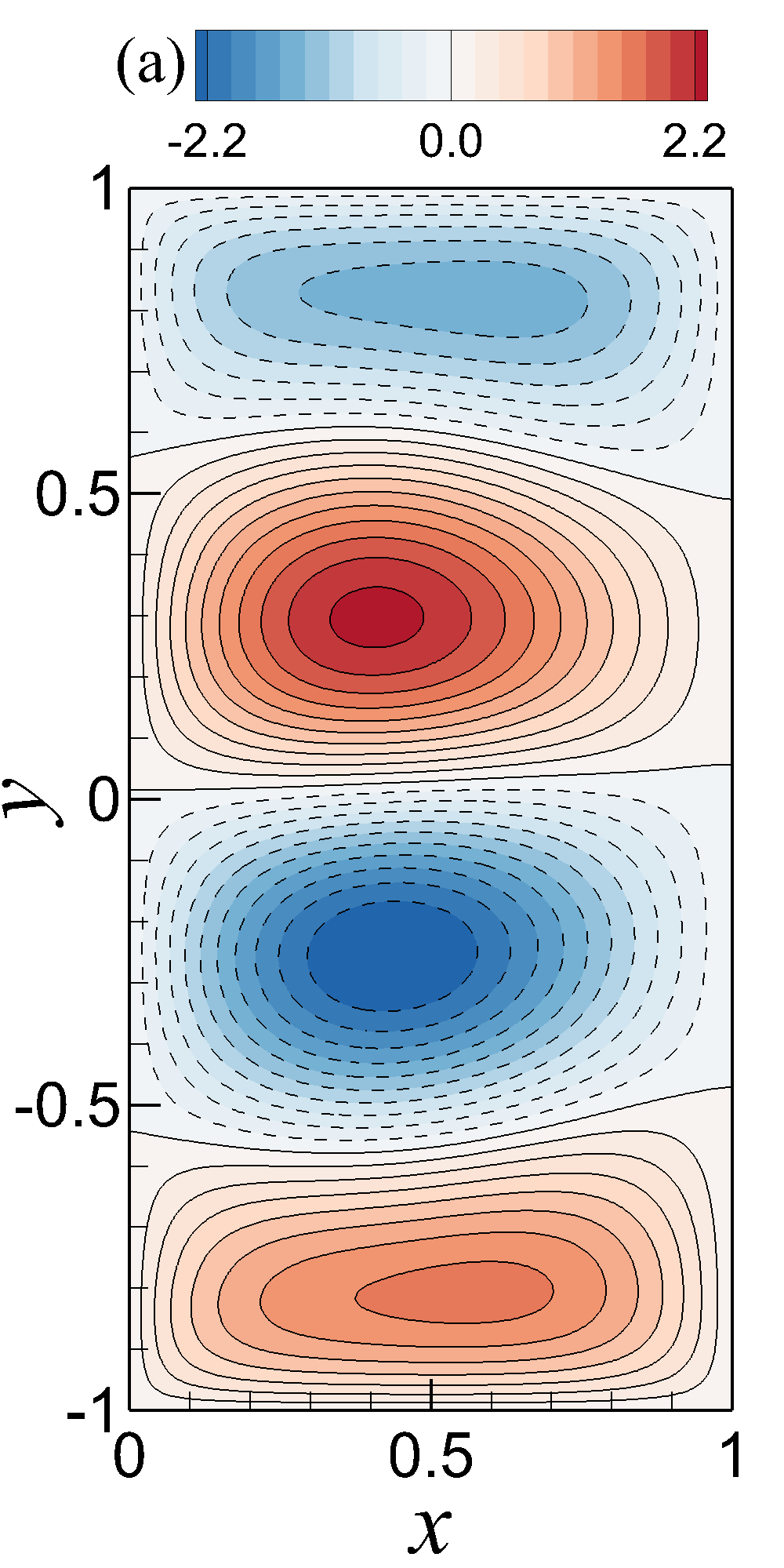}}
\subfigure{\includegraphics[width=0.2\textwidth]{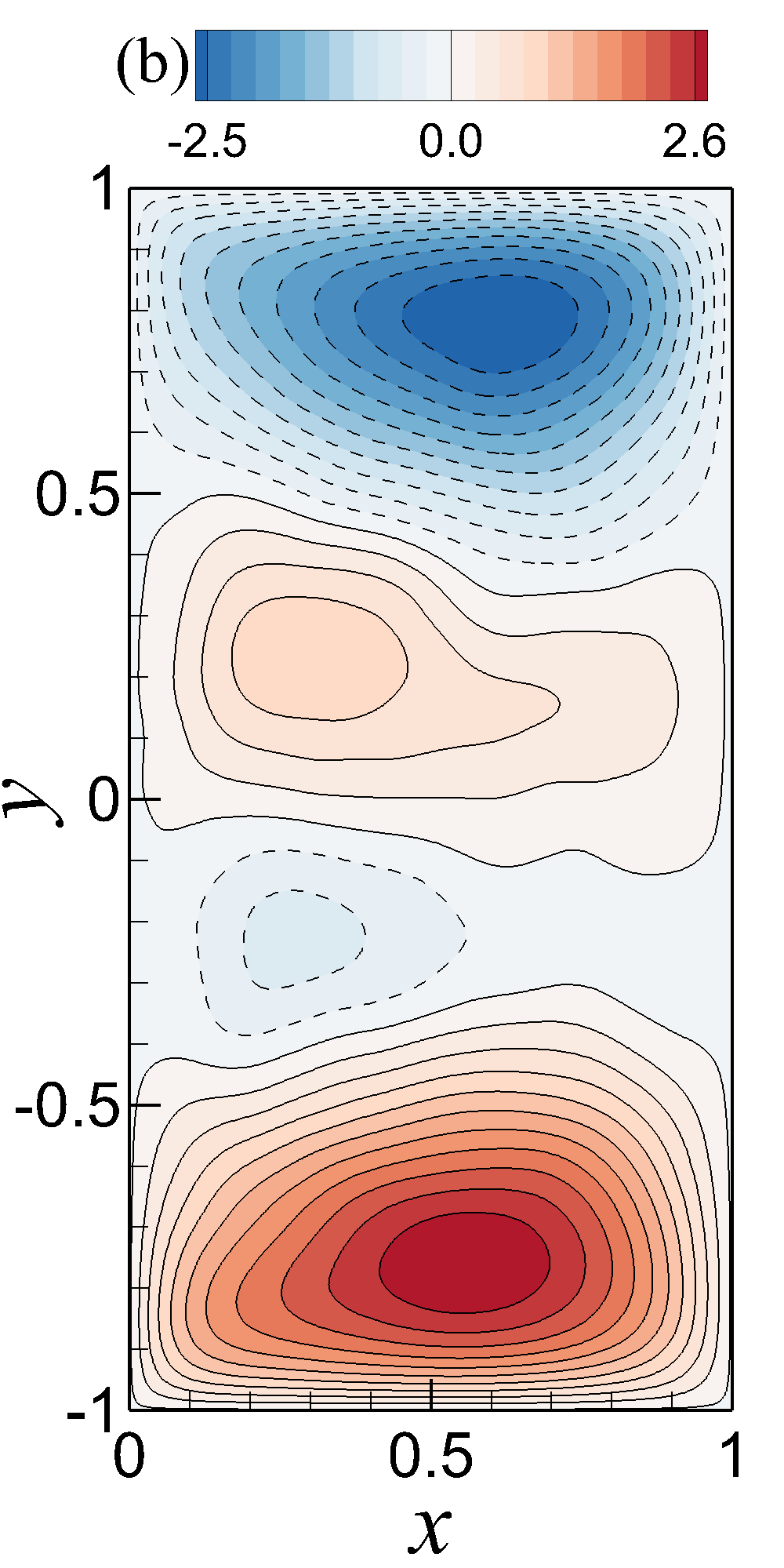}}
\subfigure{\includegraphics[width=0.2\textwidth]{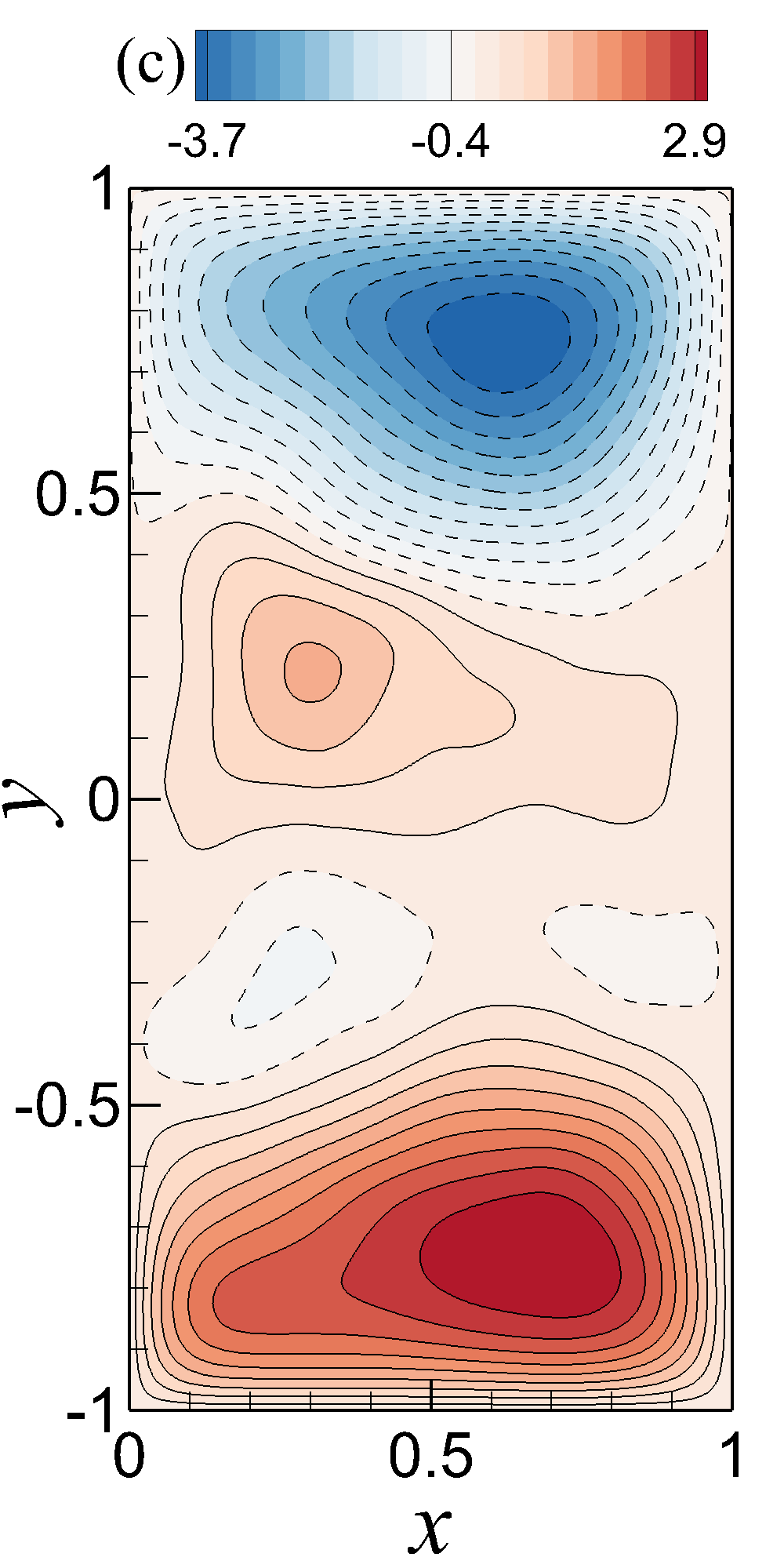}}
\subfigure{\includegraphics[width=0.2\textwidth]{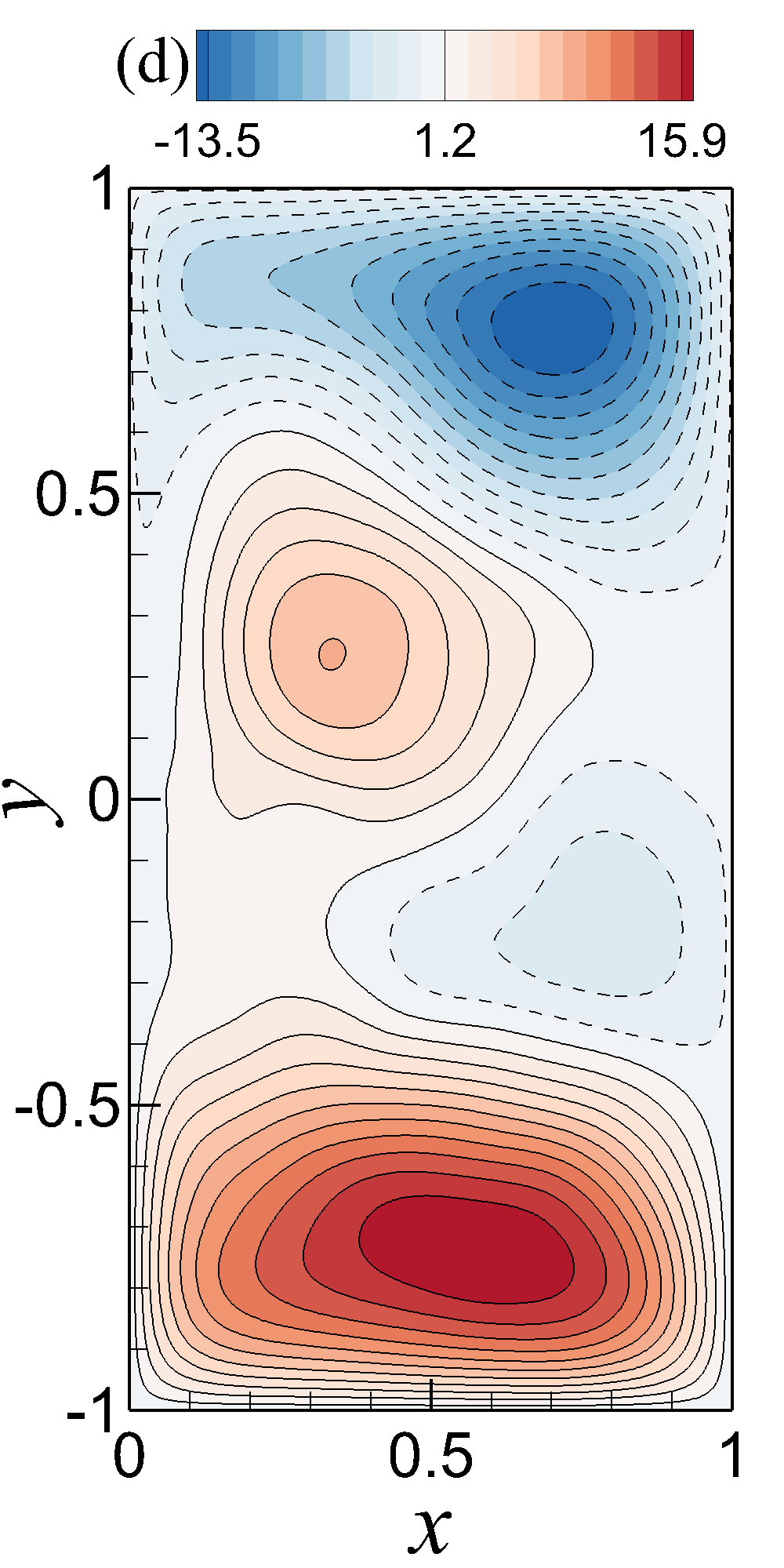}}
}\\
\mbox{
\subfigure{\includegraphics[width=0.2\textwidth]{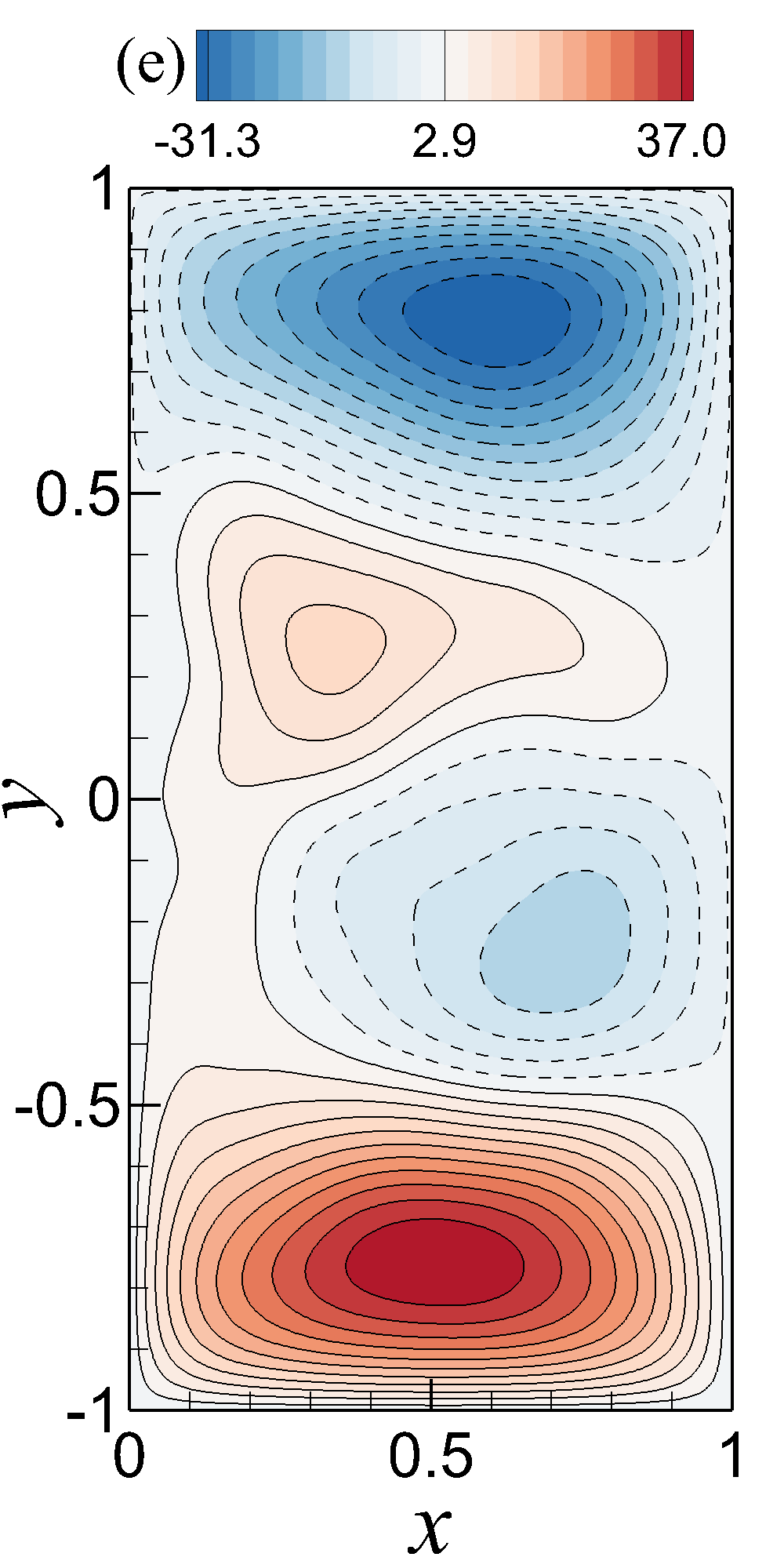}}
\subfigure{\includegraphics[width=0.2\textwidth]{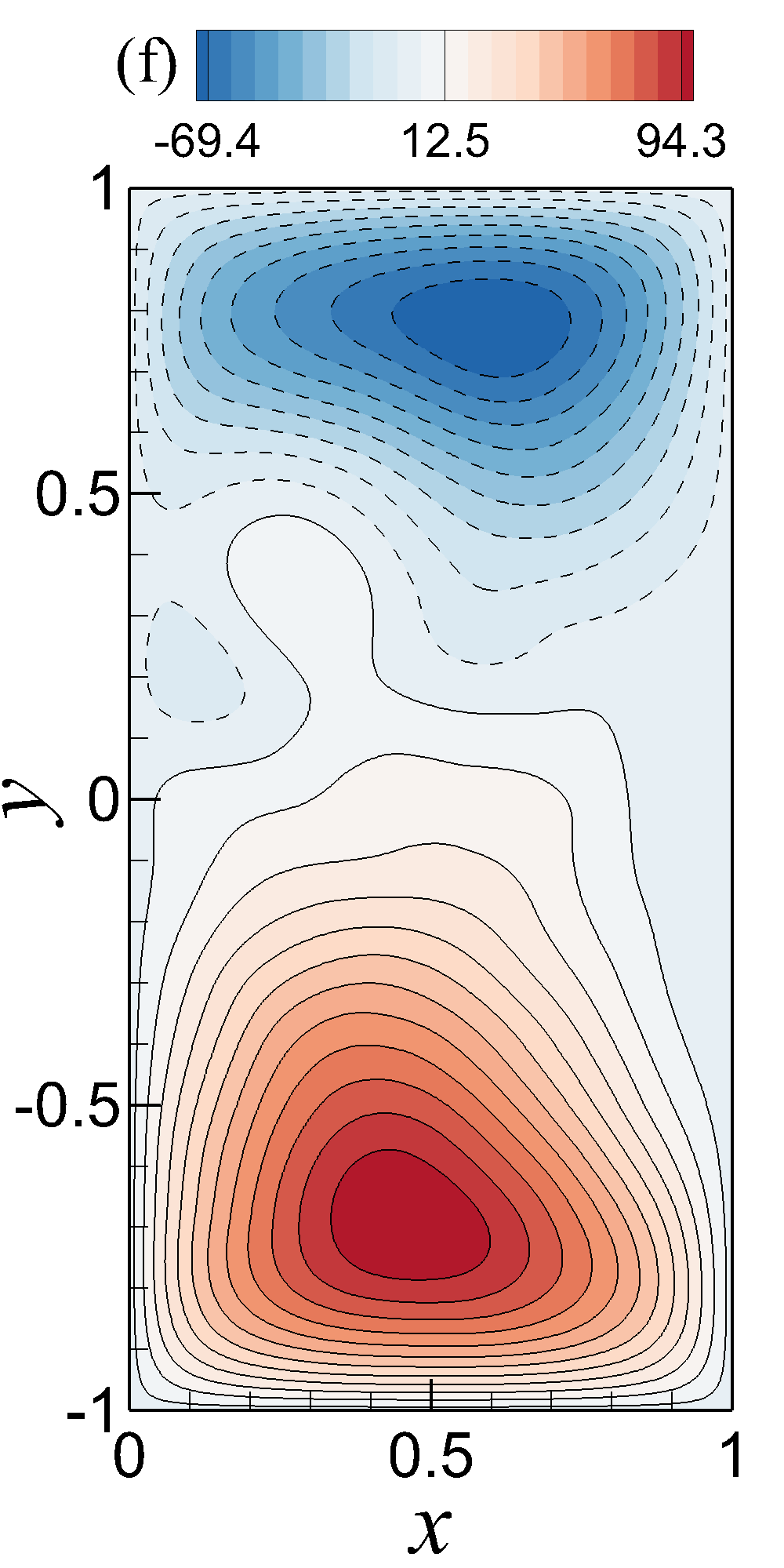}}
\subfigure{\includegraphics[width=0.2\textwidth]{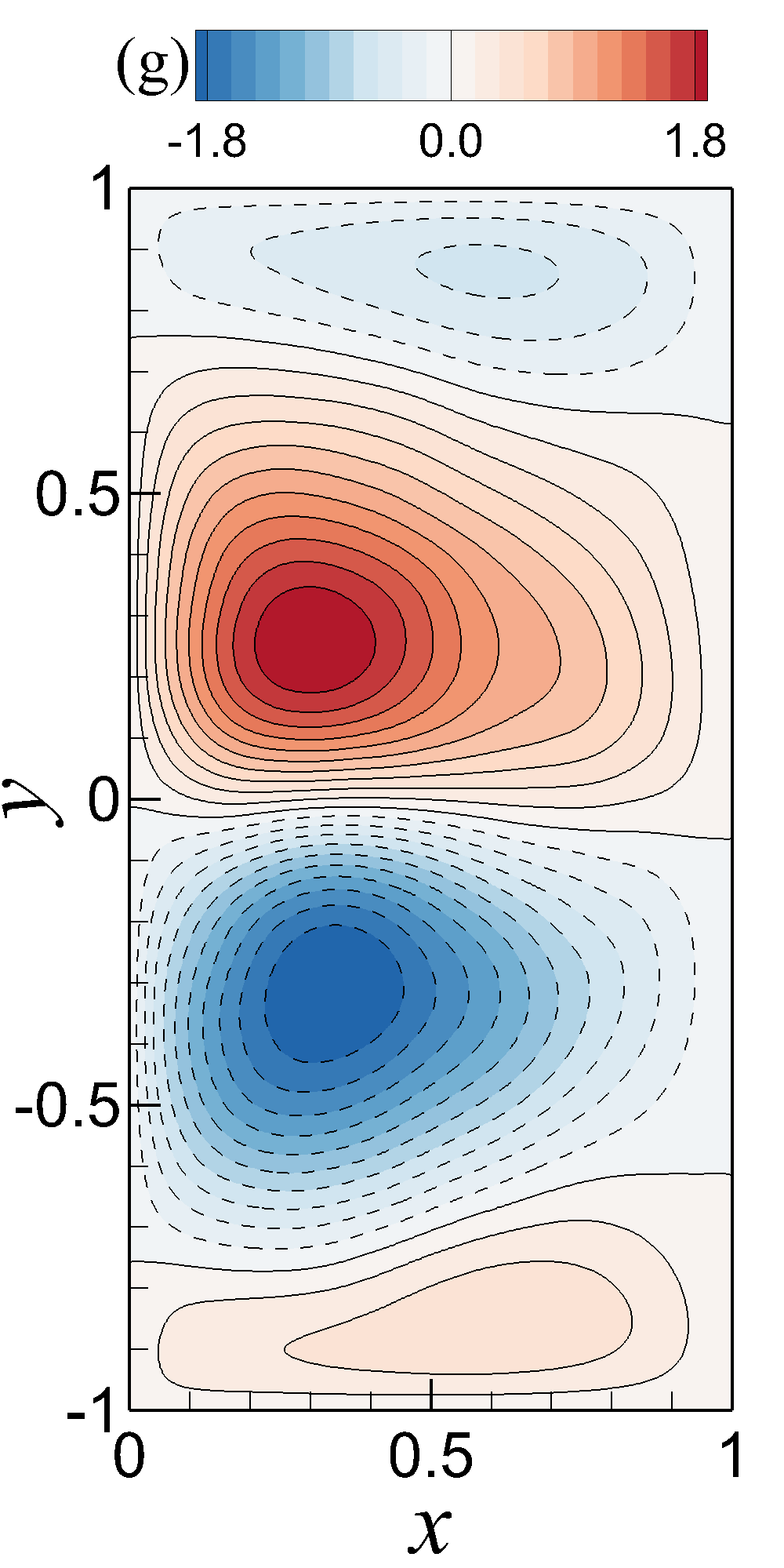}}
\subfigure{\includegraphics[width=0.2\textwidth]{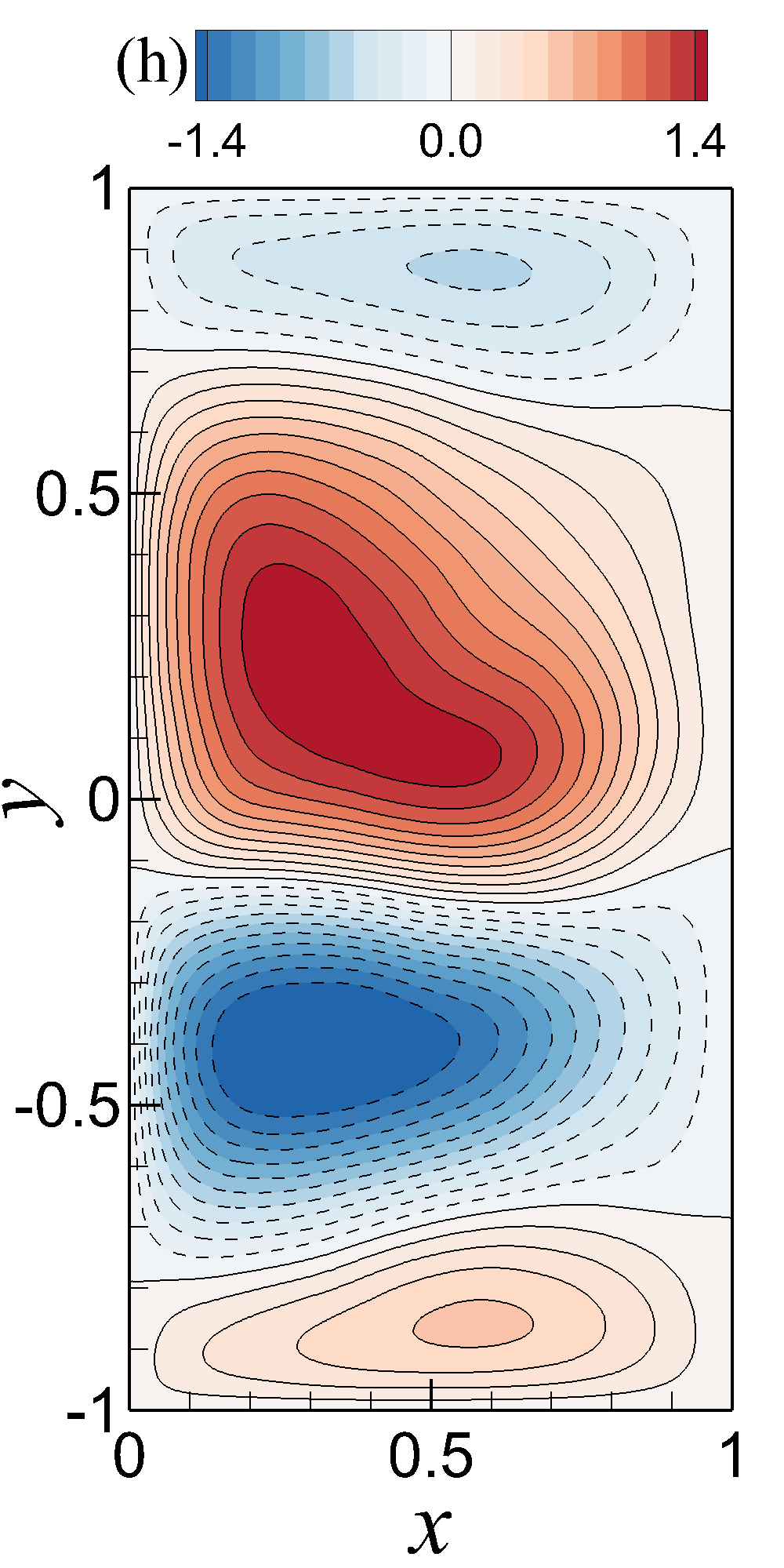}}
}
\caption{Mean stream function contours for Experiment IV showing the extrapolatory predictive performance at $\mbox{Re}=450$, and $\mbox{Ro}=0.0036$ (i.e, using POD basis functions and mean fields associated with the training data obtained at $\mbox{Re}=200$, and $\mbox{Ro}=0.0016$). (a) FOM at a resolution of $256 \times 512$; (b) ROM-G with $R=50$ modes; (c) ROM-G with $R=40$ modes; (d) ROM-G with $R=30$ modes; (e) ROM-G with $R=20$ modes; (f) ROM-G with $R=10$ modes; (g) proposed ROM-D with $R=20$ modes and $\Delta R=3$; (h) proposed ROM-D with $R=10$ modes and $\Delta R=3$. Note that $\Delta R = R - \tilde{R}$.}
\label{fig:12}
\end{figure*}

\begin{figure*}[htbp]
\centering
\mbox{
\subfigure{\includegraphics[width=0.9\textwidth]{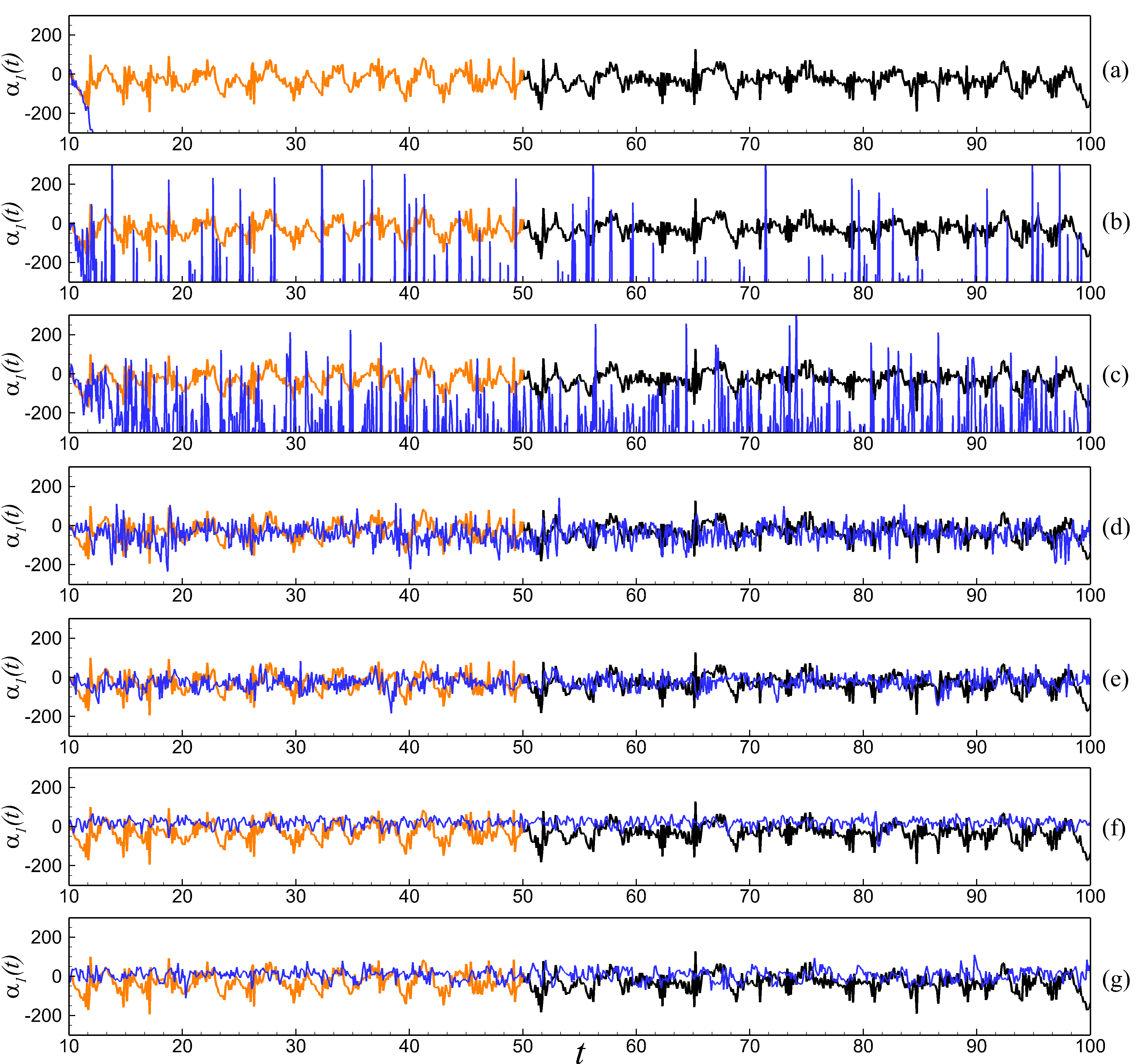}}
}
\caption{Time series of the first modal coefficient for Experiment IV showing the extrapolatory predictive performance at $\mbox{Re}=450$, and $\mbox{Ro}=0.0036$ (i.e, using POD basis functions and mean fields associated with the training data obtained at $\mbox{Re}=200$, and $\mbox{Ro}=0.0016$). (a) ROM-G with $R=10$ modes; (b) ROM-G with $R=20$ modes; (c) ROM-G with $R=30$ modes; (d) ROM-G with $R=40$ modes; (e) ROM-G with $R=50$ modes; (f) proposed ROM-D with $R=10$ modes and $\Delta R=3$; (g) proposed ROM-D with $R=20$ modes and $\Delta R=3$. Note that $\Delta R = R - \tilde{R}$. True projection data is underlined in each figure with orange (training zone) and black (extended zone). }
\label{fig:13}
\end{figure*}

Our final experiment is based on the prediction at (Re $=450$, Ro $=0.0036$) flow configuration using the data snapshots at (Re $=200$, Ro $=0.0016$). Following the similar analyses to Experiment III, here we put an effort to understand the extrapolatory predictive behavior of the ROM-D model for the opposite test condition than the previous experiment. In FIG.~\ref{fig:12}, we can see in the mean streamfunction contours that the ROM-D model is showing a good prediction of the FOM solution for both $R = 10$ and $R = 20$ by capturing the four-gyre circulation pattern. Additionally, ROM-G with higher $R$ also displays an accurate prediction of the true solution. The time series evolution plots in FIG.~\ref{fig:13} shows that the ROM-D solutions are showing a slight phase shift from true projection states. However, the amplitude of the fluctuation for ROM-D with $R = 10$ is comparatively smaller, unlike the scenario in Experiment III, than the true projection fluctuation amplitudes. Though we observe a very good prediction of the true solution using ROM-G with $R = 50$, we can see the the ROM-D predictions for lower $R$ values are impressive compared to the ROM-G solutions for $R = 10$, $R = 20$ and $R = 30$. 

\section{Summary and Conclusions}
\label{sec:con}

In this work, we put forth a dynamic closure modeling framework for reduced order models (ROM-D) based on the idea of test truncation, analogous to the test filtering in dynamic eddy viscosity model in LES, in order to stabilize the ROM for systems with complex flow dynamics. Previously, it has been shown that the stabilization of ROM for large scale turbulent flows can be achieved using an eddy viscosity parameter with an optimal value that can improve the predictive performance of the model significantly \cite{san2015stabilized}. Inspired by the LES and ROM analogy, we devise the proposed ROM-D framework which takes into account the stabilization parameter in an automated fashion at each time step and dynamically stabilize the system of any flow condition without any external computation of optimal eddy viscosity. Through a series of numerical experiments, we show that the predictive performance of the ROM-D framework is not only promising but also consistent for different flow conditions. All performance evaluations of the ROM-D framework are done with respect to the standard ROM-G models with various levels of complexity. As a benchmark test case for our numerical investigations, we consider the BVE describing the single-layer QG ocean model. For data snapshots collection and comparison purpose, we use the Munk layer resolving FOM simulation results obtained at $256 \times 512$ resolution. 

We perform our numerical assessments based upon two different flow conditions which are (i) Re $=450$, Ro $=0.0036$, and (ii) Re $=200$, Ro $=0.0016$. Based on the existing literature and our findings in FOM simulations, the BVE model driven by two-gyre wind forcing shows a four-gyre circulation pattern in time mean once the model reaches the turbulent equilibrium state under these conditions. Physically, this means the balance between the wind stress curl forcing and the eddy flux of potential vorticity. For this reason, we investigate both mean streamfunction contour and time series evolution (after the statistically steady state achieved) plots to get a clear understanding on the simulation results in capturing the four-gyre pattern or attaining the statistically steady state. Owing to the wide range of temporal and spatial scales of the QG test problem, it is observed in the POD analysis that $50$ POD modes ($R = 50$) capture around $80 \%$ of total energy of the system for lower (Re, Ro) combination whereas capture around $85 \%$ of total energy of the system for higher (Re, Ro) combination. For the experiments on both flow conditions, we come to the main conclusion that the ROM-D model with $R = 10$ predicts the true solution with a same order accuracy as a ROM-G solution with $R = 80$. This implies considerable savings in terms of both storage needs and computing time.  

Since the ROM-D model includes dissipative contributions from truncated modes through the stabilization parameter, it is expected to capture a greater percentage of the energy in the system using lower $R$ values in ROM-D model. As a result, we can see a huge reduction in overall computational overhead for equally accurate solution using ROM-D model. Moreover, the higher value of $R$ combined with different values of $\Delta R$ in ROM-D exhibits more gain in accuracy of the predictions. To demonstrate the robustness of the ROM-D model, we also perform the sensitivity analysis on both flow conditions which display a consistent prediction for $R = 20$ irrespective to different $\Delta R$ values. Finally, we test the extrapolatory predictive performances for both ROM frameworks which reveal a better prediction of FOM solution by the ROM-D model than the ROM-G model for same value of $R$. Based on the numerical experiments and analyses above, needless to say, the ROM-D framework has a great potential for efficient model order reduction of complex turbulent flow problems.


\begin{acknowledgements}
This material is based upon work supported by the U.S. Department of Energy, Office of Science, Office of Advanced Scientific Computing Research under Award Number DE-SC0019290. O.S. gratefully acknowledges their support. Direct numerical simulations for this project were performed using resources of the Oklahoma State University High Performance Computing Center. Disclaimer: This report was prepared as an account of work sponsored by an agency of the United
States Government. Neither the United States Government nor any agency thereof, nor any of their
employees, makes any warranty, express or implied, or assumes any legal liability or responsibility for the accuracy, completeness, or usefulness of any information, apparatus, product, or process disclosed, or represents that its use would not infringe privately owned rights. Reference herein to any specific commercial product, process, or service by trade name, trademark, manufacturer, or otherwise does not necessarily constitute or imply its endorsement, recommendation, or favoring by the United States Government or any agency thereof. The views and opinions of authors expressed herein do not necessarily state or reflect those of the United States Government or any agency thereof.
\end{acknowledgements}

\section*{Appendix}

\subsection{Time Advancement Scheme}

We apply a conservative finite difference formulation for our numerical framework in the current study. Using method of lines, we cast the governing equation given by Eq.~(\ref{eq:nbve}) in the following semi-discretized ordinary differential equations form:
\begin{align}\label{ODE1}
  \frac{d \omega_{i,j}}{dt} = \pounds_{i,j},
\end{align}
where subscripts $i$ and $j$ represent the discrete spatial indices in $x$- and $y$-directions, respectively. Here, $\pounds_{i,j}$ denotes the discrete spatial derivative operators. For our time advancement scheme, we apply a third order Runge-Kutta scheme given as \cite{gottlieb1998total}:
\begin{eqnarray} \label{RK3}
&\omega_{i,j}^{(1)} = \omega_{i,j}^{(n)} + \Delta t \pounds_{i,j}^{(n)}, \nonumber \\
&\omega_{i,j}^{(2)}  = \frac{3}{4}  \omega_{i,j}^{(n)} + \frac{1}{4} \omega_{i,j}^{(1)} + \frac{1}{4} \Delta t \pounds_{i,j}^{(1)}, \\
&\omega_{i,j}^{(n+1)} =  \frac{1}{3} \omega_{i,j}^{(n)} + \frac{2}{3} \omega_{i,j}^{(2)} + \frac{2}{3} \Delta t \pounds_{i,j}^{(2)}, \nonumber 
\end{eqnarray}
where $\Delta t = 2.5\times10^{-5}$ for our FOM simulations to satisfy the numerical stability criteria through the Courant--Freidrichs--Lewy (CFL) number. However, in the time integration of ROMs, we use the same algorithm to compute $\alpha_{k}^{(n+1)}$ from $\alpha_{k}^{(n)}$ 
\begin{eqnarray} \label{RK3r}
&\alpha_{k}^{(1)} = \alpha_{k}^{(n)} + \Delta t \mathfrak{R}_{k}^{(n)}, \nonumber \\
&\alpha_{k}^{(2)}  = \frac{3}{4}  \alpha_{k}^{(n)} + \frac{1}{4} \alpha_{k}^{(1)} + \frac{1}{4} \Delta t \mathfrak{R}_{k}^{(1)}, \\
&\alpha_{k}^{(n+1)} =  \frac{1}{3} \alpha_{k}^{(n)} + \frac{2}{3} \alpha_{k}^{(2)} + \frac{2}{3} \Delta t \mathfrak{R}_{k}^{(2)}, \nonumber 
\end{eqnarray}
where $\mathfrak{R}_{k}$ refers to the right-hand-side of Eq.~(\ref{eq:srom2}), we specify larger time step, $\Delta t = 2.5\times10^{-4}$, since there is no such CFL stability constraints in the time advancement of ROMs. Although not shown here, we have verified that stable and accurate solutions can still be obtained by much bigger $\Delta t$ using the ROM-D model.

\subsection{Numerical Discretizations}

The source term, $\pounds_{i,j}$ in Eq.~(\ref{ODE1}) includes nonlinear convective terms as well as the linear rotational and diffusive terms which can be written as:
\begin{align}\label{eq:disc}
\pounds_{i,j} &=- J(\omega_{i,j},\psi_{i,j}) \nonumber \\ &+\frac{1}{\mbox{Ro}}\frac{\partial \psi_{i,j}}{\partial x} + \frac{1}{\mbox{Re}}\nabla^2 \omega_{i,j} + \frac{1}{\mbox{Ro}}\sin(\pi y_{i,j}),
\end{align}
where we use the standard second-order central finite difference schemes for linear terms. Therefore, the derivative operators in Eq.~(\ref{eq:disc}) can be written in following discrete form:
\begin{align}\label{eq:numdiff}
\frac{\partial \psi_{i,j}}{\partial x} &= \frac{\psi_{i+1,j} - \psi_{i-1,j}}{2 \Delta x}, \\
\nabla^2 \omega_{i,j} &= \frac{\omega_{i+1,j} - 2\omega_{i,j} + \omega_{i-1,j}}{\Delta x^2} \nonumber \\
&+ \frac{\omega_{i,j+1} - 2\omega_{i,j} + \omega_{i,j-1}}{\Delta y^2},
\end{align}
where $\Delta x$ and $\Delta y$ are the step sizes in $x$- and $y$-directions, respectively. For the modeling of nonlinear term, we use second-order Arakawa scheme \cite{arakawa1966computational} to avoid computational instabilities arising from nonlinear interactions for the Jacobian term, $J(\omega_{i,j}, \psi_{i,j})$ in Eq.~(\ref{eq:disc}), defined as
\begin{equation}
  J(\omega_{i,j}, \psi_{i,j}) = \frac{1}{3}(J_1 + J_2 + J_3),
\end{equation}
where the discrete parts of the Jacobians have the following forms:
\begin{align}
J_1 =& \frac{1}{4 {\Delta x \Delta y}} [(\omega_{i+1,j} - \omega_{i-1,j})(\psi_{i,j+1} - \psi_{i,j-1}) \nonumber \\ &- (\omega_{i,j+1} - \omega_{i,j-1})(\psi_{i+1,j} - \psi_{i-1,j})],
\end{align}

\begin{align}
J_2 =& \frac{1}{4 {\Delta x \Delta y}} [\omega_{i+1,j}(\psi_{i+1,j+1} - \psi_{i+1,j-1}) \nonumber \\
&- \omega_{i-1,j}(\psi_{i-1,j+1} - \psi_{i-1,j-1}) - \omega_{i,j+1}(\psi_{i+1,j+1} \nonumber \\
&- \psi_{i-1,j+1}) - \omega_{i,j-1}(\psi_{i+1,j-1} - \psi_{i-1,j-1})],
\end{align}

\begin{align}
J_3 =& \frac{1}{4 {\Delta x \Delta y}} [\omega_{i+1,j+1}(\psi_{i,j+1} - \psi_{i+1,j}) \nonumber \\
&- \omega_{i-1,j-1}(\psi_{i-1,j} - \psi_{i,j-1}) - \omega_{i-1,j+1}(\psi_{i,j+1}\nonumber \\
&- \psi_{i-1,j}) - \omega_{i+1,j-1}(\psi_{i+1,j} - \psi_{i,j-1})].
\end{align}

We refer the article by San and Illiescu \cite{san2015stabilized} for reader's interest on the derivation of higher-order Arakawa schemes. In addition, we require to solve an elliptic Poisson equation at each substep in the time integration to find the streamfunction from the updated vorticity values which is the most computationally heavy part of our QG model solver. Since our computational domain is simple and uniform in this study, we utilize the fast Fourier transform (FFT) method to solve the Poisson equation. In our framework, not to deviate from the main focus of this paper, we use the FFT algorithm given by Press et al. \cite{press1992numerical} to compute the forward and inverse sine transforms. The formulation of our FFT algorithm can be found in the referred article \cite{san2015stabilized}. We must note here that even though an FFT based Poisson solver reduces the overall computational burden of elliptic equations, it is limited to simple geometries on structured grids. Alternatively, a multigrid Poisson solver might be a better option for more complex basin problems which works well in complex geometries \cite{san2013coarse}. 

\subsection{Numerical Integration}

As stated before, we take the inner products of two quantities to compute the POD modes in our ROM frameworks which is done by taking the
integral of the product over the domain $\Omega$. For the two-dimensional numerical integration over $\Omega$, we use the fourth-order accurate Simpson's $1/3^{rd}$ rule \cite{moin2010fundamentals,hoffman2001numerical}. To illustrate the integration technique, we can consider $f(x, y) = u(x, y)v(x, y)$ so that it can be written as follows
\begin{align}
\langle u;v \rangle &= \int_{\Omega}{f(x,y)dxdy} \nonumber \\
&=\dfrac{\Delta y}{3}\sum_{j=1}^{N_y/2}{\left( \hat{f}_{2j-1} + 4\hat{f}_{2j} + \hat{f}_{2j+1}   \right)},
\end{align}
where
\begin{align}
\hat{f}_{j} & =  \dfrac{\Delta x}{3} \sum_{i=1}^{N_x/2}{\left( f_{2i-1,j} + 4f_{2i,j} + 4f_{2i+1,j}  \right)}, \\ j & =1,2, ..., N_y \nonumber.
\end{align}
Here, $N_x$ and $N_y$ are the number grid points in both $x$- and $y$-directions, respectively. For valid implementation of Simpson's $1/3^{rd}$ rule, the number of intervals in the $x$- and $y$-directions should be even. Also, the uniform interval sizes in each direction is assumed in the above formulae.

\subsection{Proper Orthogonal Decomposition Modes}

A number of snapshots of the 2D vorticity field, denoted as $\omega(x,y,t_n)$, are stored at consecutive times $t_n$ for $n=1,2,\dots,N$. The time-averaged field, called ``base flow", can be computed as
\begin{equation}
\bar{\omega}(x,y)=\dfrac{1}{N}\sum_{n=1}^{N}\omega(x,y,t_n).
\end{equation}
The mean-subtracted snapshots, also called as anomaly or fluctuation fields, are then computed as the difference between the instantaneous field and the mean field
\begin{equation}
\omega'(x,y,t_n)=\omega(x,y,t_n)-\bar{\omega}(x,y).
\end{equation}
This subtraction has been common in ROM community, and it guarantees that ROM solution would satisfy the same boundary conditions as full order model \cite{chen2012variants}. This anomaly field procedure can be also interpreted as a mapping of snapshot data to its origin.  

For generating the POD modes, we are following the standard method of snapshots proposed by Sirovich \cite{sirovich1987turbulence} as an efficient numerical procedure to save time in solving the eigenvalue problem necessary for POD, when the data dimension is much bigger than number of snapshots. First, an $N\times N$ snapshot data matrix $\mathbf{A}=[a_{ij}]$  is computed from the inner product of mean-subtracted snapshots
\begin{equation}
a_{ij}=\langle \omega'(x,y,t_i); \omega'(x,y,t_j)\rangle
\end{equation}
Then, an eigen decomposition of $\mathbf{A}$ is performed as
\begin{equation}
\mathbf{A} \mathbf{V} = \mathbf{V} \mathbf{\Lambda}
\end{equation} 
where $\mathbf{\Lambda}$ is a diagonal matrix whose entries are the eigenvalues $\lambda_k$ of $\mathbf{A}$, and $\mathbf{V}$ is a matrix whose columns $\mathbf{v}_k$ are the corresponding eigenvectors. In our computations, we use the eigensystem solver based on the Jacobi transformations since $\mathbf{A}$ is a symmetric positive definite matrix \cite{press1992numerical}. It should be noted that eigenvalues need to be arranged in a descending order (i.e., $\lambda_1\ge\lambda_2\ge\dots\ge\lambda_N$), for proper selection of the POD modes. The POD modes $\phi_{k}$ are then computed as
\begin{equation}
\phi_{k}(x,y)=\dfrac{1}{\sqrt{\lambda_k}}\sum_{n=1}^{N} v^{n}_{k} \omega'(x,y,t_n)
\end{equation}
where $v^{n}_{k}$ is the $n$th component of the eigenvector $\mathbf{v}_k$. The scaling factor, $1/\sqrt{\lambda_k}$, is to guarantee the orthonormality of POD modes, i.e., $\langle \phi_i; \phi_j\rangle = \delta_{ij}$.

\bibliography{reference}
\end{document}